\documentclass[nonacm,screen]{acmart}

\AtBeginDocument{%
 \providecommand\BibTeX{{%
    \normalfont B\kern-0.5em{\scshape i\kern-0.25em b}\kern-0.8em\TeX}}}

\PassOptionsToPackage{hyphens}{url}
\PassOptionsToPackage{breaklinks = true,unicode = true,urlcolor = blue,colorlinks = true,citecolor = red,linkcolor = red}{hyperref}

\usepackage{algorithmic}
\usepackage{graphicx}
\usepackage{textcomp}
\usepackage{bmpsize}
\usepackage{xcolor}
\usepackage{lipsum}

\usepackage{caption} 
\usepackage{booktabs} 

\usepackage{color,colortbl}
\usepackage{xcolor}

\PassOptionsToPackage{breaklinks = true,unicode = true,urlcolor = blue,colorlinks = true,citecolor = red,linkcolor = red}{hyperref}
\PassOptionsToPackage{hyphens}{url}

\usepackage{graphicx}

\usepackage{multirow}
\usepackage{array}
\newcolumntype{P}[1]{>{\centering\arraybackslash}p{#1}} 

\usepackage{enumitem}
\usepackage{tabularx}
\usepackage{balance}
\usepackage{epstopdf}

\usepackage{subcaption}
\usepackage{amsfonts}

\usepackage[nameinlink,capitalize]{cleveref} 

\newcommand{\paragraphb}[1]{\medskip\noindent{\bf #1.} }

\newcommand{\ignore}[1]{}

\newcommand{\pr}[1]{\mathrm{Pr}\{#1\}} 

\usepackage{tikz}
\newcommand{\pie}[1]{%
\begin{tikzpicture}
\draw (0,0) circle (1ex);\fill (1ex,0) arc (0:#1:1ex) -- (0,0) -- cycle;\end{tikzpicture}}

\setcopyright{none}

\acmJournal{CSUR}

\begin{document}

\title{Generative Models for Security: Attacks, Defenses, and Opportunities}

\author{Luke A. Bauer}
\affiliation{%
  \institution{University of Florida}
   \country{USA}
}
\email{lukedrebauer@ufl.edu}

\author{Vincent Bindschaedler}
\affiliation{%
  \institution{University of Florida}
  \country{USA}
}
\email{vbindsch@cise.ufl.edu}

\renewcommand{\shortauthors}{Bauer and Bindschaedler}

\begin{abstract}

Generative models learn the distribution of data from a sample dataset and can then generate new data instances. Recent advances in deep learning has brought forth improvements in generative model architectures, and some state-of-the-art models can (in some cases) produce outputs realistic enough to fool humans. 

We survey recent research at the intersection of security and privacy and generative models. In particular, we discuss the use of generative models in adversarial machine learning, in helping automate or enhance existing attacks, and as building blocks for defenses in contexts such as intrusion detection, biometrics spoofing, and malware obfuscation. We also describe the use of generative models in diverse applications such as fairness in machine learning, privacy-preserving data synthesis, and steganography. Finally, we discuss new threats due to generative models: the creation of synthetic media such as deepfakes that can be used for disinformation.

\end{abstract}

\maketitle

\section{Introduction}
Generative models learn to characterize the distribution of data using only samples from it and then generate new data instances from this distribution. Although generative models are not new, the deep learning revolution has reinvigorated research into generative model architectures, and deep generative models using state-of-the-art architectures can now produce output that is sometimes indistinguishable from real-world data. Along with this, comes a host of new issues and opportunities relating to security and privacy. 

This paper provides a comprehensive survey of research at the intersection of generative models and security and privacy. In particular, we describe the recent use of generative models in adversarial machine learning. We also discuss applications such as producing adversarial examples without perturbations, steganography, and privacy-preserving data synthesis. We show how a number of attacks and defenses for various cybersecurity problems such as password generation, intrusion, and malware detection, can benefit from generative models because of their ability to learn the distribution of the training data. By characterizing the data distribution, generative models help practitioners better understand phenomena they are studying and supplement existing attack and defense techniques. Finally, we discuss the extent to which deep generative models present new threats when used to produce synthetic media.

\begin{figure}[tp]
    \centering
        \includegraphics[width=0.95\linewidth]{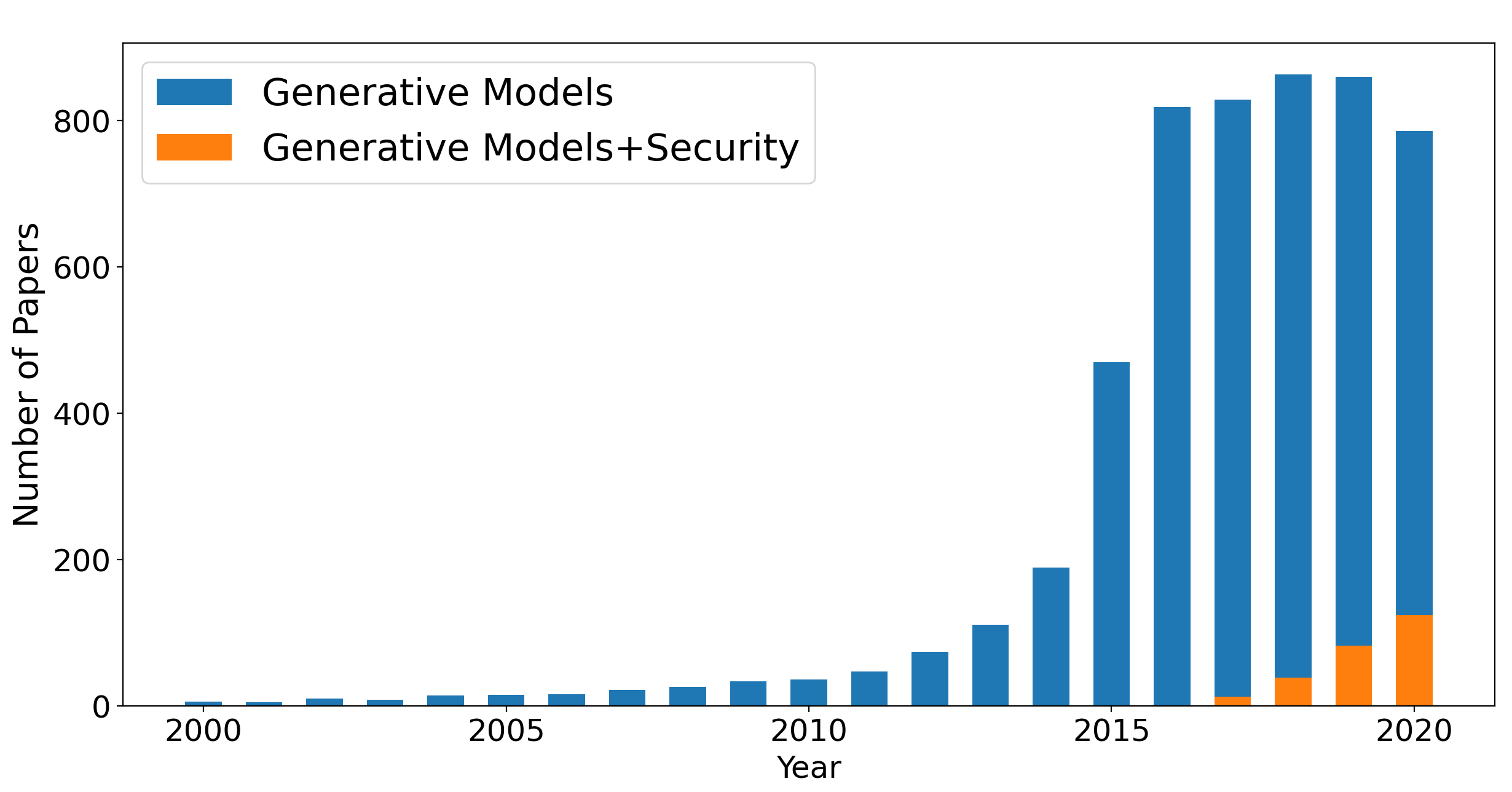}
        \caption{Number of papers (published and pre-prints) on the topic of generative models, and on the topic of generative models {\em and} security from 2000 to 2020 according to Google Scholar. A significant increase of work on generative models can be seen starting in 2015. In 2017, there is also a significant increase of work on generative models {\em and} security.
        Methodology: we gathered the list of papers from Google Scholar using the Publish or Perish program~\cite{harzing2020publish}. Concretely, we performed two queries. Query 1 searches for terms relating to generative models (i.e.: Generative Model or RNN or LSTM or Variational Auto-encoder or GAN) in the body of papers, whereas query 2 searches for terms relating to security/privacy (i.e.: Security or Privacy or Attacks or Defenses) in the title of papers. The first query then provides number of generative model papers for each year, whereas the intersection of the results of both queries provides the number of generative model {\em and} security for each year.
        }
        \label{fig:papersontopic}
\end{figure}

The increased intensity of work at the intersection of generative models and security/privacy is evidenced by a growing body of literature. This is illustrated in \cref{fig:papersontopic}, which shows the number of papers (published and pre-prints) on this topic from 2000 to 2020.

This survey is structured as follows. \cref{sec:background} provides a brief overview of generative model architectures, as well as a discussion of metrics used to evaluate generative models.  \cref{sec:intrusion}, we survey the use of generative models for intrusion detection, malware detection, and biometric spoofing. In~\cref{sec:datasynthesis}, we describe how generative models can be used to create synthetic datasets.
In~\cref{sec:advml}, we discuss generative models in adversarial machine learning. We then describe attacks on generative models~(\cref{sec:attacksongen}). In~\cref{sec:steganography}, we discuss applications of generative models to steganography, whereas in~\cref{sec:fairness} we discuss their application to fairness problems. We then discuss the emerging threat of synthetic media and disinformation~(\cref{sec:deepfakes}). Finally, in~\cref{sec:discussion}, we highlight some open research problems. We conclude in~\cref{sec:conclusions}.

\section{Background: Neural Networks and Generative Models}\label{sec:background}
This section provides a brief overview of neural network architectures, generative models, and performance metrics for generative models.

\subsection{Neural Networks \& Deep Learning}\label{sec:background:neuralnets}

\subsubsection{Neural Networks}
Neural networks are machine learning (ML) models loosely based on how brains work. A neural network consists of one or more layers of neurons or units. A neuron/unit is associated with a set of weights for its connections, a bias term, and an activation function. Each unit computes a linear combination of its inputs and weights and then outputs the result of the activation function (typically a non-linear function) on this combination~\cite{krose1993introduction}. Neural networks that have arbitrarily many layers can approximate any function to any degree of precision~\cite{scarselli1998universal}. Layers between the input layer and output layers are called hidden layers and a neural network with more than one hidden layer is called a deep neural network. 

To train a neural network, all its parameters (i.e., weights and biases) are initialized randomly, then through the training process, the parameters are tuned using backpropagation~\cite{krose1993introduction}. There are different choices for activation function for each unit or layer, loss function (training objective), and ways to connect the units of one layer to those of another layer, leading to virtually endless possibilities for neural network architectures. 

\subsubsection{Architectures}
The simplest neural network architecture is composed of fully-connected (sometimes called dense) layers where the information only flows forward from the input layer to the output layer (through the hidden layers). These are called feed-forward neural networks~\cite{hassoun1995fundamentals}. 

Recurrent Neural Networks (RNNs) have layers that connect the output of units at one time step to its input at the next step, thereby creating some kind of hidden state or memory~\cite{mikolov2010recurrent}. This enables feeding the network sequences in such a way that the information from earlier elements of the sequence can be used when processing later elements. For example, this means that an RNN processing text sentences can remember and use information from the first word of the sentence while processing the last word. In practice, simple RNNs architectures have trouble retaining information for long periods of time~\cite{gers1999learning}. Long Short Term Memory (LSTM) cells were created to mitigate this problem and allow information to be retained over longer time frames~\cite{pascanu2013difficulty,rumelhart1986learning}. Recurrent neural networks based on LSTMs cells and alternatives such as Gated Recurrent Units~\cite{cho2014learning} (GRUs) are well-suited to process time-series data or other sequential data such as natural language text and speech. 

Convolutional Neural Networks (CNNs) use a combination of convolutional layers and sampling/pooling layers. The convolutional layers are made up of filters that convolve over the data using a sliding window and extract features into a set of feature maps~\cite{collobert2008unified}. Pooling layers downsample, reduce or otherwise aggregate the information of a feature map. CNNs are well-suited for data where spatial information is related such as images (e.g., nearby pixels of an image often contain similar information).

\subsection{Generative Models}\label{sec:background:genmodels}
Generative models provide a way to sample new data from a distribution learned from existing ``training'' data. More formally, constructing a generative model involves estimating an underlying distribution $p$ that approximates some true (unknown) distribution $P$ of interest. For this, one uses a training dataset $X$, where each $x \in X$ is a data point assumed to have been sampled i.i.d according to $P$. 

The problem of constructing (i.e., training) the generative model is difficult because it is an underspecified problem. Indeed, there are innumerable distributions that are compatible with the training dataset. As a consequence, generative model architectures are distinguished by the assumptions used to make constructing the generative model tractable. 

To estimate the underlying distribution $p$, some models (e.g., autoregressive models) explicitly characterize this distribution for any data point, so that given any data point $x$ in the support of $p$, they can calculate $p(x)$ (i.e., the probability that $x$ is sampled from $p$) exactly. In contrast, other models (e.g., variational autoencoders) learn $p$ only implicitly, for example by introducing a ``latent'' variable $z$ and then modeling only the probability of $x$ conditional on $z$, i.e., $p(x|z)$. The latent variable is then assigned a known prior distribution $q$ (e.g., isotropic Gaussian), so that one can sample $z$ according to $q$ and then $x$ according to $p(x|z)$.

In this survey, we focus on state-of-the-art generative models based on neural network architectures (sometimes called ``deep'' generative models). This includes generative adversarial networks, autoregressive models, and variational autoencoders. More traditional types of models such as probabilistic graphical models (e.g., Bayesian networks), Gaussian mixture models, and Markov chains or hidden Markov models can also be used as generative models. However, they are not the focus of this survey.

\subsubsection{Autoregressive Generative Models}\label{sec:background:genmodels:ar}
Autoregressive models view each data point $x$ as a tuple $x=(x_1, x_2, \ldots, x_m)$, which allows them to process sequential data, as well as view inherently non-sequential data (e.g., an image) as a sequence (e.g., the sequence of pixels that the image is composed off). The key idea is to capture the underlying distribution $p$ using the chain rule of probability~\cite{schum2001evidential}, so that: $p(x) = \prod_{i=1}^{m} \pr{x_i | x_{i-1}, \ldots, x_2, x_1}$. As a result, autoregressive models can directly estimate the density $p(x)$ for any data point $x$.

Despite the sequential nature of the internal representation of the density, autoregressive models can be built using recurrent architectures, but also feed-forward neural networks such as CNNs with masked convolutions~\cite{jain2020locally}. 
To maintain the autoregressive property, it suffices to ensure that the distribution over the $x_i$ depends only on $x_j$ for $j < i$.

Prominent examples of autoregressive models include PixelRNN~\cite{oord2016pixelrnn} and PixelCNN~\cite{van2016conditional}, Wavenet~\cite{oord2016wavenet}, and many others~\cite{chen2018pixelsnail,you2018graphrnn,vasquez2019melnet,yang2019xlnet, salimans2017pixelcnn++}. The PixelRNN and PixelCNN models produce images and are constructed by training an autoregressive model using a stack of LSTMs for PixelRNN and convolutional networks for PixelCNN. Both models sample an image pixel by pixel, with each pixel depending only on previously generated pixels. Wavenet produces state-of-the-art audio samples. It uses a CNN to process the training data. To generate audio, each sample produced is then fed back into the model and used to determine the next sample.

\subsubsection{Sequential Generative Models}\label{sec:background:genmodels:seq}
Recurrent neural networks such as those based on LSTMs and GRUs are often used in supervised learning tasks such as classification or labeling tasks. But, due to their inherently sequential nature, when the models are made to learn to predict the next element of a sequence, they can be viewed as generative models in the sense that given a starting input $x_1$, they can ``generate'' the sequence $x_2$, $x_3$, etc. However, RNNs and LSTMs are often used as components of larger generative models instead of by themselves~\cite{liu2018table,yu2019conditional,mogren2016c}. %

An alternative to RNNs and LSTMs for problems that deal with sequences is the transformer architecture~\cite{vaswani2017attention,wolf2020transformers}. Transformer models use stacked encoding/decoding layers and self-attention~\cite{vaswani2017attention} to allow for parallelization and remembering information over long input sequences \cite{,devlin2018bert,dai2019transformer,keskar2019ctrl,dehghani2018universal,raffel2019exploring,parmar2018image,o2016radio}. The Transformer architecture has become the state-of-the-art architecture for natural language processing~\cite{wolf2020transformers}. 

Both LSTMs and Transformers are often used for tasks involving natural language. For both text and speech they are used for generation, translation, and understanding. In particular, LSTMs have been used to generate text \cite{graves2013generating}, speech recognition \cite{graves2013hybrid}, and machine translation~\cite{ahmed2017weighted}. Transformers have been used to generate text~\cite{radford2019language}, for language understanding~\cite{devlin2018bert}, machine translation~\cite{ahmed2017weighted}, text to speech~\cite{ren2019fastspeech}, audio speech generation~\cite{li2019neural} and even some image generation tasks~\cite{parmar2018image}. Some models are bi-directional~\cite{graves2013hybrid,devlin2018bert}.

\subsubsection{Autoencoders}
Autoencoders (AE) learn to approximately ``copy'' their input to their output. To do so, the autoencoder architecture combines an encoder and a decoder (both neural networks) to efficiently represent inputs as {\em codes} or {\em latent variables} in a latent space. Because the latent space is typical of lower dimensionality than the input space, the encoding process must discard extraneous information in the input, preserving only the most important information for the reconstruction. Autoencoders have applications to dimensionality reduction~\cite{makhzani2015adversarial,sakurada2014anomaly} and feature learning~\cite{yousefi2017autoencoder}, but they can also be used as generative models or to learn generative models. For example, sampling latent space points (from an adequate distribution) and feeding the samples to the decoder yields new data points from the autoencoder's learned data distribution.

To ensure that autoencoders are able to generate samples not in the training dataset, are not overfitted, and have a tractable, continuous, and interpretable latent space, they need to be regularized. There are several types of autoencoders using various regularization techniques. For example, denoising autoencoders are trained by adding noise to their training data~\cite{vincent2010stacked}, whereas contractive auto-encoders are regularized by adding a penalty to the reconstruction loss~\cite{rifai2011contractive}. A prominent alternative are variational autoencoders, proposed by Kingma and Welling~\cite{kingma2013auto}. Variational AutoEncoders (VAEs) model the latent space as isotropic Gaussian. To approximate the posterior distribution of the encoder, variational inference~\cite{blei2017variational} is used.

VAEs benefit from relatively straightforward training. The encoder can be used for continual updates of the latent space or dimensionality reduction. The decoder allows for sampling from different parts of the latent space. This can be leveraged to reveal relationships between data points and/or produce samples with a specific relationship to other samples. VAEs are often used for image generation~\cite{razavi2019generating}, and can be used to edit existing photos~\cite{engel2017latent}. However, some authors have expressed concern about the blurry output of VAEs~\cite{zhao2017towards}, and while some mitigation methods have been proposed, this problem persists in many cases~\cite{dosovitskiy2016generating}. VAEs have also been used for text \cite{hu2017toward} and video generation~\cite{he2018probabilistic}. There are many variants of VAEs such as $\beta$-VAE~\cite{higgins2017beta} or IWAE~\cite{burda2015importance}. Significant differences between autoencoder architectures are the underlying neural network architecture used for the encoder and decoder (e.g., CNN or RNN), and also the loss function used during training. Another notable type of autoencoders are adversarial autoencoders (AAE)~\cite{makhzani2015adversarial}, which use generative adversarial networks for variational inference.

\subsubsection{Generative Adversarial Networks}
Generative adversarial networks (GANs) were proposed by Goodfellow et al. \cite{goodfellow2014generative} in 2014. GANs are made up of two separate feed-forward neural networks that are set up to be adversaries of each other. The first network is a generator model that produces samples from an estimation of the training distribution. The second network is a discriminator that attempts to determine if a given sample is from the training data (returns 1) or produced by the generator (returns 0). The generator uses the predictions from the discriminator to determine the quality of its samples, while the discriminator uses the generator's samples to refine its results. Both networks are trained simultaneously. 

In practice, GANs can be difficult to train. The generator and the discriminator must learn at a similar pace or the process will fail. If the discriminator becomes significantly better than the generator, or vice-versa, the gradient will diminish until the model is unable to improve. In addition, even if the two networks are balanced in their learning, the GAN may still suffer mode collapse~\cite{bau2019seeing}, a condition where the model always produces a few samples of high scoring output rather than reflecting the diversity in the training data.

\subsubsection{GAN Architectures}
There are numerous GAN variants. Most significant alternative GAN architectures target the generator or the loss function~\cite{pan2019recent}. 
GANs initially used a simple feed-forward network as a generator, but later implementations replaced this with different models including a CNN~\cite{radford2015unsupervised}, an autoencoder~\cite{makhzani2015adversarial}, and a VAE~\cite{larsen2016vaegan}. Other variations make changes to the loss function~\cite{goodfellow2014generative,mao2017least,arjovsky2017wasserstein} to improve and stabilize the training process. Finally, some variations seek to add features to GANs. For example, Zhu et al.~\cite{zhu2017unpaired} allow for image-to-image translation without the need for training images that have been translated through alternative methods. Gomez et al.~\cite{gomez2017reversible} make GANs reversible. Donahue et al.~\cite{donahue2016adversarial} propose a GAN that can project data back into the latent space. 

\subsubsection{Applications \& Data Types}
The nature and architecture of these generative models lend themselves to some data types and applications/tasks better than others.

\paragraphb{Text and natural language}
Transformers are generally regarded as the state-of-the-art text generator models. Examples include BERT~\cite{devlin2018bert}, OpenAI's GPT-2~\cite{radford2019language} and GPT-3~\cite{brown2020language}, and XLNet~\cite{yang2019xlnet}, which is an autoregressive model used for language understanding. There have also been several proposals applying GANs to text generation~\cite{zhang2017adversarial,chen2018adversarial}, but the GAN architecture can be challenging to apply to sets or sequences of words. That said, several papers have proposed ways to improve performance on this type of data \cite{kusner2016gans,xu2019modeling}. Finally, VAEs have also been applied to text~\cite{hu2017toward}, but their performance is not competitive with transformers.  

\paragraphb{Audio and speech}
A popular audio generator is Wavenet~\cite{oord2016wavenet}. Transformers have also seen success in music generation~\cite{huang2018music}.  GANs have also been applied to audio~\cite{donahue2018adversarial,fedus2018maskgan}. In particular, Bi\'nkowski et al. propose GAN-TTS, a generative adversarial network for text-to-speech synthesis~\cite{binkowski2019high}. GANs have also been used to generate music~\cite{kumar2019melgan,engel2018gansynth}.

\paragraphb{Images and videos}
GANs are well-suited for image generation~\cite{karras2017progressive,brock2018large,isola2017image} and editing~\cite{antipov2017face,zhu2016generative}. Autoregressive models are also able to produce competitive images through PixelCNN~\cite{van2016conditional,salimans2017pixelcnn++}. VAEs are also often used for image generation~\cite{razavi2019generating,yan2016attribute2image,hou2017deep,dosovitskiy2016generating} and to making small edits to existing photos~\cite{engel2017latent}. LSTMs~\cite{gregor2015draw} and transformers~\cite{parmar2018image} have also been used to generate images. Finally, both VAEs~\cite{he2018probabilistic} and GANs~\cite{tulyakov2018mocogan} have been applied to video generation. GANs have also been applied to video prediction~\cite{liang2017dual,bansal2018recycle,walker2017pose}.

\subsection{Evaluating Generative Models}\label{sec:background:output}
The performance of generative models is difficult to evaluate. A classifier can simply be evaluated on how many examples it correctly classifies; but there is no obvious one-size-fits-all metric to quantify the quality of model-generated data such as images, audio, or text. For generative models that have a tractable probability distribution, we can quantify the likelihood of producing a specific output, but this is not always indicative of output quality. Further, the notion of quality may be task-dependent, so that no single metric can be applied in all cases and multiple metrics may be necessary to evaluate different aspects of the model. Borji~\cite{borji2019pros} gives a comprehensive comparison of many different evaluation metrics for GANs, mainly focusing on image generation. There is also evidence that metrics estimating output likelihood often do not match up with human judgement~\cite{theis2015note}. For example, factors such as background color can have a significant impact on likelihood. Also, images with (relatively) low likelihood may nevertheless appear to be much clearer and more realistic (to humans) than images with high likelihood. Additionally, measures of quality often do not take output diversity into account. Theis et al.~\cite{theis2015note} suggest that the best evaluation method depends on the model, type of data, and the goals of the user.

In this section, we provide a brief taxonomy of existing metrics. We classify metrics by what they aim to measure and summarize each class in a table. The classes are output quality, model quality, comparison against training data, and human evaluation. The tables include image, audio, and text columns to indicate metrics that are often used to compare output of a model, such as between epochs or different versions of the same model. Metrics useful for model comparison can be used to compare two or more models attempting to generate the same distribution. Finally, metrics that evaluate output diversity quantify the extent to which the model is able to produce diverse outputs. For example, models that output data points that all look similar, or models that suffer from mode collapse would score poorly on output diversity metrics. Note that different aspects are not always mutually exclusive: methods that evaluate image quality can in some cases be used to determine model quality, and high-quality models usually produce high-quality output. To show this, we use the following symbols: \pie{360} what it is intended to measure; \pie{180} what it can be adapted to measure; and \pie{0} and what it cannot measure. 

\subsubsection{Output Quality Metrics}
\cref{tab:output} shows metrics that are evaluated by solely examining the output of the generative models. Methods such as asking survey participants to rate the quality of output or low-level image statistics are often used to evaluate model output against images that are not generated. These types of methods are important since the end goal of many generative models is to produce the most realistic output possible. It should be noted that the text comparison methods are intended for use with machine translation, but can also be used for evaluating generated text. 

\begin{table}[ht!]
\centering
\caption{\small Output quality metrics. These metrics solely look at the output of a generative model and can be used to compare images, audio, text, and the models themselves. We do not include the diversity column here since none of the metrics in this table are useful for determining output diversity of a model. \label{tab:output}}
\small
\begin{tabular}{p{.45\linewidth}||c|c|c|c|c|c|}
\cline{2-5}

 \phantom{Output Quality Metrics} & Image &  Audio & Text &  Model\\
\hline\hline

  Image Quality Measurements~\cite{wang2004image,odena2017conditional,regmi2018cross}&\pie{360} & \pie{0} & \pie{0} & \pie{180}\\

\hline
  Low-level Image Statistics~\cite{zeng2017statistics,karras2017progressive}&\pie{360} & \pie{0} & \pie{0} & \pie{180}\\

  \hline
  Inter/Intra Track Evaluation~\cite{dong2018musegan}&\pie{0} & \pie{360} & \pie{0} &\pie{180}\\

  \hline
  BLEU~\cite{papineni2002bleu}&\pie{0} & \pie{0} & \pie{360} & \pie{180}\\

\hline
  ROGUE~\cite{lin2004automatic}&\pie{0} & \pie{0} & \pie{360} & \pie{180}\\

\hline
  METEOR~\cite{denkowski2011meteor}&\pie{0} & \pie{0} & \pie{360} & \pie{180}\\

\hline

\end{tabular}
\end{table}

\subsubsection{Model Quality}
\cref{tab:model} shows metrics that are used to evaluate the quality of a model itself rather than only its output. These metrics often identify desirable qualities of a model such as producing diverse output or having a discriminator able to function well as a classifier. Since these metrics are examining the model rather than the output, they may not always correlate with outputs that humans think of as high quality. For instance, an image may have a high likelihood of being produced by the model, but it may look ``off" or unintelligible to a person. These metrics are still useful for comparing different outputs from a single model or evaluating a certain aspect such as output sharpness. Many of these metrics directly measure model output diversity.
A model may produce a perfect picture, but if it always produces the same picture it is not useful as a generative model. This is especially important for evaluating GANs as they are susceptible to mode collapse. Finally, methods such as tournament win rate and generative adversarial metric are useful for comparing competing models.

\begin{table}[ht!]
\caption{\small Model quality metrics. These metrics examine the internals and the output of a generative model to measure desirable qualities. They can be used to compare images, audio, text, the models themselves, and output diversity.  }
\label{tab:model}
\small
\centering
\begin{tabular}{p{.475\linewidth}||c|c|c|c|c|c|}
\cline{2-6}

  \phantom{Model Quality Metrics} & Image &  Audio &  Text &  Model & Diversity\\
\hline\hline
Average Log Likelihood~\cite{theis2015note,goodfellow2014generative} &\pie{360} & \pie{360} & \pie{360} & \pie{0}&\pie{0}\\

\hline
  Inception Scores~\cite{salimans2016improved} &\pie{360} & \pie{0} & \pie{0} & \pie{180}&\pie{360}\\

  \hline
Generative Adversarial Metric~\cite{im2016generating}&\pie{0} & \pie{0} & \pie{0} & \pie{360}&\pie{0}\\

\hline
  Tournament Win Rate and Skill Rating~\cite{olsson2018skill}&\pie{0} & \pie{0} & \pie{0} & \pie{360}&\pie{0}\\

  \hline
  Normalized Relative Discriminative Score~\cite{zhang2018decoupled}&\pie{0} & \pie{0} & \pie{0} & \pie{360}&\pie{0}\\

  \hline
  Birthday Paradox Test~\cite{arora2017gans}&\pie{0} & \pie{0} & \pie{0} & \pie{0}&\pie{360}\\

  \hline
  Number of Statistically Different Bins~\cite{richardson2018gans}&\pie{0} & \pie{0} & \pie{0} & \pie{0}&\pie{360}\\

  \hline
   Precision, Recall, and F1 Score~\cite{lucic2018gans}&\pie{180} & \pie{0} & \pie{0} & \pie{360}&\pie{360}\\

  \hline
  Examining Internals of Networks~\cite{yu2017unsupervised,chen2016infogan}&\pie{360} & \pie{0} & \pie{0} & \pie{360}&\pie{180}\\

   \hline

Perplexity~\cite{jelinek1977perplexity}&\pie{0} & \pie{0} & \pie{360} & \pie{0}&\pie{0}\\

\hline
  Classification Performance~\cite{radford2015unsupervised}&\pie{360} & \pie{180} & \pie{180} & \pie{360}&\pie{0}\\

 \hline

\end{tabular}
\end{table}

\subsubsection{Comparison Against Training Data}
\cref{tab:training} shows metrics that compare model output against training samples of real data. These metrics are mainly used to ensure that the model approximates the unknown training data distribution as closely as possible. This means that the model both produces output that is similar to training data, and it produces a full distribution rather than just a few high scoring samples. These metrics do not directly examine the internals of the model, allowing them to be used to compare models with different architectures trained on the same data. For example, a VAE can be compared against a GAN.  

\begin{table}[ht!]
\caption{\small Comparison against training data metrics. These metrics compare generative model output against training data to ensure that the model is learning the training data distribution. They can be used to compare images, audio, text, the models themselves, and output diversity.}
\label{tab:training}
\small
\centering
\begin{tabular}{p{.45\linewidth}||c|c|c|c|c|c|}
\cline{2-6}

 \phantom{Comparison Against Training Data Metrics} & Image &  Audio &  Text &  Model & Diversity\\
\hline\hline
  Fr\'echet Inception Distance~\cite{heusel2017gans} & \pie{360} & \pie{0} & \pie{0} & \pie{180} & \pie{360}\\

\hline
Maximum Mean Discrepancy~\cite{fortet1953convergence} & \pie{180} & \pie{180} & \pie{180} & \pie{360} & \pie{0}\\

\hline
Reconstruction Error~\cite{xiang2017effects} & \pie{0} & \pie{0} & \pie{0} & \pie{360}&\pie{0}\\

  \hline
  Wasserstein Distance~\cite{arjovsky2017wasserstein}&\pie{180} & \pie{0} & \pie{0} & \pie{180}&\pie{360}\\

  \hline
  Boundary Distortion~\cite{santurkar2017classification}&\pie{0} & \pie{0} & \pie{0} & \pie{180}&\pie{360}\\

  \hline

  Classifier Two-sample Tests~\cite{lehmann2006testing}&\pie{360} & \pie{360} & \pie{360} & \pie{180}&\pie{180}\\

  \hline

  Image Retrieval Performance~\cite{wang2016ensembles}&\pie{360} & \pie{0} & \pie{0} & \pie{180}&\pie{180}\\

  \hline
Fr\'echet Audio Distance~\cite{kilgour2018frchet}&\pie{0} & \pie{360} & \pie{0} & \pie{180}&\pie{360}\\

\hline

\end{tabular}
\end{table}

\subsubsection{Human Evaluation}
\cref{tab:human} shows metrics based on human evaluation. Human evaluation, though expensive to conduct, is an important aspect of generative model evaluation. For some applications, the most important factor is whether humans can distinguish between the generated output and real samples. However, there are several challenges with human evaluation. Humans are inconsistent, so ratings can vary greatly between individuals. Another consideration is the cost and effort of recruiting participants and obtaining meaningful data from this. As a result, human evaluation should (in most cases) not be the only metric used, but applications that entail a generative model whose outputs will be shown to humans would benefit from some human evaluation.

\begin{table}[ht!]
\caption{\small Human evaluation metrics. These metrics have humans directly evaluate the quality of generative model output. They can be used to compare images, audio, text, the models themselves, and output diversity.}
\label{tab:human}
\small
\centering
\begin{tabular}{p{.45\linewidth}||c|c|c|c|c|c|}
\cline{2-6}

 \phantom{Human Evaluation Metrics} & Image &  Audio& Text &  Model & Diversity\\
 \hline\hline
Nearest Neighbors~\cite{wang2009mean}&\pie{360} & \pie{360} & \pie{360} & \pie{180}&\pie{180}\\

\hline

  Rapid Scene Categorization~\cite{touretzky1996advances}&\pie{360} & \pie{0} & \pie{0} & \pie{180}&\pie{0}\\

\hline

  Rating and Preference Judgement~\cite{zhang2017stackgan}&\pie{360} & \pie{360} & \pie{360} & \pie{180}&\pie{0}\\
  
 \hline
  Realism Measurement~\cite{zhou2019hype}&\pie{360} & \pie{180} & \pie{180} & \pie{180}&\pie{0}\\
  
   \hline
  Timed Realism Measurement~\cite{zhou2019hype}&\pie{360} & \pie{180} & \pie{180} & \pie{180}&\pie{0}\\

\hline
\end{tabular}
\end{table}

\section{Intrusion, Biometrics, and Malware}\label{sec:intrusion}
Because generative models are able to learn detailed representations of data, attackers can create realistic-looking data that may be able to bypass defenses while still retaining malicious functionality. At the same time, defenders can use generative models to learn the distribution of ``normal'' or benign data and attempt to leverage this to identify ``anomalous'' objects that do not fit the distribution such as zero-day malware. In this section, we provide a brief overview of research on intrusion and anomaly detection, as well as biometrics and malware attacks and defenses that leverage generative models.

\subsection{Intrusion Attacks}
Generative models can be used to learn the distribution of normal traffic, and then automate the generation of attacks that look benign. For example, Feng et al.~\cite{feng2017deep} use deep learning to automatically learn what normal traffic entering the system looks like, and then generate samples from this distribution. Lin et al.~\cite{lin2018idsgan} use GAN's training method to train a generator against a discriminator, creating adversarial traffic samples that can evade the target detector. These models can be used to generate traffic to evade Intrusion Detection Systems (IDS) protecting Industrial Control Systems (ICS). 

While attacks on ICS, most notably Stuxnet~\cite{falliere2011w32}, are well-established, they remain difficult to perform because they require knowledge of both the ICS and the IDS. Generative models change the equation because they simplify these attacks by analyzing the structure traffic going through the IDS and crafting messages that are able to bypass the system by appearing benign, while still containing the malicious code.

\subsection{Anomaly Detection}\label{sec:intrusion:anomaly}
Generative models' ability to learn data distributions can also be used to detect attacks because it can produce a model of normal use or operation of the system. Any input that does not fit this ``normal'' model can then be rejected or flagged for further analysis. An advantage of this method is that it can readily detect never seen before malicious input. While most classifiers can be trained to identify malicious input versus benign input, this requires samples of benign and malicious input.

For example, Kim et al.~\cite{kim2018zero} use deep autoencoders and a GAN to learn the features of a zero-day malware sample allowing for sample generation while also training the GAN's discriminator. The discriminator's learning can then be transferred to the detector to improve the detection of the new malware. In a similar vein, Zheng et al.~\cite{zheng2018generative} use a deep denoising autoencoder to detect fraud in bank transfers. Another example is Chandy et al.'s~\cite{chandy2019cyberattack} proposal to use a variational autoencoder to detect cyberattacks on a town's water distribution system.

A different methodology to achieve a similar goal is to use generative models to support other detectors. An example of this is Bot-GAN~\cite{yin2018enhancing}. The idea is to use an existing botnet detector with a GAN continuously producing new samples of botnet activity. This allows the detector to become more robust, since it has many more samples to train on.

\subsection{Biometric Spoofing}
Biometrics use a person's unique attributes, such as fingerprints, voice, or even handwriting, to identify an individual. Biometrics are used to protect many systems from smartphones to buildings. However, generative models can be used to create new biometrics samples to break into systems. 

GANs can be used to simply generate synthetic fingerprints~\cite{minaee2018finger}, but producing random fingerprints is not typically sufficient to break biometrics systems. More importantly, by sampling from the latent space generative models can generate masterprints~\cite{bontrager2018deepmasterprints}. Masterprints are special fingerprints that incorrectly match with many different fingerprints, undermining the security of fingerprint-protected systems.

Generative models can also spoof handwriting~\cite{lopresti2005effectiveness}, and keystroke latency~\cite{monaco2015spoofing}.

\subsection{Malware Obfuscation and Detection}
Malware are malicious programs that seek to compromise computer systems to exfiltrate data or otherwise cause harm. Ideally, malware can be detected before it compromises a system. But novel malware and obfuscated malware can be difficult to detect. Generative models can be used to obfuscate malware in new and unpredictable ways~\cite{hu2017generating}, but they can also be used as malware detectors~\cite{park2019generative,kim2018zero,kim2018detecting,lutz2019malware}.

Generative models can help improve malware detection by augmenting the available amount of training data. In particular, generative models can be used to simply generate malware, creating samples to bolster detectors~\cite{kim2018detecting,lutz2019malware}. For more refined detectors they can be used to create clusters of similar malware around likely malware base forms, which augment the detector so that it can identify the different classes of malware even if they are changed slightly~\cite{park2019generative,kim2018zero}.

An example of the use of a generative model to avoid detection is MalGAN~\cite{hu2017generating}. MalGAN produces malware examples that are able to bypass many malware detectors. The idea is to train a GAN with an estimation of the target detector as the adversarial network and a generator which adds and removes malware features. This creates adversarial malware samples that are better able to avoid the target detector, compared to just adding random variation.
\section{Data Synthesis}\label{sec:datasynthesis}
The ability of generative models to produce new samples from the distribution of existing data is useful by itself for several security and privacy related applications.
In particular, generative models can be used to produce synthetic datasets that can be shared when the original (sensitive) data cannot itself be shared or publicly released for legal, policy, or privacy reasons. Furthermore, the ability to create large datasets enables the study of data which may impact security policies. For example, generating realistic passwords can help practitioners study password management policies or design defenses for data breaches.

\subsection{Privacy-Preserving Data Synthesis}
Organizations owning large datasets of sensitive individual data such as medical records databases may wish to share this data with outside parties. Due to privacy concerns captured in organizational policies or legal considerations, such data sharing is often impossible. Organizations could of course release only aggregate statistics of the data in an anonymized way, but this limits the type of analyzes that can be performed.

Generative models offer an attractive alternative solution: train a generative model on the sensitive dataset then uses it to produce a synthetic dataset that can be shared or published. To preserve utility, we need to ensure that the synthetic dataset reflects the statistical properties of the original dataset. Note that just because the data is ``synthetic'' does not mean it automatically safeguards privacy; we need to ensure that the synthetic dataset does not inadvertently leak sensitive information from the original dataset~\cite{stadler2021synthetic}. So, research in the past decade has focused on ways to create the synthetic dataset in such a way that the process satisfies $\varepsilon$-differential privacy~\cite{dwork2008differential}, a well-established notion for privacy protection.

Methods to achieve differential privacy need to be tailored to the specifics of the generative model used, but the process typically involves adding random noise (from a carefully chosen distribution and in a specific way) during the model's training process. In particular, one of the first lines of work for this used Bayesian networks as generative models~\cite{bindschaedler2017plausible,zhang2017privbayes,ping2017datasynthesizer,zhang2020privsyn}. For example, PrivBayes~\cite{zhang2017privbayes} shows how to construct the Bayesian network and estimate its probabilities from data such that the process satisfies $\varepsilon$-differential privacy.

More recently, generative models based on neural networks have been used to create synthetic data~\cite{xie2018differentially,acs2018differentially,xu2019ganobfuscator,triastcyn2018generating,beaulieu2019privacy,frigerio2019differentially,tantipongpipat2019differentially,abay2018privacy,alzantotdifferential,torkzadehmahani2019dp,triastcyn2020generating,rosenblatt2020differentially}. These papers use a variety of model architecture, including GANs and variations of GANs such as W-GAN \cite{alzantotdifferential}, CGANs \cite{torkzadehmahani2019dp}, or autoencoder GANs \cite{tantipongpipat2019differentially}. Variational Autoencoders have also been used \cite{acs2018differentially}. There are various techniques to achieve differential privacy (or related notions), but a common approach is to train the model using DP-SGD~\cite{abadi2016deep,nasr2020improving}, which adds carefully crafted Gaussian noise to the gradient during the stochastic gradient descent (SGD) process that tunes the parameters. Some of the other methods include average KL-Privacy~ \cite{triastcyn2018generating}, differentially private clusters~\cite{xie2018differentially}.

A different but related line of work for the semi-supervised setting is Private Aggregation of Teacher Ensembles (PATE)~\cite{papernot2016semi,papernot2018scalable,long2019scalable,jordon2018pate}. The idea is to partition the original (sensitive) dataset into disjoint subsets, and then train a ``teacher'' model on each subset. Teacher models are trained without consideration of privacy. But the predictions of all teacher models are aggregated with random noise to achieve differential privacy. Finally, a privacy-preserving ``student'' model can be obtained from a public dataset of unlabeled data, through knowledge transfer from the teacher models. While the PATE approach does not directly provide a synthetic dataset or a generative model, it can be extended to do so as described by Jordon et al.~\cite{jordon2018pate} or Long et al.~\cite{long2019scalable}.

\subsection{Understanding Data}
Generative models can also help better understand data that influence security decisions. PassGAN~\cite{hitaj2019passgan} is a GAN that is able to generate a large number of potential passwords. Previous methods of guessing passwords would often consist of human-generated rules, such as replacing certain letters with numbers, or limited n-gram Markov models. While effective, they are limited and unable to discover passwords that do not fit the specific rules in use. Although PassGAN was presented as a password guessing attack method, its existence shows how generative models can be used to develop an understanding of the data being examined, without humans having to go through an enormous amount of data to identify patterns. 

Models like PassGAN and more work recent~\cite{nam2020recurrent} can be used to learn common password patterns and steer users away from easily guessable passwords. Another potential application is to improve honeywords~\cite{juels2013honeywords}. Honeywords are passwords of fake accounts setup to detect compromises. 

Augenstein et al.~\cite{augenstein2019generative} shows how differentially private generative models, as previously described, can be used to examine federated data while still maintaining the privacy of the entries in the database. They use generative models to study private federated data and debug other models trained with the data. This involves creating two private generative models using federated data, one with the best samples and one with the worst, and compare outputs.
With this technique, the authors were able to debug a specific pixel inversion issue, showing that generative models can be used to better understand data humans cannot realistically study. 

\section{Adversarial Machine Learning}\label{sec:advml}
Adversarial machine learning is the study of attacks against and defenses for machine learning models and the associated threat models~\cite{huang2011adversarial}. Attacks against machine learning can be classified into one of three categories: exploratory, evasion, or poisoning~\cite{chakraborty2018adversarial}. Exploratory attacks attempt to extract information from a model, such as data used to train the model~\cite{shokri2017membership}, the model parameters~\cite{fredrikson2015model}, or its hyperparameters~\cite{wang2018stealing}. Such attacks can violate the privacy of individuals whose data is part of the training set or steal confidential and valuable information about machine learning engineering from the model's developers. Evasion attacks create malicious inputs  (often called adversarial examples) that result in the model making a false prediction \cite{papernot2017practical,goodfellow2014explaining,athalye2018synthesizing}. This can be used by attackers in various settings such as to evade detection by a model trying to detect or filter out attacks. The goal of poisoning attacks is to corrupt the training data so that the attackers can control the model's future predictions under certain scenarios \cite{chen2017targeted,liu2017trojaning,gu2019badnets}. Both evasion and poisoning attacks can have potentially disastrous real-world effects if they are perpetrated against some systems. For instance, a self-driving car could be made to ignore a stop sign~\cite{song2018physical} when provided with specially crafted malicious inputs. Finally, there exist defenses against these attacks. Also, within all of these categories, there are different attack scenarios and threat models such as whether the attacker has black-box or white-box access to the model.

In this section, we survey the use of generative models within adversarial machine learning. This includes the use of generative models to help craft adversarial examples (\cref{sec:advml:advexamples}), as a defense against adversarial examples~(\cref{sec:advml:def}), and for poisoning attacks~(\cref{sec:advml:poison}).

\subsection{Generating Adversarial Examples}\label{sec:advml:advexamples}
Adversarial examples are maliciously crafted inputs that force a machine learning model to produce specific decisions or outputs that are counter to human expectations. In particular, if the model is a classifier, a benign input belonging to class A is maliciously perturbed in a way that a human would consider it to still be of class A, but such that the model classifies it as class B. For example, an adversary may create an ``adversarial stop sign'' by adding a small perturbation to an existing stop sign in a way that is imperceptible or inconspicuous to humans but causes a self-driving car to ignore the stop sign~\cite{eykholt2018robust}. 

There are several methods for creating adversarial examples depending on the constraints of the problem and the threat model. However, one of the most common methods is to treat the task of crafting an adversarial example as an optimization problem. This typically involves starting with a benign example and taking steps in the opposite direction of the gradient of the loss of the classifier (with respect to the input) on the desired target class~\cite{goodfellow2014explaining,szegedy2013intriguing}. 

Generative models provide a different way to craft adversarial examples. The idea is to use a generative model to sample adversarial perturbations~\cite{xiao2018generating} or adversarial examples directly~\cite{song2018constructing,zhao2017generating,jalal2017robust,dolaadvflow}. For example, this can be accomplished by training a generative model on the original (benign) data and then finding adversarial examples in the latent space of the generative model. There are several search strategies over the latent space, but they involve two steps. First, one finds a normal/benign data point in the latent space. Then one searches near this data point (in the latent space) to find another data point which (in the original space) is classified differently by the target model. This method allows attackers to generate completely new adversarial images without explicitly constructing perturbations. 

A key advantage of this technique is that it can be applied to domains where valid perturbations are difficult to find. For instance, Hu et al.~\cite{hu2017generating} uses a GAN to generate malware that looks benign to a target discriminator. In a separate paper, the same authors use a generative RNN to creates adversarial examples for RNN-based malware detectors~\cite{hu2017black}. Also, some types of data, such as text and code, do not easily lend themselves to perturbations.  For example, adversarial text can be created by switching out words~\cite{ebrahimi2017hotflip}, but this is more difficult to accomplish without changing the meaning of the sentence, compared to images where the pixel values can often be shifted by a small amount while resulting in (almost) imperceptible changes. 

A line of work that may be of particular interest to cybersecurity practitioners is the use of GANs to generate domain names that are able to bypass detectors~\cite{anderson2016deepdga,corley2019domaingan}. Domain names generation algorithms are often used to generate domain names for command and control servers. Corley et al.~\cite{corley2019domaingan} train three different GANs capable of producing outputs that can evade state-of-the-art classifiers meant to detect domain generation algorithms. They conclude that, compared to using traditional domain generation algorithms, the use of GANs significantly improves the ability to evade detection.

Finally, the ability to produce adversarial examples without producing an explicit perturbation allows one to bypass some adversarial example detectors. This is because some adversarial example detectors are specifically designed to identify and remove perturbations of benign inputs~\cite{samangouei2018defense,song2017pixeldefend}.

\subsection{Defenses Against Adversarial Examples}\label{sec:advml:def}
Generative models can also be used to defend against adversarial examples. To improve the robustness of a classifier, Lee et al. \cite{lee2017generative} propose using the GAN training method as a trainer for the classifier, where the classifier works as the discriminator and an adversarial example generator works as the generator. This method makes the model more robust against adversarial examples. A similar effect can be achieved by simply adding adversarial examples to the training dataset. This is known as adversarial training~\cite{madry2018towards} and it has been evaluated in several papers~\cite{anderson2016deepdga,jalal2017robust}.

Zhang et al. ~\cite{zhang2020understanding} used the assumption that the underlying data distribution of a set of images is captured by generative models to study model robustness and interactions with adversarial examples. By comparing the robustness of their generative model and a theoretical limit of robustness they are better able to understand how models are vulnerable to adversarial examples. While the main purpose of the paper is to display the disparity in model robustness versus the robustness of the actual image space, they propose that this method could also be used to bolster defensive classifiers by showing where adversarial examples are likely to be created.

Alternatively, GANs can create a ``denoised'' version of any input that can then be fed into the classifier \cite{samangouei2018defense,song2017pixeldefend,jin2019ape,bao2018featurized}. This can eliminate many adversarial examples by removing the adversarial perturbation, leaving behind the original input which can be correctly classified. A prominent instance of this method is Defense-GAN \cite{samangouei2018defense} which uses a WGAN to take in an input image and generates a new image that retains all of the pertinent information while eliminating the extraneous artifacts that have the potential to be malicious. A similar effect can be accomplished through dimensionality reduction with a simple autoencoder, but GANs are better able to completely recreate a clean image, rather than simply removing many of the finer details. This method increases a model's resistance to adversarial examples, while being flexible, since it is not necessary to make any changes to the classifier itself; GANs just need to be run over any potential input first. Ironically, adversarial examples that are produced by generative models (discussed in~\cref{sec:advml:advexamples}) may defeat this type of defense because they do not rely on explicit perturbations.

\subsection{Poisoning Attacks}\label{sec:advml:poison}
The goal of poisoning attacks is to corrupt the target model \cite{chen2017targeted,liu2017trojaning,gu2019badnets}. In particular, a subset of poisoning attacks, backdoor attacks add a ``trigger'' (e.g., a small pattern or even a single pixel of a specific color in the context of images) to any data point to force the classifier to make a specific decision. Backdoor attacks accomplish this goal by adding the trigger to many training images and labeling them all as a specific target class, regardless of their original class labels. During training the model learns to associate the trigger with the target class. 

For backdoor attacks, a trigger does not need to be generated, it can be added to any image. In addition, the trigger need not be hidden. It can be as obvious as the attacker desires. That said, Turner et al.~\cite{turner2019label} propose label-consistent backdoor attacks that use generative models to create adversarial samples that look more similar to their original image, while still forcing the model to learn the trigger. This is useful since these examples are less likely to be filtered out of the training data. If the image is too different from the target class, a simple classifier run over the training data will pick it out and remove the image. 

In addition, generative models can be used to speed up the generation of poisoned samples and to mount attacks on federated learning. Yang et al.~\cite{yang2017generative} used a GAN to generate poisoned samples over 200 times faster than traditional gradient methods. Zhang et al.~\cite{zhang2019poisoning} used GANs to attack federated learning networks.
\section{Attacks Against Generative Models}\label{sec:attacksongen}
In addition to adversarial machine learning attacks performed using generative models, there are attacks used against generative models themselves. Indeed, generative models have been shown to be vulnerable to the three main categories of adversarial ML attacks: exploratory, evasion, and poisoning attacks. 
However, it should be pointed out that evasion attacks on generative models focus on causing the model to generate specific outputs. 

\subsection{Exploratory Attacks}
A prominent type of exploratory attack on machine learning is membership inference attacks. In a membership inference attack, the attacker aims to determine whether a specific individual's data was part of the target model's training data~\cite{shokri2017membership}. These attacks exploit the fact that machine learning models such as classifiers often behave differently when asked to make a prediction for a data point from their training dataset compared to for a data point not part of the training dataset (but from the same distribution as the training dataset). This behavior is related to overfitting, although recent work suggests that models may be vulnerable even in the absence of overfitting~\cite{long2018understanding,long2020pragmatic,leino2020stolen}.

Membership inference attacks on generative models seek to exploit the same effect. In particular, some membership inference attacks have been developed against GANs~\cite{hayes2019logan} and generative models~\cite{hilprecht2019monte}. Additionally, VAEs are vulnerable to reconstruction attacks~\cite{hilprecht2019monte}.

\subsection{Adversarial Examples}
Generative models are susceptible to adversarial examples. Concretely, adversarial examples attacks have been demonstrated against generative models in natural language processing (NLP) and image compression. 

In the NLP context, Wallace et al.~\cite{wallace2019universal} show that there exist universal adversarial triggers. These are sequences of tokens that can be appended to any input to cause the model to output specific predictions. More specifically, Wallace et al. were able to append certain words which caused models such as GPT-2 to output offensive text. Interestingly many of these triggers worked on all of the models tested.  

In the image compression context, Kos et al.~\cite{kos2018adversarial} propose a scenario where the encoder and decoder of a VAE or VAE-GAN have been separated. The encoder is being used by a sender to encode an image into the latent space representation $z$, which is transported to a receiver who will attempt to decode $z$ and reproduce the original image. They argue that an attacker could potentially trick the sender into choosing to send an adversarial image that looks benign but returns an entirely different image when compressed and then decoded. For instance, a picture of a cat could be slightly perturbed to decode to an image asking for money.  

\subsection{Poisoning Attacks}
Wallace et al.~\cite{wallace2020customizing} show how poisoning attacks can be applied to NLP translation models. They focus on classification models, but the authors show that the method also works on translation models. The idea is to insert a small number of triggers into the model, none of which actually contain the trigger phrase, making it difficult to determine which training samples are causing the error. Although this work only applies this technique to classification and translation models, it shows the potential vulnerability of generative models to poisoning. 

\section{Steganography}\label{sec:steganography}
Steganography is the process of hiding secret messages inside non-secret mediums to evade detection. The idea is not to simply hide the message content, as one would with encryption. Instead, the goal is to hide the very fact that a message is being sent.

\subsection{Steganography without Generative Models}
Steganography hides the message in some inconspicuous ``cover'' medium such as an image or audio sample. While there are a number of ways to perform steganography, a popular method is through embedding. Distortion can be added to a cover, which the recipient can decode to retrieve the secret message. This method is often used with images as cover ~\cite{bender1996techniques,pevny2010using,holub2014universal,holub2012designing}, but can also use audio~\cite{mat2011review} or text \cite{bennett2004linguistic,por2008whitesteg} as cover. However, even small distortions to an image or audio can be detected~\cite{fridrich2012rich,pibre2016deep,qian2015deep}. This issue is even more prevalent with text since many changes are evident to humans~\cite{bennett2004linguistic}.

\subsection{Steganography with Generative Models}
Methods using generative models for steganography can be classified into three categories: (1) {\em cover creation}, (2) {\em cover modification}, and (3) {\em generative steganography}. We illustrate this by taxonomizing recent work across these categories and types of data in~\cref{tab:stega}. 

The first category is cover creation. Here generative models are used to generate a cover media (e.g., an image, text, audio sample) into which the message can be embedded using other methods~\cite{shi2017ssgan,volkhonskiy2020steganographic} that may not rely on generative models. An advantage of this method is that it creates new cover media, which prevents detectors from comparing a potentially pre-existing version of the cover against a steganographic version. However, the distortion of the cover from embedding the secret message may still be discoverable. Remark that, the cover media may look conspicuous depending on the context, or the quality of the generative model used.

The second category involves modifying existing cover media and embedding the message in it. Methods in this category start from an initial (typically not model generated) cover media, which the secret message is then embedded into using a generative model through a variety of methods. Similar to more traditional steganography through embedding, the message can be transformed into a vector or modification matrix and then added to the cover through generated distortion~\cite{cui2019image,yang2018spatial,baluja2017hiding}. Here GANs' adversarial training is useful since the generator can add the distortion while the discriminator can try and detect steganography. This allows for more refined embedding since the model learns how to avoid the detector while still embedding the secret message. Other steganographic methods embed a secret sample of the same data type as the cover media. Instead of a text message, an entire image or audio sample can be transformed into a new form or vector and then hidden within a cover sample. This has been used in both images \cite{zhang2019invisible}, and audio \cite{ye2019heard,kreuk2019hide,jiang2020smartsteganogaphy}.

Finally, the last category involves the use of generative models to directly produce the cover media with the secret message embedded in it. The idea is to use bits of the secret message to directly sample from the probability distribution of the generative model. The receiver can then recover the message by examining the steganographic image \cite{hu2018novel,liu2017coverless,li2020generative} or audio sample \cite{chen2021distribution,yang2018aag} and inverting the choices made in the generation of the sample. 

It is worth noting that similar techniques have been used to produce text steganographically~\cite{moraldo2014approach,ziegler2019neural,fang2017generating}. In these cases, a language model (e.g., a Markov chain, a Transformer, or LSTM) is used to define the probability distribution of text. The secret message is encoded into samples of the probability distribution of the language model. The exact encoding method varies, with two popular algorithms being Huffman coding~\cite{moffat2019huffman} and arithmetic coding~\cite{witten1987arithmetic}. The receiver can then invert the sampling from the probability distributions and retrieve the secret message. It is worth noting that text usually has much lower capacity than images or audio. Also, text cannot be easily distorted without it being noticeable; adding out of place letters or punctuation is conspicuous, so generative steganography makes text significantly more viable as a steganographic carrier. 

It is also possible to perform generative steganography by training generative models to embed a message or secret sample into future samples it generates. For instance, a cycle-GAN can be trained so that all images it generates have a secret image hidden in the signal of the new image \cite{chu2017cyclegan}. A similar method is used by Zhang et al.~\cite{zhang2019generative} where the message is embedded in certain parts of the image which are then retained in future generated images.

\begin{table}[thb]
    \caption{\small Methods using generative models for steganography. Columns represent the different ways the models can be used. Rows represent the data types being generated.}
    \label{tab:stega}
    \small
    \centering
    \begin{tabular}{l||l|l|l|}
    \cline{2-4}
     & \multicolumn{3}{c|}{Method} \\ \cline{2-4}
    ~ & Cover Creation & Cover Modification & Generative Steganography \\ \hline \hline
    Images & \cite{cui2019image, shi2017ssgan,volkhonskiy2020steganographic} & \cite{shi2017ssgan,yang2018spatial,baluja2017hiding,zhang2019invisible} & \cite{hu2018novel,liu2017coverless,li2020generative,chu2017cyclegan,zhang2019generative} \\ \hline
    Audio &\textbf{N/A} &   \cite{ye2019heard,kreuk2019hide,jiang2020smartsteganogaphy} & \cite{chen2021distribution,yang2018aag} \\ \hline
    Text & \textbf{N/A} & \textbf{N/A} & \cite{moraldo2014approach,ziegler2019neural,fang2017generating,yang2018rnn,yang2020vae,zhou2021linguistic,zhang2021provably,yang2021linguistic,yang2019gan} \\ \hline
    \end{tabular}
\end{table}
\subsection{Detecting Steganography}
There are a number of papers for detecting steganography performed through embedding~\cite{yang2019ts,wen2019convolutional}. The idea is to detect differences in latent spaces and probability distributions between the images or text that contains messages compared to the version with nothing embedded. However, there is a dearth of steganalysis papers focused on detecting generative steganography. Many generative steganography papers measure detectability using measures such as perplexity (for text)~\cite{yang2019gan}, bits per sample (for audio)~\cite{chen2021distribution}, human comparison (all)~\cite{kreuk2019hide,luo2017text}, or discriminator models trained on innocent vs steganographic samples~\cite{hayes2017generating}.
Moreover, steganographic systems can potentially be thwarted if one can easily remove the secret message. For example, one could use GANs to resample suspicious images, thereby likely making the secret message unrecoverable in the process.

\section{Fairness}\label{sec:fairness}
There is a rich literature of proposed notions of fairness and their formalization and application to machine learning. These include notions of individual fairness such as fairness through awareness~\cite{dwork2012fairness} and counterfactual fairness~\cite{kusner2017counterfactual} and also notions of group fairness such as statistical parity~\cite{dwork2012fairness}, conditional statistical parity~\cite{corbett2017algorithmic}, and equalized odds~\cite{hardt2016equality}. However, many of these notions  conflict~\cite{dwork2012fairness} and there is little to no consensus as to what notion (if any) is the correct one.

An important motivation for the study of fairness in machine learning is the recognition that machine learning can be applied in situations where the stakes are high such as crime sentencing~\cite{angwin2016machine,dieterich2016compas} and loan approval~\cite{kallus2018residual,hardt2016equality}. The concern is that due to training bias, bias inherited in the data~\cite{kallus2018residual}, or other reasons, models trained may be biased. There have been many attempts to create truly fair models, from removing sensitive attributes from models representations~\cite{zemel2013learning,louizos2015variational} to optimizing models to ignore sensitive attributes~\cite{calders2009building,woodworth2017learning}. 

Generative models provide ways of generating fair data rather than having to train fair models or modifying an existing model to guarantee some fairness criterion. There are several proposals to generate fair datasets. One is FairGAN by Xu et al.~\cite{xu2018fairgan}. FairGAN can remove sensitive attributes and recreate the data without them. This has been used with text datasets to recreate the whole dataset and maintain relationships while removing any indication of sensitive features. In a similar vein, Sattigeri et al.~\cite{sattigeri2019fairness} propose removing sensitive features such as skin color from image databases. The hope is that as generative models improve, these methods can be used to completely replace datasets with sanitized versions. This would ensure that decisions made with that data are fair, without having to specifically train a fair model.  

A separate line of work explores the use of generative models to detect unfair models. While detecting unfair models can be done without generative models, one way to check for fairness is to use a generative model to synthesize data points with and without sensitive information and then feed them to the potentially unfair model. If the model returns different results, then it is unfair. This idea has been applied to both text~\cite{black2020fliptest} and images~\cite{denton2019detecting}, by changing sensitive attributes such as race, skin color, age, and others and evaluating model output for different values of these attributes.

\section{Threat of Generated Information}\label{sec:deepfakes}
Generative models can be used to falsify information or generate fake information that is then presented as legitimate in an attempt to influence people's thinking or decisions. A prominent example is deepfakes~\cite{fallis2020epistemic}, which are realistic-looking media (i.e., audio, photos, or videos) that depict situations or events that did not occur such as famous people saying things they did not say. The concern surrounding deepfakes is its potential use as disinformation: e.g., that unsuspecting individuals may believe that a deepfake is genuine and share it~\cite{ahmed2021inadvertently} on social media thereby potentially influencing public opinion~\cite{toews_2020} or maybe even the outcome of elections~\cite{solsman}. A related concern is that deepfakes have the potential to undermine the public's trust in information~\cite{futurism_trust}. 

It is also worth noting that the use of this technology for certain purposes has already been outlawed in several US states~\cite{curtis_2019,korosec_2019} due to its potential to cause harm, for example through the creation of revenge and celebrity porn.

\subsection{Creation of Deepfakes and Text-based Synthetic Media}
The use of generative models to modify images or videos has legitimate applications. Early examples of this technology focused on making small alterations to images, such as making a person look older~\cite{antipov2017face} or change the color of a purse~\cite{zhu2016generative}. These techniques were then further utilized to alter images more substantially, for example by pasting one person's face onto an image or video of someone else~\cite{nirkin2019fsgan}, creating an entirely new (fictional) people~\cite{karras2019style}, or even lips-syncing entirely new audio~\cite{suwajanakorn2017synthesizing}. 

These techniques are the basis for the most famous example of this threat, deepfakes. Deepfakes were first created by a Reddit user named Deepfake. There are several methods to create deepfakes, but an early one involved using autoencoders trained on facial images. The encoder extracts features from the face and the decoder reconstructs the face. By making two encoder-decoder pairs it is possible to swap the encoders and thereby swap faces. When the output of the encoder for face A is passed to the decoder of face B, the features of face A are essentially ``pasted'' onto face B \cite{nguyen2019deep}. While theoretically used by amateurs as a proof of concept, the threat of this technology has been demonstrated on politicians~\cite{verge2018deepfake}, celebrities~ \cite{mallenbaum_2019}, and in revenge porn~\cite{wang_2019}.

A popular method to generate deepfakes is DeepFaceLab~\cite{petrov2020deepfacelab}. This framework allows users to create their own deepfake models and customize them with different model architectures, loss functions, and targets. It is popular because it achieves high-quality results while requiring a comparatively small amount of user input. This is in contrast to earlier methods such as Synthesizing Obama~\cite{suwajanakorn2017synthesizing}, which required the manual construction of a full 3-D model of Obama's face.

Audio can also be faked as part of a video deepfake, or a standalone audio deepfake can be produced. Audio deepfakes are often produced through voice conversion~\cite{chen2014voice,arik2018neural} or impersonation~\cite{gao2018voice}. This process takes as input a sentence by the attacker and outputs an audio sample of the target uttering that sentence. While there are a variety of models and methods used, the general idea is to train a GAN or autoencoder to replicate the speaker's pitch, tone, and more features, while ensuring that the result matches the desired sentence(s). These techniques have already been used in several real-world voice impersonation attacks where companies were tricked into forwarding money to attackers~\cite{statt_2019}.

The threat of model-generated information presented as genuine is not limited to visual or audio media. For example, advances in language modeling have resulted in models that produce increasingly realistic outputs. For example, when OpenAI's GPT-2~ \cite{solaiman2019release} was released, the authors were concerned about malicious use of their work~\cite{clark_2020}. To prevent their model from being used for spam and disinformation, they initially released a smaller version of their model that only had about 124 million parameters. Several months later, they released a 355M version, then 774M, and finally they released the full 1.5 billion parameters model in November of 2019~\cite{solaiman_2020}. The time between each release allowed for detectors and other defensive measures to be developed. 

Since then GPT-3 has also been developed. While the model has not been open-sourced, an API makes it available to the public. Samples from it~\cite{gpt3_examples} alone, which can fool fool humans, have sparked concerns~\cite{das_2020}. For instance, a blog received 26 thousand visitors before someone evoked the possibility that its articles were generated using GPT-3~\cite{gpt3blogfool}. In contrast to deepfakes, which are enabled by generative models, text-based disinformation can easily be produced and disseminated by humans without the use of language models such as GPT-2 and GPT-3.

\subsection{Detecting Synthetic Media}
Several techniques have been proposed to detect deepfakes and generative models' output more generally. These often fall into two broad categories: (1) identify artifacts commonly present in generative models' output, or (2) detect if a possibly synthetic image came from a known generative model.

An example of a technique that falls in the first category is Li et al.~\cite{li2018ictu}. The idea is to use the fact that many deepfake generators do not properly replicate blinking. Indeed, in many generated videos the subject of the video did not blink at all or if they did it was rare. While this detection method worked for the models they examined (LSTM-RNN), better generative models were then designed to take blinking into account. Since then, there have been similar proposals that use politician facial patterns~\cite{agarwal2019protecting}, eye gaze~\cite{demir2021deep}, or general biological signals~\cite{ciftci2019fakecatcher} to identify deepfakes. 

The second category attempts to leverage knowledge of the model to detect synthetic media. This is related to the attribution problem, which involves attributing synthetic media to the specific model that produced it. Detectors such as GLTR~\cite{gehrmann-etal-2019-gltr} use the likelihood of text under GPT-2 to determine if that text was generated using GPT-2. Several papers have produced methods of inverting images back to the model~\cite{albright2019source}, or looking for specific telltale artifacts left behind by certain models~\cite{yu2019attributing}. Zhang et al.~\cite{zhang2020not} uses entropy measures to estimate if an image came from one of their known models. Chai et al.~\cite{chai2020makes} developed a method to split an image up into patches and use them to identify places where a synthetic image is most detectable. This allows defenders to quickly identify artifacts left behind by new models. A major downside of the attribution approach is that the model used must be known. While this assumption may be true for adversaries that use public models, an adversary who trains their own model may be able to avoid detection.

Additionally, some methods do not neatly fall into the two aforementioned categories. For example, Nataraj et al.~\cite{nataraj2019detecting} propose splitting an image into co-occurrence matrices they are able to detect if it was generated with a GAN. Wang et al.~\cite{wang2020cnn} show that a detector trained on one image generation model is transferable to many other image generators while still maintaining high accuracy. 
\section{Discussion \& Open Problems}\label{sec:discussion}
There are already many applications of generative models to security and privacy. It is expected that as research into generative model architectures leads to improvements, more security-related applications will emerge. In addition, improvements will also enhance existing applications. For example, generative models producing more realistic outputs will improve the security of steganography techniques (\cref{sec:steganography}), and the quality of synthetic datasets produced for privacy~(\cref{sec:datasynthesis}). At the same time, these improvements will also increase the quality of deepfakes and synthetic media, thus making them harder to detect.

Beyond improvements in generative models architectures and techniques, there are important open research problems that deserve discussion.

\paragraphb{Evaluating generative models}
As we discussed in~\cref{sec:background:output}, 
there is no one-size-fits-all metric and each specific application of generative techniques often requires its own sets of metrics and evaluation methodology. A metric or singular process that would be able to evaluate output realism, similarity to the training dataset, and diversity would vastly simplify the model training process. This problem is exacerbated when considering generative models in a security context, due to the lack of security metrics for some applications. 

For example, when generative models are used for image-based steganography~(\cref{sec:steganography}), the goal of the evaluation methodology may be to quantify the ability of an adversary to distinguish between images that have been produced using the generative model (and thus embed some hidden message) and ``normal'' images. But this depends on the adversarial threat model (i.e., on the assumed capabilities and knowledge of the adversary). Are the images being compared against samples in a dataset of ``normal'' images? Are they being evaluated by a human? In the first case, image comparison metrics are needed, while in the second human evaluation metrics would be more accurate.

In contrast, when applying generative models to produce adversarial examples~(\cref{sec:advml:advexamples}), important factors may include how well the attacking model represents the target's latent space, as well as the output diversity/stability. A model that scores high on these metrics would be able to generate a large number of adversarial examples, even if they may not be the most realistic. A possible direction for future research is to develop metrics and methodologies for evaluating generative models within the context of specific security-related applications.

\paragraphb{Detecting and attributing synthetic media to generative models}
A related open problem is the detection of synthetic media and its attribution to specific generative models. It seems natural to think that the more realistic a generative model's output is, the harder it becomes to distinguish it from real data. However, this is not necessarily true. For instance, take a generative model that reproduces {\em exactly} a photo from its training data and only ever produces this specific output; the model is easy to detect. It is also clear what it cannot produce (i.e., everything else). 

As described in~\cref{sec:deepfakes}, there are several techniques to detect deepfakes on the basis of ``artifacts'' left during the generation process. However, the limitation of these approaches is that one can (in principle) always train a better generative model that does not produce these artifacts. Thus, defenses in this vein are chasing an ever-moving target. Future research may investigate the possibility of detecting entire classes of generative techniques by the kind of artifacts they may produce, or examine the relationship between model complexity and the kinds of outputs it can produce. 
\section{Conclusions}\label{sec:conclusions}
Generative models have many applications. In this paper, we present a comprehensive survey of their application to security-related problems. We describe the different architectures of generative models and how they apply to different types of data or applications. We taxonomize metrics of generative models quality. We then turn our study to examining the use of generative models to enhance, automate, or otherwise facilitate attacks. We also survey their defensive uses and survey recent research on the use of generative techniques for privacy-preserving data synthesis, steganography, and fairness. We discuss the new threats of generative models being used to produce deepfakes and other types of synthetic media for the purpose of disinformation. Finally, we highlight some open problems and possible research directions.

{
    \bibliographystyle{ACM-Reference-Format}
    \bibliography{references}


\begin{thebibliography}{293}


\ifx \showCODEN    \undefined \def \showCODEN     #1{\unskip}     \fi
\ifx \showDOI      \undefined \def \showDOI       #1{#1}\fi
\ifx \showISBNx    \undefined \def \showISBNx     #1{\unskip}     \fi
\ifx \showISBNxiii \undefined \def \showISBNxiii  #1{\unskip}     \fi
\ifx \showISSN     \undefined \def \showISSN      #1{\unskip}     \fi
\ifx \showLCCN     \undefined \def \showLCCN      #1{\unskip}     \fi
\ifx \shownote     \undefined \def \shownote      #1{#1}          \fi
\ifx \showarticletitle \undefined \def \showarticletitle #1{#1}   \fi
\ifx \showURL      \undefined \def \showURL       {\relax}        \fi
\providecommand\bibfield[2]{#2}
\providecommand\bibinfo[2]{#2}
\providecommand\natexlab[1]{#1}
\providecommand\showeprint[2][]{arXiv:#2}

\bibitem[\protect\citeauthoryear{Abadi, Chu, Goodfellow, McMahan, Mironov,
  Talwar, and Zhang}{Abadi et~al\mbox{.}}{2016}]%
        {abadi2016deep}
\bibfield{author}{\bibinfo{person}{Martin Abadi}, \bibinfo{person}{Andy Chu},
  \bibinfo{person}{Ian Goodfellow}, \bibinfo{person}{H~Brendan McMahan},
  \bibinfo{person}{Ilya Mironov}, \bibinfo{person}{Kunal Talwar}, {and}
  \bibinfo{person}{Li Zhang}.} \bibinfo{year}{2016}\natexlab{}.
\newblock \showarticletitle{Deep learning with differential privacy}. In
  \bibinfo{booktitle}{\emph{Proceedings of the 2016 ACM SIGSAC conference on
  computer and communications security}}. \bibinfo{pages}{308--318}.
\newblock


\bibitem[\protect\citeauthoryear{Abay, Zhou, Kantarcioglu, Thuraisingham, and
  Sweeney}{Abay et~al\mbox{.}}{2018}]%
        {abay2018privacy}
\bibfield{author}{\bibinfo{person}{Nazmiye~Ceren Abay}, \bibinfo{person}{Yan
  Zhou}, \bibinfo{person}{Murat Kantarcioglu}, \bibinfo{person}{Bhavani
  Thuraisingham}, {and} \bibinfo{person}{Latanya Sweeney}.}
  \bibinfo{year}{2018}\natexlab{}.
\newblock \showarticletitle{Privacy preserving synthetic data release using
  deep learning}. In \bibinfo{booktitle}{\emph{Joint European Conference on
  Machine Learning and Knowledge Discovery in Databases}}. Springer,
  \bibinfo{pages}{510--526}.
\newblock


\bibitem[\protect\citeauthoryear{Acs, Melis, Castelluccia, and
  De~Cristofaro}{Acs et~al\mbox{.}}{2018}]%
        {acs2018differentially}
\bibfield{author}{\bibinfo{person}{Gergely Acs}, \bibinfo{person}{Luca Melis},
  \bibinfo{person}{Claude Castelluccia}, {and} \bibinfo{person}{Emiliano
  De~Cristofaro}.} \bibinfo{year}{2018}\natexlab{}.
\newblock \showarticletitle{Differentially private mixture of generative neural
  networks}.
\newblock \bibinfo{journal}{\emph{IEEE Transactions on Knowledge and Data
  Engineering}} \bibinfo{volume}{31}, \bibinfo{number}{6}
  (\bibinfo{year}{2018}), \bibinfo{pages}{1109--1121}.
\newblock


\bibitem[\protect\citeauthoryear{Agarwal, Farid, Gu, He, Nagano, and
  Li}{Agarwal et~al\mbox{.}}{2019}]%
        {agarwal2019protecting}
\bibfield{author}{\bibinfo{person}{Shruti Agarwal}, \bibinfo{person}{Hany
  Farid}, \bibinfo{person}{Yuming Gu}, \bibinfo{person}{Mingming He},
  \bibinfo{person}{Koki Nagano}, {and} \bibinfo{person}{Hao Li}.}
  \bibinfo{year}{2019}\natexlab{}.
\newblock \showarticletitle{Protecting world leaders against deep fakes}. In
  \bibinfo{booktitle}{\emph{Proceedings of the IEEE Conference on Computer
  Vision and Pattern Recognition Workshops}}. \bibinfo{pages}{38--45}.
\newblock


\bibitem[\protect\citeauthoryear{Ahmed, Keskar, and Socher}{Ahmed
  et~al\mbox{.}}{2017}]%
        {ahmed2017weighted}
\bibfield{author}{\bibinfo{person}{Karim Ahmed},
  \bibinfo{person}{Nitish~Shirish Keskar}, {and} \bibinfo{person}{Richard
  Socher}.} \bibinfo{year}{2017}\natexlab{}.
\newblock \showarticletitle{Weighted transformer network for machine
  translation}.
\newblock \bibinfo{journal}{\emph{arXiv preprint arXiv:1711.02132}}
  (\bibinfo{year}{2017}).
\newblock


\bibitem[\protect\citeauthoryear{Ahmed}{Ahmed}{2021}]%
        {ahmed2021inadvertently}
\bibfield{author}{\bibinfo{person}{Saifuddin Ahmed}.}
  \bibinfo{year}{2021}\natexlab{}.
\newblock \showarticletitle{Who inadvertently shares deepfakes? Analyzing the
  role of political interest, cognitive ability, and social network size}.
\newblock \bibinfo{journal}{\emph{Telematics and Informatics}}
  \bibinfo{volume}{57} (\bibinfo{year}{2021}), \bibinfo{pages}{101508}.
\newblock


\bibitem[\protect\citeauthoryear{Albright and McCloskey}{Albright and
  McCloskey}{2019}]%
        {albright2019source}
\bibfield{author}{\bibinfo{person}{Michael Albright} {and}
  \bibinfo{person}{Scott McCloskey}.} \bibinfo{year}{2019}\natexlab{}.
\newblock \showarticletitle{Source generator attribution via inversion}. In
  \bibinfo{booktitle}{\emph{CVPR Workshop on Media Forensics}}.
  \bibinfo{pages}{96--103}.
\newblock


\bibitem[\protect\citeauthoryear{Alzantot and Srivastava}{Alzantot and
  Srivastava}{2019}]%
        {alzantotdifferential}
\bibfield{author}{\bibinfo{person}{Moustafa Alzantot} {and}
  \bibinfo{person}{Mani Srivastava}.} \bibinfo{year}{2019}\natexlab{}.
\newblock \bibinfo{title}{Differential Privacy Synthetic Data Generation using
  WGANs}.
\newblock
  \bibinfo{howpublished}{\url{https://github.com/nesl/nist_differential_privacy_synthetic_data_challenge}}.
\newblock


\bibitem[\protect\citeauthoryear{Anderson, Woodbridge, and Filar}{Anderson
  et~al\mbox{.}}{2016}]%
        {anderson2016deepdga}
\bibfield{author}{\bibinfo{person}{Hyrum~S Anderson}, \bibinfo{person}{Jonathan
  Woodbridge}, {and} \bibinfo{person}{Bobby Filar}.}
  \bibinfo{year}{2016}\natexlab{}.
\newblock \showarticletitle{DeepDGA: Adversarially-tuned domain generation and
  detection}. In \bibinfo{booktitle}{\emph{Proceedings of the 2016 ACM Workshop
  on Artificial Intelligence and Security}}. \bibinfo{pages}{13--21}.
\newblock


\bibitem[\protect\citeauthoryear{Angwin, Larson, Mattu, and Kirchner}{Angwin
  et~al\mbox{.}}{2016}]%
        {angwin2016machine}
\bibfield{author}{\bibinfo{person}{Julia Angwin}, \bibinfo{person}{Jeff
  Larson}, \bibinfo{person}{Surya Mattu}, {and} \bibinfo{person}{Lauren
  Kirchner}.} \bibinfo{year}{2016}\natexlab{}.
\newblock \showarticletitle{Machine bias}.
\newblock \bibinfo{journal}{\emph{ProPublica, May}}  \bibinfo{volume}{23}
  (\bibinfo{year}{2016}), \bibinfo{pages}{2016}.
\newblock


\bibitem[\protect\citeauthoryear{Antipov, Baccouche, and Dugelay}{Antipov
  et~al\mbox{.}}{2017}]%
        {antipov2017face}
\bibfield{author}{\bibinfo{person}{Grigory Antipov}, \bibinfo{person}{Moez
  Baccouche}, {and} \bibinfo{person}{Jean-Luc Dugelay}.}
  \bibinfo{year}{2017}\natexlab{}.
\newblock \showarticletitle{Face aging with conditional generative adversarial
  networks}. In \bibinfo{booktitle}{\emph{2017 IEEE international conference on
  image processing (ICIP)}}. IEEE, \bibinfo{pages}{2089--2093}.
\newblock


\bibitem[\protect\citeauthoryear{Arik, Chen, Peng, Ping, and Zhou}{Arik
  et~al\mbox{.}}{2018}]%
        {arik2018neural}
\bibfield{author}{\bibinfo{person}{Sercan~O Arik}, \bibinfo{person}{Jitong
  Chen}, \bibinfo{person}{Kainan Peng}, \bibinfo{person}{Wei Ping}, {and}
  \bibinfo{person}{Yanqi Zhou}.} \bibinfo{year}{2018}\natexlab{}.
\newblock \showarticletitle{Neural voice cloning with a few samples}.
\newblock \bibinfo{journal}{\emph{arXiv preprint arXiv:1802.06006}}
  (\bibinfo{year}{2018}).
\newblock


\bibitem[\protect\citeauthoryear{Arjovsky, Chintala, and Bottou}{Arjovsky
  et~al\mbox{.}}{2017}]%
        {arjovsky2017wasserstein}
\bibfield{author}{\bibinfo{person}{Martin Arjovsky}, \bibinfo{person}{Soumith
  Chintala}, {and} \bibinfo{person}{L{\'e}on Bottou}.}
  \bibinfo{year}{2017}\natexlab{}.
\newblock \showarticletitle{Wasserstein generative adversarial networks}. In
  \bibinfo{booktitle}{\emph{Proceedings of the 34th International Conference on
  Machine Learning-Volume 70}}. \bibinfo{pages}{214--223}.
\newblock


\bibitem[\protect\citeauthoryear{Arora and Zhang}{Arora and Zhang}{2017}]%
        {arora2017gans}
\bibfield{author}{\bibinfo{person}{Sanjeev Arora} {and} \bibinfo{person}{Yi
  Zhang}.} \bibinfo{year}{2017}\natexlab{}.
\newblock \showarticletitle{Do gans actually learn the distribution? an
  empirical study}.
\newblock \bibinfo{journal}{\emph{arXiv preprint arXiv:1706.08224}}
  (\bibinfo{year}{2017}).
\newblock


\bibitem[\protect\citeauthoryear{Athalye, Engstrom, Ilyas, and Kwok}{Athalye
  et~al\mbox{.}}{2018}]%
        {athalye2018synthesizing}
\bibfield{author}{\bibinfo{person}{Anish Athalye}, \bibinfo{person}{Logan
  Engstrom}, \bibinfo{person}{Andrew Ilyas}, {and} \bibinfo{person}{Kevin
  Kwok}.} \bibinfo{year}{2018}\natexlab{}.
\newblock \showarticletitle{Synthesizing robust adversarial examples}. In
  \bibinfo{booktitle}{\emph{International conference on machine learning}}.
  PMLR, \bibinfo{pages}{284--293}.
\newblock


\bibitem[\protect\citeauthoryear{Augenstein, McMahan, Ramage, Ramaswamy,
  Kairouz, Chen, Mathews, et~al\mbox{.}}{Augenstein et~al\mbox{.}}{2019}]%
        {augenstein2019generative}
\bibfield{author}{\bibinfo{person}{Sean Augenstein}, \bibinfo{person}{H~Brendan
  McMahan}, \bibinfo{person}{Daniel Ramage}, \bibinfo{person}{Swaroop
  Ramaswamy}, \bibinfo{person}{Peter Kairouz}, \bibinfo{person}{Mingqing Chen},
  \bibinfo{person}{Rajiv Mathews}, {et~al\mbox{.}}}
  \bibinfo{year}{2019}\natexlab{}.
\newblock \showarticletitle{Generative models for effective ml on private,
  decentralized datasets}.
\newblock \bibinfo{journal}{\emph{arXiv preprint arXiv:1911.06679}}
  (\bibinfo{year}{2019}).
\newblock


\bibitem[\protect\citeauthoryear{Baluja}{Baluja}{2017}]%
        {baluja2017hiding}
\bibfield{author}{\bibinfo{person}{Shumeet Baluja}.}
  \bibinfo{year}{2017}\natexlab{}.
\newblock \showarticletitle{Hiding images in plain sight: Deep steganography}.
  In \bibinfo{booktitle}{\emph{Advances in Neural Information Processing
  Systems}}. \bibinfo{pages}{2069--2079}.
\newblock


\bibitem[\protect\citeauthoryear{Bansal, Ma, Ramanan, and Sheikh}{Bansal
  et~al\mbox{.}}{2018}]%
        {bansal2018recycle}
\bibfield{author}{\bibinfo{person}{Aayush Bansal}, \bibinfo{person}{Shugao Ma},
  \bibinfo{person}{Deva Ramanan}, {and} \bibinfo{person}{Yaser Sheikh}.}
  \bibinfo{year}{2018}\natexlab{}.
\newblock \showarticletitle{Recycle-gan: Unsupervised video retargeting}. In
  \bibinfo{booktitle}{\emph{Proceedings of the European conference on computer
  vision (ECCV)}}. \bibinfo{pages}{119--135}.
\newblock


\bibitem[\protect\citeauthoryear{Bao, Liang, and Wang}{Bao
  et~al\mbox{.}}{2018}]%
        {bao2018featurized}
\bibfield{author}{\bibinfo{person}{Ruying Bao}, \bibinfo{person}{Sihang Liang},
  {and} \bibinfo{person}{Qingcan Wang}.} \bibinfo{year}{2018}\natexlab{}.
\newblock \showarticletitle{Featurized bidirectional gan: Adversarial defense
  via adversarially learned semantic inference}.
\newblock \bibinfo{journal}{\emph{arXiv preprint arXiv:1805.07862}}
  (\bibinfo{year}{2018}).
\newblock


\bibitem[\protect\citeauthoryear{Bau, Zhu, Wulff, Peebles, Strobelt, Zhou, and
  Torralba}{Bau et~al\mbox{.}}{2019}]%
        {bau2019seeing}
\bibfield{author}{\bibinfo{person}{David Bau}, \bibinfo{person}{Jun-Yan Zhu},
  \bibinfo{person}{Jonas Wulff}, \bibinfo{person}{William Peebles},
  \bibinfo{person}{Hendrik Strobelt}, \bibinfo{person}{Bolei Zhou}, {and}
  \bibinfo{person}{Antonio Torralba}.} \bibinfo{year}{2019}\natexlab{}.
\newblock \showarticletitle{Seeing what a {GAN} cannot generate}. In
  \bibinfo{booktitle}{\emph{Proceedings of the IEEE/CVF International
  Conference on Computer Vision}}. \bibinfo{pages}{4502--4511}.
\newblock


\bibitem[\protect\citeauthoryear{Beaulieu-Jones, Wu, Williams, Lee, Bhavnani,
  Byrd, and Greene}{Beaulieu-Jones et~al\mbox{.}}{2019}]%
        {beaulieu2019privacy}
\bibfield{author}{\bibinfo{person}{Brett~K Beaulieu-Jones},
  \bibinfo{person}{Zhiwei~Steven Wu}, \bibinfo{person}{Chris Williams},
  \bibinfo{person}{Ran Lee}, \bibinfo{person}{Sanjeev~P Bhavnani},
  \bibinfo{person}{James~Brian Byrd}, {and} \bibinfo{person}{Casey~S Greene}.}
  \bibinfo{year}{2019}\natexlab{}.
\newblock \showarticletitle{Privacy-preserving generative deep neural networks
  support clinical data sharing}.
\newblock \bibinfo{journal}{\emph{Circulation: Cardiovascular Quality and
  Outcomes}} \bibinfo{volume}{12}, \bibinfo{number}{7} (\bibinfo{year}{2019}),
  \bibinfo{pages}{e005122}.
\newblock


\bibitem[\protect\citeauthoryear{Bender, Gruhl, Morimoto, and Lu}{Bender
  et~al\mbox{.}}{1996}]%
        {bender1996techniques}
\bibfield{author}{\bibinfo{person}{Walter Bender}, \bibinfo{person}{Daniel
  Gruhl}, \bibinfo{person}{Norishige Morimoto}, {and} \bibinfo{person}{Anthony
  Lu}.} \bibinfo{year}{1996}\natexlab{}.
\newblock \showarticletitle{Techniques for data hiding}.
\newblock \bibinfo{journal}{\emph{IBM systems journal}} \bibinfo{volume}{35},
  \bibinfo{number}{3.4} (\bibinfo{year}{1996}), \bibinfo{pages}{313--336}.
\newblock


\bibitem[\protect\citeauthoryear{Bennett}{Bennett}{2004}]%
        {bennett2004linguistic}
\bibfield{author}{\bibinfo{person}{Krista Bennett}.}
  \bibinfo{year}{2004}\natexlab{}.
\newblock \showarticletitle{Linguistic steganography: Survey, analysis, and
  robustness concerns for hiding information in text}.
\newblock  (\bibinfo{year}{2004}).
\newblock


\bibitem[\protect\citeauthoryear{Bindschaedler, Shokri, and
  Gunter}{Bindschaedler et~al\mbox{.}}{2017}]%
        {bindschaedler2017plausible}
\bibfield{author}{\bibinfo{person}{Vincent Bindschaedler},
  \bibinfo{person}{Reza Shokri}, {and} \bibinfo{person}{Carl~A Gunter}.}
  \bibinfo{year}{2017}\natexlab{}.
\newblock \showarticletitle{Plausible Deniability for Privacy-Preserving Data
  Synthesis}.
\newblock \bibinfo{journal}{\emph{Proceedings of the VLDB Endowment}}
  \bibinfo{volume}{10}, \bibinfo{number}{5} (\bibinfo{year}{2017}).
\newblock


\bibitem[\protect\citeauthoryear{Bi{\'n}kowski, Donahue, Dieleman, Clark,
  Elsen, Casagrande, Cobo, and Simonyan}{Bi{\'n}kowski et~al\mbox{.}}{2019}]%
        {binkowski2019high}
\bibfield{author}{\bibinfo{person}{Miko{\l}aj Bi{\'n}kowski},
  \bibinfo{person}{Jeff Donahue}, \bibinfo{person}{Sander Dieleman},
  \bibinfo{person}{Aidan Clark}, \bibinfo{person}{Erich Elsen},
  \bibinfo{person}{Norman Casagrande}, \bibinfo{person}{Luis~C Cobo}, {and}
  \bibinfo{person}{Karen Simonyan}.} \bibinfo{year}{2019}\natexlab{}.
\newblock \showarticletitle{High Fidelity Speech Synthesis with Adversarial
  Networks}. In \bibinfo{booktitle}{\emph{International Conference on Learning
  Representations}}.
\newblock


\bibitem[\protect\citeauthoryear{Black, Yeom, and Fredrikson}{Black
  et~al\mbox{.}}{2020}]%
        {black2020fliptest}
\bibfield{author}{\bibinfo{person}{Emily Black}, \bibinfo{person}{Samuel Yeom},
  {and} \bibinfo{person}{Matt Fredrikson}.} \bibinfo{year}{2020}\natexlab{}.
\newblock \showarticletitle{FlipTest: fairness testing via optimal transport}.
  In \bibinfo{booktitle}{\emph{Proceedings of the 2020 Conference on Fairness,
  Accountability, and Transparency}}. \bibinfo{pages}{111--121}.
\newblock


\bibitem[\protect\citeauthoryear{Blei, Kucukelbir, and McAuliffe}{Blei
  et~al\mbox{.}}{2017}]%
        {blei2017variational}
\bibfield{author}{\bibinfo{person}{David~M Blei}, \bibinfo{person}{Alp
  Kucukelbir}, {and} \bibinfo{person}{Jon~D McAuliffe}.}
  \bibinfo{year}{2017}\natexlab{}.
\newblock \showarticletitle{Variational inference: A review for statisticians}.
\newblock \bibinfo{journal}{\emph{Journal of the American statistical
  Association}} \bibinfo{volume}{112}, \bibinfo{number}{518}
  (\bibinfo{year}{2017}), \bibinfo{pages}{859--877}.
\newblock


\bibitem[\protect\citeauthoryear{Bontrager, Roy, Togelius, Memon, and
  Ross}{Bontrager et~al\mbox{.}}{2018}]%
        {bontrager2018deepmasterprints}
\bibfield{author}{\bibinfo{person}{Philip Bontrager}, \bibinfo{person}{Aditi
  Roy}, \bibinfo{person}{Julian Togelius}, \bibinfo{person}{Nasir Memon}, {and}
  \bibinfo{person}{Arun Ross}.} \bibinfo{year}{2018}\natexlab{}.
\newblock \showarticletitle{Deepmasterprints: Generating masterprints for
  dictionary attacks via latent variable evolution}. In
  \bibinfo{booktitle}{\emph{2018 IEEE 9th International Conference on
  Biometrics Theory, Applications and Systems (BTAS)}}. IEEE,
  \bibinfo{pages}{1--9}.
\newblock


\bibitem[\protect\citeauthoryear{Borji}{Borji}{2019}]%
        {borji2019pros}
\bibfield{author}{\bibinfo{person}{Ali Borji}.}
  \bibinfo{year}{2019}\natexlab{}.
\newblock \showarticletitle{Pros and cons of gan evaluation measures}.
\newblock \bibinfo{journal}{\emph{Computer Vision and Image Understanding}}
  \bibinfo{volume}{179} (\bibinfo{year}{2019}), \bibinfo{pages}{41--65}.
\newblock


\bibitem[\protect\citeauthoryear{Brock, Donahue, and Simonyan}{Brock
  et~al\mbox{.}}{2018}]%
        {brock2018large}
\bibfield{author}{\bibinfo{person}{Andrew Brock}, \bibinfo{person}{Jeff
  Donahue}, {and} \bibinfo{person}{Karen Simonyan}.}
  \bibinfo{year}{2018}\natexlab{}.
\newblock \showarticletitle{Large Scale GAN Training for High Fidelity Natural
  Image Synthesis}. In \bibinfo{booktitle}{\emph{International Conference on
  Learning Representations}}.
\newblock


\bibitem[\protect\citeauthoryear{Brown, Mann, Ryder, Subbiah, Kaplan, Dhariwal,
  Neelakantan, Shyam, Sastry, Askell, et~al\mbox{.}}{Brown
  et~al\mbox{.}}{2020}]%
        {brown2020language}
\bibfield{author}{\bibinfo{person}{Tom~B Brown}, \bibinfo{person}{Benjamin
  Mann}, \bibinfo{person}{Nick Ryder}, \bibinfo{person}{Melanie Subbiah},
  \bibinfo{person}{Jared Kaplan}, \bibinfo{person}{Prafulla Dhariwal},
  \bibinfo{person}{Arvind Neelakantan}, \bibinfo{person}{Pranav Shyam},
  \bibinfo{person}{Girish Sastry}, \bibinfo{person}{Amanda Askell},
  {et~al\mbox{.}}} \bibinfo{year}{2020}\natexlab{}.
\newblock \showarticletitle{Language models are few-shot learners}.
\newblock \bibinfo{journal}{\emph{arXiv preprint arXiv:2005.14165}}
  (\bibinfo{year}{2020}).
\newblock


\bibitem[\protect\citeauthoryear{Burda, Grosse, and Salakhutdinov}{Burda
  et~al\mbox{.}}{2016}]%
        {burda2015importance}
\bibfield{author}{\bibinfo{person}{Yuri Burda}, \bibinfo{person}{Roger~B
  Grosse}, {and} \bibinfo{person}{Ruslan Salakhutdinov}.}
  \bibinfo{year}{2016}\natexlab{}.
\newblock \showarticletitle{Importance Weighted Autoencoders}. In
  \bibinfo{booktitle}{\emph{ICLR (Poster)}}.
\newblock


\bibitem[\protect\citeauthoryear{Calders, Kamiran, and Pechenizkiy}{Calders
  et~al\mbox{.}}{2009}]%
        {calders2009building}
\bibfield{author}{\bibinfo{person}{Toon Calders}, \bibinfo{person}{Faisal
  Kamiran}, {and} \bibinfo{person}{Mykola Pechenizkiy}.}
  \bibinfo{year}{2009}\natexlab{}.
\newblock \showarticletitle{Building classifiers with independency
  constraints}. In \bibinfo{booktitle}{\emph{2009 IEEE International Conference
  on Data Mining Workshops}}. IEEE, \bibinfo{pages}{13--18}.
\newblock


\bibitem[\protect\citeauthoryear{Chai, Bau, Lim, and Isola}{Chai
  et~al\mbox{.}}{2020}]%
        {chai2020makes}
\bibfield{author}{\bibinfo{person}{Lucy Chai}, \bibinfo{person}{David Bau},
  \bibinfo{person}{Ser-Nam Lim}, {and} \bibinfo{person}{Phillip Isola}.}
  \bibinfo{year}{2020}\natexlab{}.
\newblock \showarticletitle{What makes fake images detectable? Understanding
  properties that generalize}. In \bibinfo{booktitle}{\emph{European Conference
  on Computer Vision}}. Springer, \bibinfo{pages}{103--120}.
\newblock


\bibitem[\protect\citeauthoryear{Chakraborty, Alam, Dey, Chattopadhyay, and
  Mukhopadhyay}{Chakraborty et~al\mbox{.}}{2018}]%
        {chakraborty2018adversarial}
\bibfield{author}{\bibinfo{person}{Anirban Chakraborty},
  \bibinfo{person}{Manaar Alam}, \bibinfo{person}{Vishal Dey},
  \bibinfo{person}{Anupam Chattopadhyay}, {and} \bibinfo{person}{Debdeep
  Mukhopadhyay}.} \bibinfo{year}{2018}\natexlab{}.
\newblock \showarticletitle{Adversarial attacks and defences: A survey}.
\newblock \bibinfo{journal}{\emph{arXiv preprint arXiv:1810.00069}}
  (\bibinfo{year}{2018}).
\newblock


\bibitem[\protect\citeauthoryear{Chandy, Rasekh, Barker, and Shafiee}{Chandy
  et~al\mbox{.}}{2019}]%
        {chandy2019cyberattack}
\bibfield{author}{\bibinfo{person}{Sarin~E Chandy}, \bibinfo{person}{Amin
  Rasekh}, \bibinfo{person}{Zachary~A Barker}, {and} \bibinfo{person}{M~Ehsan
  Shafiee}.} \bibinfo{year}{2019}\natexlab{}.
\newblock \showarticletitle{Cyberattack detection using deep generative models
  with variational inference}.
\newblock \bibinfo{journal}{\emph{Journal of Water Resources Planning and
  Management}} \bibinfo{volume}{145}, \bibinfo{number}{2}
  (\bibinfo{year}{2019}), \bibinfo{pages}{04018093}.
\newblock


\bibitem[\protect\citeauthoryear{Chen, Zhou, Zhao, Chen, Zhang, and Yu}{Chen
  et~al\mbox{.}}{2021}]%
        {chen2021distribution}
\bibfield{author}{\bibinfo{person}{Kejiang Chen}, \bibinfo{person}{Hang Zhou},
  \bibinfo{person}{Hanqing Zhao}, \bibinfo{person}{Dongdong Chen},
  \bibinfo{person}{Weiming Zhang}, {and} \bibinfo{person}{Nenghai Yu}.}
  \bibinfo{year}{2021}\natexlab{}.
\newblock \showarticletitle{Distribution-Preserving Steganography Based on
  Text-to-Speech Generative Models}.
\newblock \bibinfo{journal}{\emph{IEEE Transactions on Dependable and Secure
  Computing}} (\bibinfo{year}{2021}).
\newblock


\bibitem[\protect\citeauthoryear{Chen, Dai, Tao, Zhang, Gan, Shen, Zhang, Wang,
  Zhang, and Carin}{Chen et~al\mbox{.}}{2018a}]%
        {chen2018adversarial}
\bibfield{author}{\bibinfo{person}{Liqun Chen}, \bibinfo{person}{Shuyang Dai},
  \bibinfo{person}{Chenyang Tao}, \bibinfo{person}{Haichao Zhang},
  \bibinfo{person}{Zhe Gan}, \bibinfo{person}{Dinghan Shen},
  \bibinfo{person}{Yizhe Zhang}, \bibinfo{person}{Guoyin Wang},
  \bibinfo{person}{Ruiyi Zhang}, {and} \bibinfo{person}{Lawrence Carin}.}
  \bibinfo{year}{2018}\natexlab{a}.
\newblock \showarticletitle{Adversarial text generation via feature-mover's
  distance}. In \bibinfo{booktitle}{\emph{Advances in Neural Information
  Processing Systems}}. \bibinfo{pages}{4666--4677}.
\newblock


\bibitem[\protect\citeauthoryear{Chen, Ling, Liu, and Dai}{Chen
  et~al\mbox{.}}{2014}]%
        {chen2014voice}
\bibfield{author}{\bibinfo{person}{Ling-Hui Chen}, \bibinfo{person}{Zhen-Hua
  Ling}, \bibinfo{person}{Li-Juan Liu}, {and} \bibinfo{person}{Li-Rong Dai}.}
  \bibinfo{year}{2014}\natexlab{}.
\newblock \showarticletitle{Voice conversion using deep neural networks with
  layer-wise generative training}.
\newblock \bibinfo{journal}{\emph{IEEE/ACM Transactions on Audio, Speech, and
  Language Processing}} \bibinfo{volume}{22}, \bibinfo{number}{12}
  (\bibinfo{year}{2014}), \bibinfo{pages}{1859--1872}.
\newblock


\bibitem[\protect\citeauthoryear{Chen, Duan, Houthooft, Schulman, Sutskever,
  and Abbeel}{Chen et~al\mbox{.}}{2016}]%
        {chen2016infogan}
\bibfield{author}{\bibinfo{person}{Xi Chen}, \bibinfo{person}{Yan Duan},
  \bibinfo{person}{Rein Houthooft}, \bibinfo{person}{John Schulman},
  \bibinfo{person}{Ilya Sutskever}, {and} \bibinfo{person}{Pieter Abbeel}.}
  \bibinfo{year}{2016}\natexlab{}.
\newblock \showarticletitle{Infogan: Interpretable representation learning by
  information maximizing generative adversarial nets}. In
  \bibinfo{booktitle}{\emph{Advances in neural information processing
  systems}}. \bibinfo{pages}{2172--2180}.
\newblock


\bibitem[\protect\citeauthoryear{Chen, Liu, Li, Lu, and Song}{Chen
  et~al\mbox{.}}{2017}]%
        {chen2017targeted}
\bibfield{author}{\bibinfo{person}{Xinyun Chen}, \bibinfo{person}{Chang Liu},
  \bibinfo{person}{Bo Li}, \bibinfo{person}{Kimberly Lu}, {and}
  \bibinfo{person}{Dawn Song}.} \bibinfo{year}{2017}\natexlab{}.
\newblock \showarticletitle{Targeted backdoor attacks on deep learning systems
  using data poisoning}.
\newblock \bibinfo{journal}{\emph{arXiv preprint arXiv:1712.05526}}
  (\bibinfo{year}{2017}).
\newblock


\bibitem[\protect\citeauthoryear{Chen, Mishra, Rohaninejad, and Abbeel}{Chen
  et~al\mbox{.}}{2018b}]%
        {chen2018pixelsnail}
\bibfield{author}{\bibinfo{person}{Xi Chen}, \bibinfo{person}{Nikhil Mishra},
  \bibinfo{person}{Mostafa Rohaninejad}, {and} \bibinfo{person}{Pieter
  Abbeel}.} \bibinfo{year}{2018}\natexlab{b}.
\newblock \showarticletitle{Pixelsnail: An improved autoregressive generative
  model}. In \bibinfo{booktitle}{\emph{International Conference on Machine
  Learning}}. \bibinfo{pages}{864--872}.
\newblock


\bibitem[\protect\citeauthoryear{Cho, Van~Merri{\"e}nboer, Gulcehre, Bahdanau,
  Bougares, Schwenk, and Bengio}{Cho et~al\mbox{.}}{2014}]%
        {cho2014learning}
\bibfield{author}{\bibinfo{person}{Kyunghyun Cho}, \bibinfo{person}{Bart
  Van~Merri{\"e}nboer}, \bibinfo{person}{Caglar Gulcehre},
  \bibinfo{person}{Dzmitry Bahdanau}, \bibinfo{person}{Fethi Bougares},
  \bibinfo{person}{Holger Schwenk}, {and} \bibinfo{person}{Yoshua Bengio}.}
  \bibinfo{year}{2014}\natexlab{}.
\newblock \showarticletitle{Learning phrase representations using RNN
  encoder-decoder for statistical machine translation}.
\newblock \bibinfo{journal}{\emph{arXiv preprint arXiv:1406.1078}}
  (\bibinfo{year}{2014}).
\newblock


\bibitem[\protect\citeauthoryear{Christian}{Christian}{2018}]%
        {futurism_trust}
\bibfield{author}{\bibinfo{person}{Jon Christian}.}
  \bibinfo{year}{2018}\natexlab{}.
\newblock \bibinfo{title}{Lawmakers: Deepfakes Could “Undermine Public
  Trust” in “Objective Depictions of Reality”}.
\newblock
  \bibinfo{howpublished}{\url{https://futurism.com/the-byte/lawmakers-deepfakes-undermine-public-trust}}.
\newblock


\bibitem[\protect\citeauthoryear{Chu, Zhmoginov, and Sandler}{Chu
  et~al\mbox{.}}{2017}]%
        {chu2017cyclegan}
\bibfield{author}{\bibinfo{person}{Casey Chu}, \bibinfo{person}{Andrey
  Zhmoginov}, {and} \bibinfo{person}{Mark Sandler}.}
  \bibinfo{year}{2017}\natexlab{}.
\newblock \showarticletitle{Cyclegan, a master of steganography}.
\newblock \bibinfo{journal}{\emph{arXiv preprint arXiv:1712.02950}}
  (\bibinfo{year}{2017}).
\newblock


\bibitem[\protect\citeauthoryear{Ciftci, Demir, and Yin}{Ciftci
  et~al\mbox{.}}{2020}]%
        {ciftci2019fakecatcher}
\bibfield{author}{\bibinfo{person}{Umur~Aybars Ciftci}, \bibinfo{person}{Ilke
  Demir}, {and} \bibinfo{person}{Lijun Yin}.} \bibinfo{year}{2020}\natexlab{}.
\newblock \showarticletitle{Fakecatcher: Detection of synthetic portrait videos
  using biological signals}.
\newblock \bibinfo{journal}{\emph{IEEE Transactions on Pattern Analysis and
  Machine Intelligence}} (\bibinfo{year}{2020}).
\newblock


\bibitem[\protect\citeauthoryear{Clark}{Clark}{2020}]%
        {clark_2020}
\bibfield{author}{\bibinfo{person}{Jack Clark}.}
  \bibinfo{year}{2020}\natexlab{}.
\newblock \bibinfo{title}{GPT-2: 6-Month Follow-Up}.
\newblock
  \bibinfo{howpublished}{\url{https://openai.com/blog/gpt-2-6-month-follow-up/}}.
\newblock


\bibitem[\protect\citeauthoryear{Collobert and Weston}{Collobert and
  Weston}{2008}]%
        {collobert2008unified}
\bibfield{author}{\bibinfo{person}{Ronan Collobert} {and}
  \bibinfo{person}{Jason Weston}.} \bibinfo{year}{2008}\natexlab{}.
\newblock \showarticletitle{A unified architecture for natural language
  processing: Deep neural networks with multitask learning}. In
  \bibinfo{booktitle}{\emph{Proceedings of the 25th international conference on
  Machine learning}}. \bibinfo{pages}{160--167}.
\newblock


\bibitem[\protect\citeauthoryear{Corbett-Davies, Pierson, Feller, Goel, and
  Huq}{Corbett-Davies et~al\mbox{.}}{2017}]%
        {corbett2017algorithmic}
\bibfield{author}{\bibinfo{person}{Sam Corbett-Davies}, \bibinfo{person}{Emma
  Pierson}, \bibinfo{person}{Avi Feller}, \bibinfo{person}{Sharad Goel}, {and}
  \bibinfo{person}{Aziz Huq}.} \bibinfo{year}{2017}\natexlab{}.
\newblock \showarticletitle{Algorithmic decision making and the cost of
  fairness}. In \bibinfo{booktitle}{\emph{Proceedings of the 23rd acm sigkdd
  international conference on knowledge discovery and data mining}}.
  \bibinfo{pages}{797--806}.
\newblock


\bibitem[\protect\citeauthoryear{Corley, Lwowski, and Hoffman}{Corley
  et~al\mbox{.}}{2019}]%
        {corley2019domaingan}
\bibfield{author}{\bibinfo{person}{Isaac Corley}, \bibinfo{person}{Jonathan
  Lwowski}, {and} \bibinfo{person}{Justin Hoffman}.}
  \bibinfo{year}{2019}\natexlab{}.
\newblock \showarticletitle{DomainGAN: Generating Adversarial Examples to
  Attack Domain Generation Algorithm Classifiers}.
\newblock \bibinfo{journal}{\emph{arXiv preprint arXiv:1911.06285}}
  (\bibinfo{year}{2019}).
\newblock


\bibitem[\protect\citeauthoryear{Cui, Zhou, Fu, Meng, Sun, and Wu}{Cui
  et~al\mbox{.}}{2019}]%
        {cui2019image}
\bibfield{author}{\bibinfo{person}{Qi Cui}, \bibinfo{person}{Zhili Zhou},
  \bibinfo{person}{Zhangjie Fu}, \bibinfo{person}{Ruohan Meng},
  \bibinfo{person}{Xingming Sun}, {and} \bibinfo{person}{QM~Jonathan Wu}.}
  \bibinfo{year}{2019}\natexlab{}.
\newblock \showarticletitle{Image steganography based on foreground object
  generation by generative adversarial networks in mobile edge computing with
  Internet of Things}.
\newblock \bibinfo{journal}{\emph{IEEE Access}}  \bibinfo{volume}{7}
  (\bibinfo{year}{2019}), \bibinfo{pages}{90815--90824}.
\newblock


\bibitem[\protect\citeauthoryear{Curtis}{Curtis}{2019}]%
        {curtis_2019}
\bibfield{author}{\bibinfo{person}{Cara Curtis}.}
  \bibinfo{year}{2019}\natexlab{}.
\newblock \bibinfo{title}{California makes deepfakes illegal to curb revenge
  porn and doctored political videos}.
\newblock
  \bibinfo{howpublished}{\url{https://thenextweb.com/tech/2019/10/07/california-makes-deepfakes-illegal-to-curb-revenge-porn-and-doctored-political-videos/}}.
\newblock


\bibitem[\protect\citeauthoryear{Dai, Yang, Yang, Carbonell, Le, and
  Salakhutdinov}{Dai et~al\mbox{.}}{2019}]%
        {dai2019transformer}
\bibfield{author}{\bibinfo{person}{Zihang Dai}, \bibinfo{person}{Zhilin Yang},
  \bibinfo{person}{Yiming Yang}, \bibinfo{person}{Jaime~G Carbonell},
  \bibinfo{person}{Quoc Le}, {and} \bibinfo{person}{Ruslan Salakhutdinov}.}
  \bibinfo{year}{2019}\natexlab{}.
\newblock \showarticletitle{Transformer-XL: Attentive Language Models beyond a
  Fixed-Length Context}. In \bibinfo{booktitle}{\emph{Proceedings of the 57th
  Annual Meeting of the Association for Computational Linguistics}}.
  \bibinfo{pages}{2978--2988}.
\newblock


\bibitem[\protect\citeauthoryear{Das}{Das}{2020}]%
        {das_2020}
\bibfield{author}{\bibinfo{person}{Sejuti Das}.}
  \bibinfo{year}{2020}\natexlab{}.
\newblock \bibinfo{title}{How OpenAI's GPT-3 Can Be Alarming For The Society}.
\newblock
  \bibinfo{howpublished}{\url{https://analyticsindiamag.com/how-openais-gpt-3-can-be-alarming-for-the-society/}}.
\newblock


\bibitem[\protect\citeauthoryear{Dehghani, Gouws, Vinyals, Uszkoreit, and
  Kaiser}{Dehghani et~al\mbox{.}}{2018}]%
        {dehghani2018universal}
\bibfield{author}{\bibinfo{person}{Mostafa Dehghani}, \bibinfo{person}{Stephan
  Gouws}, \bibinfo{person}{Oriol Vinyals}, \bibinfo{person}{Jakob Uszkoreit},
  {and} \bibinfo{person}{{\L}ukasz Kaiser}.} \bibinfo{year}{2018}\natexlab{}.
\newblock \showarticletitle{Universal transformers}.
\newblock \bibinfo{journal}{\emph{arXiv preprint arXiv:1807.03819}}
  (\bibinfo{year}{2018}).
\newblock


\bibitem[\protect\citeauthoryear{Demir and Ciftci}{Demir and Ciftci}{2021}]%
        {demir2021deep}
\bibfield{author}{\bibinfo{person}{Ilke Demir} {and} \bibinfo{person}{Umur~A
  Ciftci}.} \bibinfo{year}{2021}\natexlab{}.
\newblock \showarticletitle{Where Do Deep Fakes Look? Synthetic Face Detection
  via Gaze Tracking}.
\newblock \bibinfo{journal}{\emph{arXiv preprint arXiv:2101.01165}}
  (\bibinfo{year}{2021}).
\newblock


\bibitem[\protect\citeauthoryear{Denkowski and Lavie}{Denkowski and
  Lavie}{2011}]%
        {denkowski2011meteor}
\bibfield{author}{\bibinfo{person}{Michael Denkowski} {and}
  \bibinfo{person}{Alon Lavie}.} \bibinfo{year}{2011}\natexlab{}.
\newblock \showarticletitle{Meteor 1.3: Automatic metric for reliable
  optimization and evaluation of machine translation systems}. In
  \bibinfo{booktitle}{\emph{Proceedings of the sixth workshop on statistical
  machine translation}}. Association for Computational Linguistics,
  \bibinfo{pages}{85--91}.
\newblock


\bibitem[\protect\citeauthoryear{Denton, Hutchinson, Mitchell, and
  Gebru}{Denton et~al\mbox{.}}{2019}]%
        {denton2019detecting}
\bibfield{author}{\bibinfo{person}{Emily Denton}, \bibinfo{person}{Ben
  Hutchinson}, \bibinfo{person}{Margaret Mitchell}, {and}
  \bibinfo{person}{Timnit Gebru}.} \bibinfo{year}{2019}\natexlab{}.
\newblock \showarticletitle{Detecting bias with generative counterfactual face
  attribute augmentation}.
\newblock \bibinfo{journal}{\emph{arXiv preprint arXiv:1906.06439}}
  (\bibinfo{year}{2019}).
\newblock


\bibitem[\protect\citeauthoryear{Devlin, Chang, Lee, and Toutanova}{Devlin
  et~al\mbox{.}}{2018}]%
        {devlin2018bert}
\bibfield{author}{\bibinfo{person}{Jacob Devlin}, \bibinfo{person}{Ming-Wei
  Chang}, \bibinfo{person}{Kenton Lee}, {and} \bibinfo{person}{Kristina
  Toutanova}.} \bibinfo{year}{2018}\natexlab{}.
\newblock \showarticletitle{Bert: Pre-training of deep bidirectional
  transformers for language understanding}.
\newblock \bibinfo{journal}{\emph{arXiv preprint arXiv:1810.04805}}
  (\bibinfo{year}{2018}).
\newblock


\bibitem[\protect\citeauthoryear{Dieterich, Mendoza, and Brennan}{Dieterich
  et~al\mbox{.}}{2016}]%
        {dieterich2016compas}
\bibfield{author}{\bibinfo{person}{William Dieterich},
  \bibinfo{person}{Christina Mendoza}, {and} \bibinfo{person}{Tim Brennan}.}
  \bibinfo{year}{2016}\natexlab{}.
\newblock \showarticletitle{COMPAS risk scales: Demonstrating accuracy equity
  and predictive parity}.
\newblock \bibinfo{journal}{\emph{Northpointe Inc}} (\bibinfo{year}{2016}).
\newblock


\bibitem[\protect\citeauthoryear{Donahue, McAuley, and Puckette}{Donahue
  et~al\mbox{.}}{2018}]%
        {donahue2018adversarial}
\bibfield{author}{\bibinfo{person}{Chris Donahue}, \bibinfo{person}{Julian
  McAuley}, {and} \bibinfo{person}{Miller Puckette}.}
  \bibinfo{year}{2018}\natexlab{}.
\newblock \showarticletitle{Adversarial Audio Synthesis}. In
  \bibinfo{booktitle}{\emph{International Conference on Learning
  Representations}}.
\newblock


\bibitem[\protect\citeauthoryear{Donahue, Kr{\"a}henb{\"u}hl, and
  Darrell}{Donahue et~al\mbox{.}}{2016}]%
        {donahue2016adversarial}
\bibfield{author}{\bibinfo{person}{Jeff Donahue}, \bibinfo{person}{Philipp
  Kr{\"a}henb{\"u}hl}, {and} \bibinfo{person}{Trevor Darrell}.}
  \bibinfo{year}{2016}\natexlab{}.
\newblock \showarticletitle{Adversarial feature learning}.
\newblock \bibinfo{journal}{\emph{arXiv preprint arXiv:1605.09782}}
  (\bibinfo{year}{2016}).
\newblock


\bibitem[\protect\citeauthoryear{Dong, Hsiao, Yang, and Yang}{Dong
  et~al\mbox{.}}{2018}]%
        {dong2018musegan}
\bibfield{author}{\bibinfo{person}{Hao-Wen Dong}, \bibinfo{person}{Wen-Yi
  Hsiao}, \bibinfo{person}{Li-Chia Yang}, {and} \bibinfo{person}{Yi-Hsuan
  Yang}.} \bibinfo{year}{2018}\natexlab{}.
\newblock \showarticletitle{Musegan: Multi-track sequential generative
  adversarial networks for symbolic music generation and accompaniment}. In
  \bibinfo{booktitle}{\emph{Thirty-Second AAAI Conference on Artificial
  Intelligence}}.
\newblock


\bibitem[\protect\citeauthoryear{Dosovitskiy and Brox}{Dosovitskiy and
  Brox}{2016}]%
        {dosovitskiy2016generating}
\bibfield{author}{\bibinfo{person}{Alexey Dosovitskiy} {and}
  \bibinfo{person}{Thomas Brox}.} \bibinfo{year}{2016}\natexlab{}.
\newblock \showarticletitle{Generating images with perceptual similarity
  metrics based on deep networks}. In \bibinfo{booktitle}{\emph{Advances in
  neural information processing systems}}. \bibinfo{pages}{658--666}.
\newblock


\bibitem[\protect\citeauthoryear{Dwork}{Dwork}{2008}]%
        {dwork2008differential}
\bibfield{author}{\bibinfo{person}{Cynthia Dwork}.}
  \bibinfo{year}{2008}\natexlab{}.
\newblock \showarticletitle{Differential privacy: A survey of results}. In
  \bibinfo{booktitle}{\emph{International conference on theory and applications
  of models of computation}}. Springer, \bibinfo{pages}{1--19}.
\newblock


\bibitem[\protect\citeauthoryear{Dwork, Hardt, Pitassi, Reingold, and
  Zemel}{Dwork et~al\mbox{.}}{2012}]%
        {dwork2012fairness}
\bibfield{author}{\bibinfo{person}{Cynthia Dwork}, \bibinfo{person}{Moritz
  Hardt}, \bibinfo{person}{Toniann Pitassi}, \bibinfo{person}{Omer Reingold},
  {and} \bibinfo{person}{Richard Zemel}.} \bibinfo{year}{2012}\natexlab{}.
\newblock \showarticletitle{Fairness through awareness}. In
  \bibinfo{booktitle}{\emph{Proceedings of the 3rd innovations in theoretical
  computer science conference}}. \bibinfo{pages}{214--226}.
\newblock


\bibitem[\protect\citeauthoryear{Ebrahimi, Rao, Lowd, and Dou}{Ebrahimi
  et~al\mbox{.}}{2018}]%
        {ebrahimi2017hotflip}
\bibfield{author}{\bibinfo{person}{Javid Ebrahimi}, \bibinfo{person}{Anyi Rao},
  \bibinfo{person}{Daniel Lowd}, {and} \bibinfo{person}{Dejing Dou}.}
  \bibinfo{year}{2018}\natexlab{}.
\newblock \showarticletitle{HotFlip: White-Box Adversarial Examples for Text
  Classification}. In \bibinfo{booktitle}{\emph{Proceedings of the 56th Annual
  Meeting of the Association for Computational Linguistics (Volume 2: Short
  Papers)}}. \bibinfo{pages}{31--36}.
\newblock


\bibitem[\protect\citeauthoryear{Engel, Agrawal, Chen, Gulrajani, Donahue, and
  Roberts}{Engel et~al\mbox{.}}{2018a}]%
        {engel2018gansynth}
\bibfield{author}{\bibinfo{person}{Jesse Engel}, \bibinfo{person}{Kumar~Krishna
  Agrawal}, \bibinfo{person}{Shuo Chen}, \bibinfo{person}{Ishaan Gulrajani},
  \bibinfo{person}{Chris Donahue}, {and} \bibinfo{person}{Adam Roberts}.}
  \bibinfo{year}{2018}\natexlab{a}.
\newblock \showarticletitle{GANSynth: Adversarial Neural Audio Synthesis}. In
  \bibinfo{booktitle}{\emph{International Conference on Learning
  Representations}}.
\newblock


\bibitem[\protect\citeauthoryear{Engel, Hoffman, and Roberts}{Engel
  et~al\mbox{.}}{2018b}]%
        {engel2017latent}
\bibfield{author}{\bibinfo{person}{Jesse Engel}, \bibinfo{person}{Matthew
  Hoffman}, {and} \bibinfo{person}{Adam Roberts}.}
  \bibinfo{year}{2018}\natexlab{b}.
\newblock \showarticletitle{Latent Constraints: Learning to Generate
  Conditionally from Unconditional Generative Models}. In
  \bibinfo{booktitle}{\emph{International Conference on Learning
  Representations}}.
\newblock


\bibitem[\protect\citeauthoryear{Examples}{Examples}{2020}]%
        {gpt3_examples}
\bibfield{author}{\bibinfo{person}{GPT3 Examples}.}
  \bibinfo{year}{2020}\natexlab{}.
\newblock \bibinfo{howpublished}{\url{https://gpt3examples.com/}}.
\newblock


\bibitem[\protect\citeauthoryear{Eykholt, Evtimov, Fernandes, Li, Rahmati,
  Xiao, Prakash, Kohno, and Song}{Eykholt et~al\mbox{.}}{2018}]%
        {eykholt2018robust}
\bibfield{author}{\bibinfo{person}{Kevin Eykholt}, \bibinfo{person}{Ivan
  Evtimov}, \bibinfo{person}{Earlence Fernandes}, \bibinfo{person}{Bo Li},
  \bibinfo{person}{Amir Rahmati}, \bibinfo{person}{Chaowei Xiao},
  \bibinfo{person}{Atul Prakash}, \bibinfo{person}{Tadayoshi Kohno}, {and}
  \bibinfo{person}{Dawn Song}.} \bibinfo{year}{2018}\natexlab{}.
\newblock \showarticletitle{Robust physical-world attacks on deep learning
  visual classification}. In \bibinfo{booktitle}{\emph{Proceedings of the IEEE
  Conference on Computer Vision and Pattern Recognition}}.
  \bibinfo{pages}{1625--1634}.
\newblock


\bibitem[\protect\citeauthoryear{Falliere, Murchu, and Chien}{Falliere
  et~al\mbox{.}}{2011}]%
        {falliere2011w32}
\bibfield{author}{\bibinfo{person}{Nicolas Falliere}, \bibinfo{person}{Liam~O
  Murchu}, {and} \bibinfo{person}{Eric Chien}.}
  \bibinfo{year}{2011}\natexlab{}.
\newblock \showarticletitle{W32. stuxnet dossier}.
\newblock \bibinfo{journal}{\emph{White paper, Symantec Corp., Security
  Response}} \bibinfo{volume}{5}, \bibinfo{number}{6} (\bibinfo{year}{2011}),
  \bibinfo{pages}{29}.
\newblock


\bibitem[\protect\citeauthoryear{Fallis}{Fallis}{2020}]%
        {fallis2020epistemic}
\bibfield{author}{\bibinfo{person}{Don Fallis}.}
  \bibinfo{year}{2020}\natexlab{}.
\newblock \showarticletitle{The Epistemic Threat of Deepfakes}.
\newblock \bibinfo{journal}{\emph{Philosophy \& Technology}}
  (\bibinfo{year}{2020}), \bibinfo{pages}{1--21}.
\newblock


\bibitem[\protect\citeauthoryear{Fedus, Goodfellow, and Dai}{Fedus
  et~al\mbox{.}}{2018}]%
        {fedus2018maskgan}
\bibfield{author}{\bibinfo{person}{William Fedus}, \bibinfo{person}{Ian
  Goodfellow}, {and} \bibinfo{person}{Andrew~M Dai}.}
  \bibinfo{year}{2018}\natexlab{}.
\newblock \showarticletitle{MaskGAN: Better Text Generation via Filling in the
  \_}. In \bibinfo{booktitle}{\emph{International Conference on Learning
  Representations}}.
\newblock


\bibitem[\protect\citeauthoryear{Feng, Li, Zhu, and Chana}{Feng
  et~al\mbox{.}}{2017}]%
        {feng2017deep}
\bibfield{author}{\bibinfo{person}{Cheng Feng}, \bibinfo{person}{Tingting Li},
  \bibinfo{person}{Zhanxing Zhu}, {and} \bibinfo{person}{Deeph Chana}.}
  \bibinfo{year}{2017}\natexlab{}.
\newblock \showarticletitle{A deep learning-based framework for conducting
  stealthy attacks in industrial control systems}.
\newblock \bibinfo{journal}{\emph{arXiv preprint arXiv:1709.06397}}
  (\bibinfo{year}{2017}).
\newblock


\bibitem[\protect\citeauthoryear{Fortet and Mourier}{Fortet and
  Mourier}{1953}]%
        {fortet1953convergence}
\bibfield{author}{\bibinfo{person}{Robert Fortet} {and} \bibinfo{person}{Edith
  Mourier}.} \bibinfo{year}{1953}\natexlab{}.
\newblock \showarticletitle{Convergence de la r{\'e}partition empirique vers la
  r{\'e}partition th{\'e}orique}. In \bibinfo{booktitle}{\emph{Annales
  scientifiques de l'{\'E}cole Normale Sup{\'e}rieure}},
  Vol.~\bibinfo{volume}{70}. \bibinfo{pages}{267--285}.
\newblock


\bibitem[\protect\citeauthoryear{Fredrikson, Jha, and Ristenpart}{Fredrikson
  et~al\mbox{.}}{2015}]%
        {fredrikson2015model}
\bibfield{author}{\bibinfo{person}{Matt Fredrikson}, \bibinfo{person}{Somesh
  Jha}, {and} \bibinfo{person}{Thomas Ristenpart}.}
  \bibinfo{year}{2015}\natexlab{}.
\newblock \showarticletitle{Model inversion attacks that exploit confidence
  information and basic countermeasures}. In
  \bibinfo{booktitle}{\emph{Proceedings of the 22nd ACM SIGSAC Conference on
  Computer and Communications Security}}. \bibinfo{pages}{1322--1333}.
\newblock


\bibitem[\protect\citeauthoryear{Fridrich and Kodovsky}{Fridrich and
  Kodovsky}{2012}]%
        {fridrich2012rich}
\bibfield{author}{\bibinfo{person}{Jessica Fridrich} {and} \bibinfo{person}{Jan
  Kodovsky}.} \bibinfo{year}{2012}\natexlab{}.
\newblock \showarticletitle{Rich models for steganalysis of digital images}.
\newblock \bibinfo{journal}{\emph{IEEE Transactions on Information Forensics
  and Security}} \bibinfo{volume}{7}, \bibinfo{number}{3}
  (\bibinfo{year}{2012}), \bibinfo{pages}{868--882}.
\newblock


\bibitem[\protect\citeauthoryear{Frigerio, de~Oliveira, Gomez, and
  Duverger}{Frigerio et~al\mbox{.}}{2019}]%
        {frigerio2019differentially}
\bibfield{author}{\bibinfo{person}{Lorenzo Frigerio},
  \bibinfo{person}{Anderson~Santana de Oliveira}, \bibinfo{person}{Laurent
  Gomez}, {and} \bibinfo{person}{Patrick Duverger}.}
  \bibinfo{year}{2019}\natexlab{}.
\newblock \showarticletitle{Differentially private generative adversarial
  networks for time series, continuous, and discrete open data}. In
  \bibinfo{booktitle}{\emph{IFIP International Conference on ICT Systems
  Security and Privacy Protection}}. Springer, \bibinfo{pages}{151--164}.
\newblock


\bibitem[\protect\citeauthoryear{Gao, Singh, and Raj}{Gao
  et~al\mbox{.}}{2018}]%
        {gao2018voice}
\bibfield{author}{\bibinfo{person}{Yang Gao}, \bibinfo{person}{Rita Singh},
  {and} \bibinfo{person}{Bhiksha Raj}.} \bibinfo{year}{2018}\natexlab{}.
\newblock \showarticletitle{Voice impersonation using generative adversarial
  networks}. In \bibinfo{booktitle}{\emph{2018 IEEE International Conference on
  Acoustics, Speech and Signal Processing (ICASSP)}}. IEEE,
  \bibinfo{pages}{2506--2510}.
\newblock


\bibitem[\protect\citeauthoryear{Gehrmann, Strobelt, and Rush}{Gehrmann
  et~al\mbox{.}}{2019}]%
        {gehrmann-etal-2019-gltr}
\bibfield{author}{\bibinfo{person}{Sebastian Gehrmann},
  \bibinfo{person}{Hendrik Strobelt}, {and} \bibinfo{person}{Alexander Rush}.}
  \bibinfo{year}{2019}\natexlab{}.
\newblock \showarticletitle{{GLTR}: Statistical Detection and Visualization of
  Generated Text}. In \bibinfo{booktitle}{\emph{Proceedings of the 57th Annual
  Meeting of the Association for Computational Linguistics: System
  Demonstrations}}. \bibinfo{publisher}{Association for Computational
  Linguistics}, \bibinfo{address}{Florence, Italy}, \bibinfo{pages}{111--116}.
\newblock
\urldef\tempurl%
\url{https://doi.org/10.18653/v1/P19-3019}
\showDOI{\tempurl}


\bibitem[\protect\citeauthoryear{Gers, Schmidhuber, and Cummins}{Gers
  et~al\mbox{.}}{2000}]%
        {gers1999learning}
\bibfield{author}{\bibinfo{person}{Felix~A Gers}, \bibinfo{person}{J{\"u}rgen
  Schmidhuber}, {and} \bibinfo{person}{Fred Cummins}.}
  \bibinfo{year}{2000}\natexlab{}.
\newblock \showarticletitle{Learning to Forget: Continual Prediction with
  LSTM}.
\newblock \bibinfo{journal}{\emph{Neural Computation}} \bibinfo{volume}{12},
  \bibinfo{number}{10} (\bibinfo{year}{2000}), \bibinfo{pages}{2451--2471}.
\newblock


\bibitem[\protect\citeauthoryear{Gomez, Ren, Urtasun, and Grosse}{Gomez
  et~al\mbox{.}}{2017}]%
        {gomez2017reversible}
\bibfield{author}{\bibinfo{person}{Aidan~N Gomez}, \bibinfo{person}{Mengye
  Ren}, \bibinfo{person}{Raquel Urtasun}, {and} \bibinfo{person}{Roger~B
  Grosse}.} \bibinfo{year}{2017}\natexlab{}.
\newblock \showarticletitle{The reversible residual network: Backpropagation
  without storing activations}. In \bibinfo{booktitle}{\emph{Advances in neural
  information processing systems}}. \bibinfo{pages}{2214--2224}.
\newblock


\bibitem[\protect\citeauthoryear{Goodfellow, Pouget-Abadie, Mirza, Xu,
  Warde-Farley, Ozair, Courville, and Bengio}{Goodfellow
  et~al\mbox{.}}{2014a}]%
        {goodfellow2014generative}
\bibfield{author}{\bibinfo{person}{Ian Goodfellow}, \bibinfo{person}{Jean
  Pouget-Abadie}, \bibinfo{person}{Mehdi Mirza}, \bibinfo{person}{Bing Xu},
  \bibinfo{person}{David Warde-Farley}, \bibinfo{person}{Sherjil Ozair},
  \bibinfo{person}{Aaron Courville}, {and} \bibinfo{person}{Yoshua Bengio}.}
  \bibinfo{year}{2014}\natexlab{a}.
\newblock \showarticletitle{Generative adversarial nets}. In
  \bibinfo{booktitle}{\emph{Advances in neural information processing
  systems}}. \bibinfo{pages}{2672--2680}.
\newblock


\bibitem[\protect\citeauthoryear{Goodfellow, Shlens, and Szegedy}{Goodfellow
  et~al\mbox{.}}{2014b}]%
        {goodfellow2014explaining}
\bibfield{author}{\bibinfo{person}{Ian~J Goodfellow}, \bibinfo{person}{Jonathon
  Shlens}, {and} \bibinfo{person}{Christian Szegedy}.}
  \bibinfo{year}{2014}\natexlab{b}.
\newblock \showarticletitle{Explaining and harnessing adversarial examples}.
\newblock \bibinfo{journal}{\emph{arXiv preprint arXiv:1412.6572}}
  (\bibinfo{year}{2014}).
\newblock


\bibitem[\protect\citeauthoryear{Graves}{Graves}{2013}]%
        {graves2013generating}
\bibfield{author}{\bibinfo{person}{Alex Graves}.}
  \bibinfo{year}{2013}\natexlab{}.
\newblock \showarticletitle{Generating sequences with recurrent neural
  networks}.
\newblock \bibinfo{journal}{\emph{arXiv preprint arXiv:1308.0850}}
  (\bibinfo{year}{2013}).
\newblock


\bibitem[\protect\citeauthoryear{Graves, Jaitly, and Mohamed}{Graves
  et~al\mbox{.}}{2013}]%
        {graves2013hybrid}
\bibfield{author}{\bibinfo{person}{Alex Graves}, \bibinfo{person}{Navdeep
  Jaitly}, {and} \bibinfo{person}{Abdel-rahman Mohamed}.}
  \bibinfo{year}{2013}\natexlab{}.
\newblock \showarticletitle{Hybrid speech recognition with deep bidirectional
  LSTM}. In \bibinfo{booktitle}{\emph{2013 IEEE workshop on automatic speech
  recognition and understanding}}. IEEE, \bibinfo{pages}{273--278}.
\newblock


\bibitem[\protect\citeauthoryear{Gregor, Danihelka, Graves, Rezende, and
  Wierstra}{Gregor et~al\mbox{.}}{2015}]%
        {gregor2015draw}
\bibfield{author}{\bibinfo{person}{Karol Gregor}, \bibinfo{person}{Ivo
  Danihelka}, \bibinfo{person}{Alex Graves}, \bibinfo{person}{Danilo~Jimenez
  Rezende}, {and} \bibinfo{person}{Daan Wierstra}.}
  \bibinfo{year}{2015}\natexlab{}.
\newblock \showarticletitle{DRAW: A Recurrent Neural Network For Image
  Generation}. In \bibinfo{booktitle}{\emph{ICML}}.
\newblock


\bibitem[\protect\citeauthoryear{Gu, Liu, Dolan-Gavitt, and Garg}{Gu
  et~al\mbox{.}}{2019}]%
        {gu2019badnets}
\bibfield{author}{\bibinfo{person}{Tianyu Gu}, \bibinfo{person}{Kang Liu},
  \bibinfo{person}{Brendan Dolan-Gavitt}, {and} \bibinfo{person}{Siddharth
  Garg}.} \bibinfo{year}{2019}\natexlab{}.
\newblock \showarticletitle{Badnets: Evaluating backdooring attacks on deep
  neural networks}.
\newblock \bibinfo{journal}{\emph{IEEE Access}}  \bibinfo{volume}{7}
  (\bibinfo{year}{2019}), \bibinfo{pages}{47230--47244}.
\newblock


\bibitem[\protect\citeauthoryear{Hardt, Price, and Srebro}{Hardt
  et~al\mbox{.}}{2016}]%
        {hardt2016equality}
\bibfield{author}{\bibinfo{person}{Moritz Hardt}, \bibinfo{person}{Eric Price},
  {and} \bibinfo{person}{Nati Srebro}.} \bibinfo{year}{2016}\natexlab{}.
\newblock \showarticletitle{Equality of opportunity in supervised learning}. In
  \bibinfo{booktitle}{\emph{Advances in neural information processing
  systems}}. \bibinfo{pages}{3315--3323}.
\newblock


\bibitem[\protect\citeauthoryear{Harzing}{Harzing}{2020}]%
        {harzing2020publish}
\bibfield{author}{\bibinfo{person}{Anne-Wil Harzing}.}
  \bibinfo{year}{2020}\natexlab{}.
\newblock \bibinfo{title}{Publish or Perish}.
\newblock
  \bibinfo{howpublished}{\url{https://harzing.com/resources/publish-or-perish}}.
\newblock


\bibitem[\protect\citeauthoryear{Hassoun et~al\mbox{.}}{Hassoun
  et~al\mbox{.}}{1995}]%
        {hassoun1995fundamentals}
\bibfield{author}{\bibinfo{person}{Mohamad~H Hassoun} {et~al\mbox{.}}}
  \bibinfo{year}{1995}\natexlab{}.
\newblock \bibinfo{booktitle}{\emph{Fundamentals of artificial neural
  networks}}.
\newblock \bibinfo{publisher}{MIT press}.
\newblock


\bibitem[\protect\citeauthoryear{Hayes and Danezis}{Hayes and Danezis}{2017}]%
        {hayes2017generating}
\bibfield{author}{\bibinfo{person}{Jamie Hayes} {and} \bibinfo{person}{George
  Danezis}.} \bibinfo{year}{2017}\natexlab{}.
\newblock \showarticletitle{Generating steganographic images via adversarial
  training}.
\newblock \bibinfo{journal}{\emph{arXiv preprint arXiv:1703.00371}}
  (\bibinfo{year}{2017}).
\newblock


\bibitem[\protect\citeauthoryear{Hayes, Melis, Danezis, and
  De~Cristofaro}{Hayes et~al\mbox{.}}{2019}]%
        {hayes2019logan}
\bibfield{author}{\bibinfo{person}{Jamie Hayes}, \bibinfo{person}{Luca Melis},
  \bibinfo{person}{George Danezis}, {and} \bibinfo{person}{Emiliano
  De~Cristofaro}.} \bibinfo{year}{2019}\natexlab{}.
\newblock \showarticletitle{LOGAN: Membership inference attacks against
  generative models}.
\newblock \bibinfo{journal}{\emph{Proceedings on Privacy Enhancing
  Technologies}} \bibinfo{volume}{2019}, \bibinfo{number}{1}
  (\bibinfo{year}{2019}), \bibinfo{pages}{133--152}.
\newblock


\bibitem[\protect\citeauthoryear{He, Lehrmann, Marino, Mori, and Sigal}{He
  et~al\mbox{.}}{2018}]%
        {he2018probabilistic}
\bibfield{author}{\bibinfo{person}{Jiawei He}, \bibinfo{person}{Andreas
  Lehrmann}, \bibinfo{person}{Joseph Marino}, \bibinfo{person}{Greg Mori},
  {and} \bibinfo{person}{Leonid Sigal}.} \bibinfo{year}{2018}\natexlab{}.
\newblock \showarticletitle{Probabilistic video generation using holistic
  attribute control}. In \bibinfo{booktitle}{\emph{Proceedings of the European
  Conference on Computer Vision (ECCV)}}. \bibinfo{pages}{452--467}.
\newblock


\bibitem[\protect\citeauthoryear{Heusel, Ramsauer, Unterthiner, Nessler, and
  Hochreiter}{Heusel et~al\mbox{.}}{2017}]%
        {heusel2017gans}
\bibfield{author}{\bibinfo{person}{Martin Heusel}, \bibinfo{person}{Hubert
  Ramsauer}, \bibinfo{person}{Thomas Unterthiner}, \bibinfo{person}{Bernhard
  Nessler}, {and} \bibinfo{person}{Sepp Hochreiter}.}
  \bibinfo{year}{2017}\natexlab{}.
\newblock \showarticletitle{Gans trained by a two time-scale update rule
  converge to a local nash equilibrium}. In \bibinfo{booktitle}{\emph{Advances
  in neural information processing systems}}. \bibinfo{pages}{6626--6637}.
\newblock


\bibitem[\protect\citeauthoryear{Higgins, Matthey, Pal, Burgess, Glorot,
  Botvinick, Mohamed, and Lerchner}{Higgins et~al\mbox{.}}{2017}]%
        {higgins2017beta}
\bibfield{author}{\bibinfo{person}{Irina Higgins}, \bibinfo{person}{Loic
  Matthey}, \bibinfo{person}{Arka Pal}, \bibinfo{person}{Christopher Burgess},
  \bibinfo{person}{Xavier Glorot}, \bibinfo{person}{Matthew Botvinick},
  \bibinfo{person}{Shakir Mohamed}, {and} \bibinfo{person}{Alexander
  Lerchner}.} \bibinfo{year}{2017}\natexlab{}.
\newblock \showarticletitle{beta-VAE: Learning Basic Visual Concepts with a
  Constrained Variational Framework.}
\newblock \bibinfo{journal}{\emph{Iclr}} \bibinfo{volume}{2},
  \bibinfo{number}{5} (\bibinfo{year}{2017}), \bibinfo{pages}{6}.
\newblock


\bibitem[\protect\citeauthoryear{Hilprecht, H{\"a}rterich, and
  Bernau}{Hilprecht et~al\mbox{.}}{2019}]%
        {hilprecht2019monte}
\bibfield{author}{\bibinfo{person}{Benjamin Hilprecht}, \bibinfo{person}{Martin
  H{\"a}rterich}, {and} \bibinfo{person}{Daniel Bernau}.}
  \bibinfo{year}{2019}\natexlab{}.
\newblock \showarticletitle{Monte carlo and reconstruction membership inference
  attacks against generative models}.
\newblock \bibinfo{journal}{\emph{Proceedings on Privacy Enhancing
  Technologies}} \bibinfo{volume}{2019}, \bibinfo{number}{4}
  (\bibinfo{year}{2019}), \bibinfo{pages}{232--249}.
\newblock


\bibitem[\protect\citeauthoryear{Hitaj, Gasti, Ateniese, and Perez-Cruz}{Hitaj
  et~al\mbox{.}}{2019}]%
        {hitaj2019passgan}
\bibfield{author}{\bibinfo{person}{Briland Hitaj}, \bibinfo{person}{Paolo
  Gasti}, \bibinfo{person}{Giuseppe Ateniese}, {and} \bibinfo{person}{Fernando
  Perez-Cruz}.} \bibinfo{year}{2019}\natexlab{}.
\newblock \showarticletitle{Passgan: A deep learning approach for password
  guessing}. In \bibinfo{booktitle}{\emph{International Conference on Applied
  Cryptography and Network Security}}. Springer, \bibinfo{pages}{217--237}.
\newblock


\bibitem[\protect\citeauthoryear{Holub and Fridrich}{Holub and
  Fridrich}{2012}]%
        {holub2012designing}
\bibfield{author}{\bibinfo{person}{Vojt{\v{e}}ch Holub} {and}
  \bibinfo{person}{Jessica Fridrich}.} \bibinfo{year}{2012}\natexlab{}.
\newblock \showarticletitle{Designing steganographic distortion using
  directional filters}. In \bibinfo{booktitle}{\emph{2012 IEEE International
  workshop on information forensics and security (WIFS)}}. IEEE,
  \bibinfo{pages}{234--239}.
\newblock


\bibitem[\protect\citeauthoryear{Holub, Fridrich, and Denemark}{Holub
  et~al\mbox{.}}{2014}]%
        {holub2014universal}
\bibfield{author}{\bibinfo{person}{Vojt{\v{e}}ch Holub},
  \bibinfo{person}{Jessica Fridrich}, {and} \bibinfo{person}{Tom{\'a}{\v{s}}
  Denemark}.} \bibinfo{year}{2014}\natexlab{}.
\newblock \showarticletitle{Universal distortion function for steganography in
  an arbitrary domain}.
\newblock \bibinfo{journal}{\emph{EURASIP Journal on Information Security}}
  \bibinfo{volume}{2014}, \bibinfo{number}{1} (\bibinfo{year}{2014}),
  \bibinfo{pages}{1}.
\newblock


\bibitem[\protect\citeauthoryear{Hou, Shen, Sun, and Qiu}{Hou
  et~al\mbox{.}}{2017}]%
        {hou2017deep}
\bibfield{author}{\bibinfo{person}{Xianxu Hou}, \bibinfo{person}{Linlin Shen},
  \bibinfo{person}{Ke Sun}, {and} \bibinfo{person}{Guoping Qiu}.}
  \bibinfo{year}{2017}\natexlab{}.
\newblock \showarticletitle{Deep feature consistent variational autoencoder}.
  In \bibinfo{booktitle}{\emph{2017 IEEE Winter Conference on Applications of
  Computer Vision (WACV)}}. IEEE, \bibinfo{pages}{1133--1141}.
\newblock


\bibitem[\protect\citeauthoryear{Hu, Wang, Jiang, Zheng, and Li}{Hu
  et~al\mbox{.}}{2018}]%
        {hu2018novel}
\bibfield{author}{\bibinfo{person}{Donghui Hu}, \bibinfo{person}{Liang Wang},
  \bibinfo{person}{Wenjie Jiang}, \bibinfo{person}{Shuli Zheng}, {and}
  \bibinfo{person}{Bin Li}.} \bibinfo{year}{2018}\natexlab{}.
\newblock \showarticletitle{A novel image steganography method via deep
  convolutional generative adversarial networks}.
\newblock \bibinfo{journal}{\emph{IEEE Access}}  \bibinfo{volume}{6}
  (\bibinfo{year}{2018}), \bibinfo{pages}{38303--38314}.
\newblock


\bibitem[\protect\citeauthoryear{Hu and Tan}{Hu and Tan}{2017a}]%
        {hu2017black}
\bibfield{author}{\bibinfo{person}{Weiwei Hu} {and} \bibinfo{person}{Ying
  Tan}.} \bibinfo{year}{2017}\natexlab{a}.
\newblock \showarticletitle{Black-box attacks against RNN based malware
  detection algorithms}.
\newblock \bibinfo{journal}{\emph{arXiv preprint arXiv:1705.08131}}
  (\bibinfo{year}{2017}).
\newblock


\bibitem[\protect\citeauthoryear{Hu and Tan}{Hu and Tan}{2017b}]%
        {hu2017generating}
\bibfield{author}{\bibinfo{person}{Weiwei Hu} {and} \bibinfo{person}{Ying
  Tan}.} \bibinfo{year}{2017}\natexlab{b}.
\newblock \showarticletitle{Generating adversarial malware examples for
  black-box attacks based on gan}.
\newblock \bibinfo{journal}{\emph{arXiv preprint arXiv:1702.05983}}
  (\bibinfo{year}{2017}).
\newblock


\bibitem[\protect\citeauthoryear{Hu, Yang, Liang, Salakhutdinov, and Xing}{Hu
  et~al\mbox{.}}{2017}]%
        {hu2017toward}
\bibfield{author}{\bibinfo{person}{Zhiting Hu}, \bibinfo{person}{Zichao Yang},
  \bibinfo{person}{Xiaodan Liang}, \bibinfo{person}{Ruslan Salakhutdinov},
  {and} \bibinfo{person}{Eric~P Xing}.} \bibinfo{year}{2017}\natexlab{}.
\newblock \showarticletitle{Toward controlled generation of text}. In
  \bibinfo{booktitle}{\emph{Proceedings of the 34th International Conference on
  Machine Learning-Volume 70}}. JMLR. org, \bibinfo{pages}{1587--1596}.
\newblock


\bibitem[\protect\citeauthoryear{Huang, Vaswani, Uszkoreit, Shazeer, Simon,
  Hawthorne, Dai, Hoffman, Dinculescu, and Eck}{Huang et~al\mbox{.}}{2018}]%
        {huang2018music}
\bibfield{author}{\bibinfo{person}{Cheng-Zhi~Anna Huang},
  \bibinfo{person}{Ashish Vaswani}, \bibinfo{person}{Jakob Uszkoreit},
  \bibinfo{person}{Noam Shazeer}, \bibinfo{person}{Ian Simon},
  \bibinfo{person}{Curtis Hawthorne}, \bibinfo{person}{Andrew~M Dai},
  \bibinfo{person}{Matthew~D Hoffman}, \bibinfo{person}{Monica Dinculescu},
  {and} \bibinfo{person}{Douglas Eck}.} \bibinfo{year}{2018}\natexlab{}.
\newblock \showarticletitle{Music transformer}.
\newblock \bibinfo{journal}{\emph{arXiv preprint arXiv:1809.04281}}
  (\bibinfo{year}{2018}).
\newblock


\bibitem[\protect\citeauthoryear{Huang, Joseph, Nelson, Rubinstein, and
  Tygar}{Huang et~al\mbox{.}}{2011}]%
        {huang2011adversarial}
\bibfield{author}{\bibinfo{person}{Ling Huang}, \bibinfo{person}{Anthony~D
  Joseph}, \bibinfo{person}{Blaine Nelson}, \bibinfo{person}{Benjamin~IP
  Rubinstein}, {and} \bibinfo{person}{J~Doug Tygar}.}
  \bibinfo{year}{2011}\natexlab{}.
\newblock \showarticletitle{Adversarial machine learning}. In
  \bibinfo{booktitle}{\emph{Proceedings of the 4th ACM workshop on Security and
  artificial intelligence}}. \bibinfo{pages}{43--58}.
\newblock


\bibitem[\protect\citeauthoryear{Im, Kim, Jiang, and Memisevic}{Im
  et~al\mbox{.}}{2016}]%
        {im2016generating}
\bibfield{author}{\bibinfo{person}{Daniel~Jiwoong Im},
  \bibinfo{person}{Chris~Dongjoo Kim}, \bibinfo{person}{Hui Jiang}, {and}
  \bibinfo{person}{Roland Memisevic}.} \bibinfo{year}{2016}\natexlab{}.
\newblock \showarticletitle{Generating images with recurrent adversarial
  networks}.
\newblock \bibinfo{journal}{\emph{arXiv preprint arXiv:1602.05110}}
  (\bibinfo{year}{2016}).
\newblock


\bibitem[\protect\citeauthoryear{Isola, Zhu, Zhou, and Efros}{Isola
  et~al\mbox{.}}{2017}]%
        {isola2017image}
\bibfield{author}{\bibinfo{person}{Phillip Isola}, \bibinfo{person}{Jun-Yan
  Zhu}, \bibinfo{person}{Tinghui Zhou}, {and} \bibinfo{person}{Alexei~A
  Efros}.} \bibinfo{year}{2017}\natexlab{}.
\newblock \showarticletitle{Image-to-image translation with conditional
  adversarial networks}. In \bibinfo{booktitle}{\emph{Proceedings of the IEEE
  conference on computer vision and pattern recognition}}.
  \bibinfo{pages}{1125--1134}.
\newblock


\bibitem[\protect\citeauthoryear{Jain, Abbeel, and Pathak}{Jain
  et~al\mbox{.}}{2020}]%
        {jain2020locally}
\bibfield{author}{\bibinfo{person}{Ajay Jain}, \bibinfo{person}{Pieter Abbeel},
  {and} \bibinfo{person}{Deepak Pathak}.} \bibinfo{year}{2020}\natexlab{}.
\newblock \showarticletitle{Locally Masked Convolution for Autoregressive
  Models}. In \bibinfo{booktitle}{\emph{Conference on Uncertainty in Artificial
  Intelligence}}. PMLR, \bibinfo{pages}{1358--1367}.
\newblock


\bibitem[\protect\citeauthoryear{Jalal, Ilyas, Daskalakis, and Dimakis}{Jalal
  et~al\mbox{.}}{2017}]%
        {jalal2017robust}
\bibfield{author}{\bibinfo{person}{Ajil Jalal}, \bibinfo{person}{Andrew Ilyas},
  \bibinfo{person}{Constantinos Daskalakis}, {and}
  \bibinfo{person}{Alexandros~G Dimakis}.} \bibinfo{year}{2017}\natexlab{}.
\newblock \showarticletitle{The robust manifold defense: Adversarial training
  using generative models}.
\newblock \bibinfo{journal}{\emph{arXiv preprint arXiv:1712.09196}}
  (\bibinfo{year}{2017}).
\newblock


\bibitem[\protect\citeauthoryear{Jelinek, Mercer, Bahl, and Baker}{Jelinek
  et~al\mbox{.}}{1977}]%
        {jelinek1977perplexity}
\bibfield{author}{\bibinfo{person}{Fred Jelinek}, \bibinfo{person}{Robert~L
  Mercer}, \bibinfo{person}{Lalit~R Bahl}, {and} \bibinfo{person}{James~K
  Baker}.} \bibinfo{year}{1977}\natexlab{}.
\newblock \showarticletitle{Perplexity—a measure of the difficulty of speech
  recognition tasks}.
\newblock \bibinfo{journal}{\emph{The Journal of the Acoustical Society of
  America}} \bibinfo{volume}{62}, \bibinfo{number}{S1} (\bibinfo{year}{1977}),
  \bibinfo{pages}{S63--S63}.
\newblock


\bibitem[\protect\citeauthoryear{Jiang, Ye, Huang, Shang, and Zheng}{Jiang
  et~al\mbox{.}}{2020}]%
        {jiang2020smartsteganogaphy}
\bibfield{author}{\bibinfo{person}{Shunzhi Jiang}, \bibinfo{person}{Dengpan
  Ye}, \bibinfo{person}{Jiaqing Huang}, \bibinfo{person}{Yueyun Shang}, {and}
  \bibinfo{person}{Zhuoyuan Zheng}.} \bibinfo{year}{2020}\natexlab{}.
\newblock \showarticletitle{SmartSteganogaphy: Light-weight generative audio
  steganography model for smart embedding application}.
\newblock \bibinfo{journal}{\emph{Journal of Network and Computer
  Applications}}  \bibinfo{volume}{165} (\bibinfo{year}{2020}),
  \bibinfo{pages}{102689}.
\newblock


\bibitem[\protect\citeauthoryear{Jin, Shen, Zhang, Dai, and Zhang}{Jin
  et~al\mbox{.}}{2019}]%
        {jin2019ape}
\bibfield{author}{\bibinfo{person}{Guoqing Jin}, \bibinfo{person}{Shiwei Shen},
  \bibinfo{person}{Dongming Zhang}, \bibinfo{person}{Feng Dai}, {and}
  \bibinfo{person}{Yongdong Zhang}.} \bibinfo{year}{2019}\natexlab{}.
\newblock \showarticletitle{APE-GAN: Adversarial perturbation elimination with
  GAN}. In \bibinfo{booktitle}{\emph{ICASSP 2019-2019 IEEE International
  Conference on Acoustics, Speech and Signal Processing (ICASSP)}}. IEEE,
  \bibinfo{pages}{3842--3846}.
\newblock


\bibitem[\protect\citeauthoryear{Jordon, Yoon, and van~der Schaar}{Jordon
  et~al\mbox{.}}{2018}]%
        {jordon2018pate}
\bibfield{author}{\bibinfo{person}{James Jordon}, \bibinfo{person}{Jinsung
  Yoon}, {and} \bibinfo{person}{Mihaela van~der Schaar}.}
  \bibinfo{year}{2018}\natexlab{}.
\newblock \showarticletitle{PATE-GAN: Generating synthetic data with
  differential privacy guarantees}. In \bibinfo{booktitle}{\emph{International
  Conference on Learning Representations}}.
\newblock


\bibitem[\protect\citeauthoryear{Juels and Rivest}{Juels and Rivest}{2013}]%
        {juels2013honeywords}
\bibfield{author}{\bibinfo{person}{Ari Juels} {and} \bibinfo{person}{Ronald~L
  Rivest}.} \bibinfo{year}{2013}\natexlab{}.
\newblock \showarticletitle{Honeywords: Making password-cracking detectable}.
  In \bibinfo{booktitle}{\emph{Proceedings of the 2013 ACM SIGSAC conference on
  Computer \& communications security}}. \bibinfo{pages}{145--160}.
\newblock


\bibitem[\protect\citeauthoryear{Kallus and Zhou}{Kallus and Zhou}{2018}]%
        {kallus2018residual}
\bibfield{author}{\bibinfo{person}{Nathan Kallus} {and} \bibinfo{person}{Angela
  Zhou}.} \bibinfo{year}{2018}\natexlab{}.
\newblock \showarticletitle{Residual unfairness in fair machine learning from
  prejudiced data}. In \bibinfo{booktitle}{\emph{International Conference on
  Machine Learning}}. PMLR, \bibinfo{pages}{2439--2448}.
\newblock


\bibitem[\protect\citeauthoryear{Karras, Aila, Laine, and Lehtinen}{Karras
  et~al\mbox{.}}{2018}]%
        {karras2017progressive}
\bibfield{author}{\bibinfo{person}{Tero Karras}, \bibinfo{person}{Timo Aila},
  \bibinfo{person}{Samuli Laine}, {and} \bibinfo{person}{Jaakko Lehtinen}.}
  \bibinfo{year}{2018}\natexlab{}.
\newblock \showarticletitle{Progressive Growing of GANs for Improved Quality,
  Stability, and Variation}. In \bibinfo{booktitle}{\emph{International
  Conference on Learning Representations}}.
\newblock


\bibitem[\protect\citeauthoryear{Karras, Laine, and Aila}{Karras
  et~al\mbox{.}}{2019}]%
        {karras2019style}
\bibfield{author}{\bibinfo{person}{Tero Karras}, \bibinfo{person}{Samuli
  Laine}, {and} \bibinfo{person}{Timo Aila}.} \bibinfo{year}{2019}\natexlab{}.
\newblock \showarticletitle{A style-based generator architecture for generative
  adversarial networks}. In \bibinfo{booktitle}{\emph{Proceedings of the IEEE
  conference on computer vision and pattern recognition}}.
  \bibinfo{pages}{4401--4410}.
\newblock


\bibitem[\protect\citeauthoryear{Keskar, McCann, Varshney, Xiong, and
  Socher}{Keskar et~al\mbox{.}}{2019}]%
        {keskar2019ctrl}
\bibfield{author}{\bibinfo{person}{Nitish~Shirish Keskar},
  \bibinfo{person}{Bryan McCann}, \bibinfo{person}{Lav~R Varshney},
  \bibinfo{person}{Caiming Xiong}, {and} \bibinfo{person}{Richard Socher}.}
  \bibinfo{year}{2019}\natexlab{}.
\newblock \showarticletitle{Ctrl: A conditional transformer language model for
  controllable generation}.
\newblock \bibinfo{journal}{\emph{arXiv preprint arXiv:1909.05858}}
  (\bibinfo{year}{2019}).
\newblock


\bibitem[\protect\citeauthoryear{Kilgour, Zuluaga, Roblek, and Sharifi}{Kilgour
  et~al\mbox{.}}{2018}]%
        {kilgour2018frchet}
\bibfield{author}{\bibinfo{person}{Kevin Kilgour}, \bibinfo{person}{Mauricio
  Zuluaga}, \bibinfo{person}{Dominik Roblek}, {and} \bibinfo{person}{Matthew
  Sharifi}.} \bibinfo{year}{2018}\natexlab{}.
\newblock \bibinfo{title}{Fréchet Audio Distance: A Metric for Evaluating
  Music Enhancement Algorithms}.
\newblock
\newblock
\showeprint[arxiv]{1812.08466}~[eess.AS]


\bibitem[\protect\citeauthoryear{Kim, Bu, and Cho}{Kim et~al\mbox{.}}{2018}]%
        {kim2018zero}
\bibfield{author}{\bibinfo{person}{Jin-Young Kim}, \bibinfo{person}{Seok-Jun
  Bu}, {and} \bibinfo{person}{Sung-Bae Cho}.} \bibinfo{year}{2018}\natexlab{}.
\newblock \showarticletitle{Zero-day malware detection using transferred
  generative adversarial networks based on deep autoencoders}.
\newblock \bibinfo{journal}{\emph{Information Sciences}}  \bibinfo{volume}{460}
  (\bibinfo{year}{2018}), \bibinfo{pages}{83--102}.
\newblock


\bibitem[\protect\citeauthoryear{Kim and Cho}{Kim and Cho}{2018}]%
        {kim2018detecting}
\bibfield{author}{\bibinfo{person}{Jin-Young Kim} {and}
  \bibinfo{person}{Sung-Bae Cho}.} \bibinfo{year}{2018}\natexlab{}.
\newblock \showarticletitle{Detecting intrusive malware with a hybrid
  generative deep learning model}. In \bibinfo{booktitle}{\emph{International
  Conference on Intelligent Data Engineering and Automated Learning}}.
  Springer, \bibinfo{pages}{499--507}.
\newblock


\bibitem[\protect\citeauthoryear{Kingma and Welling}{Kingma and
  Welling}{2013}]%
        {kingma2013auto}
\bibfield{author}{\bibinfo{person}{Diederik~P Kingma} {and}
  \bibinfo{person}{Max Welling}.} \bibinfo{year}{2013}\natexlab{}.
\newblock \showarticletitle{Auto-encoding variational bayes}.
\newblock \bibinfo{journal}{\emph{arXiv preprint arXiv:1312.6114}}
  (\bibinfo{year}{2013}).
\newblock


\bibitem[\protect\citeauthoryear{Korosec}{Korosec}{2019}]%
        {korosec_2019}
\bibfield{author}{\bibinfo{person}{Kirsten Korosec}.}
  \bibinfo{year}{2019}\natexlab{}.
\newblock \bibinfo{title}{'Deepfake' revenge porn is now illegal in Virginia}.
\newblock
  \bibinfo{howpublished}{\url{https://techcrunch.com/2019/07/01/deepfake-revenge-porn-is-now-illegal-in-virginia/}}.
\newblock


\bibitem[\protect\citeauthoryear{Kos, Fischer, and Song}{Kos
  et~al\mbox{.}}{2018}]%
        {kos2018adversarial}
\bibfield{author}{\bibinfo{person}{Jernej Kos}, \bibinfo{person}{Ian Fischer},
  {and} \bibinfo{person}{Dawn Song}.} \bibinfo{year}{2018}\natexlab{}.
\newblock \showarticletitle{Adversarial examples for generative models}. In
  \bibinfo{booktitle}{\emph{2018 IEEE Security and Privacy Workshops (SPW)}}.
  IEEE, \bibinfo{pages}{36--42}.
\newblock


\bibitem[\protect\citeauthoryear{Kreuk, Adi, Raj, Singh, and Keshet}{Kreuk
  et~al\mbox{.}}{2019}]%
        {kreuk2019hide}
\bibfield{author}{\bibinfo{person}{Felix Kreuk}, \bibinfo{person}{Yossi Adi},
  \bibinfo{person}{Bhiksha Raj}, \bibinfo{person}{Rita Singh}, {and}
  \bibinfo{person}{Joseph Keshet}.} \bibinfo{year}{2019}\natexlab{}.
\newblock \showarticletitle{Hide and speak: Deep neural networks for speech
  steganography}.
\newblock \bibinfo{journal}{\emph{arXiv preprint arXiv:1902.03083}}
  (\bibinfo{year}{2019}).
\newblock


\bibitem[\protect\citeauthoryear{Kr{\"o}se, Krose, van~der Smagt, and
  Smagt}{Kr{\"o}se et~al\mbox{.}}{1993}]%
        {krose1993introduction}
\bibfield{author}{\bibinfo{person}{Ben Kr{\"o}se}, \bibinfo{person}{Ben Krose},
  \bibinfo{person}{Patrick van~der Smagt}, {and} \bibinfo{person}{Patrick
  Smagt}.} \bibinfo{year}{1993}\natexlab{}.
\newblock \showarticletitle{An introduction to neural networks}.
\newblock  (\bibinfo{year}{1993}).
\newblock


\bibitem[\protect\citeauthoryear{Kumar, Kumar, de~Boissiere, Gestin, Teoh,
  Sotelo, de~Br{\'e}bisson, Bengio, and Courville}{Kumar et~al\mbox{.}}{2019}]%
        {kumar2019melgan}
\bibfield{author}{\bibinfo{person}{Kundan Kumar}, \bibinfo{person}{Rithesh
  Kumar}, \bibinfo{person}{Thibault de Boissiere}, \bibinfo{person}{Lucas
  Gestin}, \bibinfo{person}{Wei~Zhen Teoh}, \bibinfo{person}{Jose Sotelo},
  \bibinfo{person}{Alexandre de Br{\'e}bisson}, \bibinfo{person}{Yoshua
  Bengio}, {and} \bibinfo{person}{Aaron Courville}.}
  \bibinfo{year}{2019}\natexlab{}.
\newblock \showarticletitle{Melgan: Generative adversarial networks for
  conditional waveform synthesis}.
\newblock \bibinfo{journal}{\emph{arXiv preprint arXiv:1910.06711}}
  (\bibinfo{year}{2019}).
\newblock


\bibitem[\protect\citeauthoryear{Kusner and Hern{\'a}ndez-Lobato}{Kusner and
  Hern{\'a}ndez-Lobato}{2016}]%
        {kusner2016gans}
\bibfield{author}{\bibinfo{person}{Matt~J Kusner} {and}
  \bibinfo{person}{Jos{\'e}~Miguel Hern{\'a}ndez-Lobato}.}
  \bibinfo{year}{2016}\natexlab{}.
\newblock \showarticletitle{Gans for sequences of discrete elements with the
  gumbel-softmax distribution}.
\newblock \bibinfo{journal}{\emph{arXiv preprint arXiv:1611.04051}}
  (\bibinfo{year}{2016}).
\newblock


\bibitem[\protect\citeauthoryear{Kusner, Loftus, Russell, and Silva}{Kusner
  et~al\mbox{.}}{2017}]%
        {kusner2017counterfactual}
\bibfield{author}{\bibinfo{person}{Matt~J Kusner}, \bibinfo{person}{Joshua~R
  Loftus}, \bibinfo{person}{Chris Russell}, {and} \bibinfo{person}{Ricardo
  Silva}.} \bibinfo{year}{2017}\natexlab{}.
\newblock \showarticletitle{Counterfactual fairness}.
\newblock \bibinfo{journal}{\emph{arXiv preprint arXiv:1703.06856}}
  (\bibinfo{year}{2017}).
\newblock


\bibitem[\protect\citeauthoryear{Larsen, Sønderby, Larochelle, and
  Winther}{Larsen et~al\mbox{.}}{2016}]%
        {larsen2016vaegan}
\bibfield{author}{\bibinfo{person}{Anders Boesen~Lindbo Larsen},
  \bibinfo{person}{Søren~Kaae Sønderby}, \bibinfo{person}{Hugo Larochelle},
  {and} \bibinfo{person}{Ole Winther}.} \bibinfo{year}{2016}\natexlab{}.
\newblock \showarticletitle{Autoencoding beyond pixels using a learned
  similarity metric}.
  \bibinfo{howpublished}{\url{http://proceedings.mlr.press/v48/larsen16.html}}.
  In \bibinfo{booktitle}{\emph{Proceedings of The 33rd International Conference
  on Machine Learning}} \emph{(\bibinfo{series}{Proceedings of Machine Learning
  Research}, Vol.~\bibinfo{volume}{48})},
  \bibfield{editor}{\bibinfo{person}{Maria~Florina Balcan} {and}
  \bibinfo{person}{Kilian~Q. Weinberger}} (Eds.). \bibinfo{publisher}{PMLR},
  \bibinfo{address}{New York, New York, USA}, \bibinfo{pages}{1558--1566}.
\newblock


\bibitem[\protect\citeauthoryear{Lee, Han, and Lee}{Lee et~al\mbox{.}}{2017}]%
        {lee2017generative}
\bibfield{author}{\bibinfo{person}{Hyeungill Lee}, \bibinfo{person}{Sungyeob
  Han}, {and} \bibinfo{person}{Jungwoo Lee}.} \bibinfo{year}{2017}\natexlab{}.
\newblock \showarticletitle{Generative adversarial trainer: Defense to
  adversarial perturbations with gan}.
\newblock \bibinfo{journal}{\emph{arXiv preprint arXiv:1705.03387}}
  (\bibinfo{year}{2017}).
\newblock


\bibitem[\protect\citeauthoryear{Lehmann and Romano}{Lehmann and
  Romano}{2006}]%
        {lehmann2006testing}
\bibfield{author}{\bibinfo{person}{Erich~L Lehmann} {and}
  \bibinfo{person}{Joseph~P Romano}.} \bibinfo{year}{2006}\natexlab{}.
\newblock \bibinfo{booktitle}{\emph{Testing statistical hypotheses}}.
\newblock \bibinfo{publisher}{Springer Science \& Business Media}.
\newblock


\bibitem[\protect\citeauthoryear{Leino and Fredrikson}{Leino and
  Fredrikson}{2020}]%
        {leino2020stolen}
\bibfield{author}{\bibinfo{person}{Klas Leino} {and} \bibinfo{person}{Matt
  Fredrikson}.} \bibinfo{year}{2020}\natexlab{}.
\newblock \showarticletitle{Stolen memories: Leveraging model memorization for
  calibrated white-box membership inference}. In \bibinfo{booktitle}{\emph{29th
  $\{$USENIX$\}$ Security Symposium ($\{$USENIX$\}$ Security 20)}}.
  \bibinfo{pages}{1605--1622}.
\newblock


\bibitem[\protect\citeauthoryear{Li, Niu, Liao, Wang, Liu, Lei, and Zhang}{Li
  et~al\mbox{.}}{2020}]%
        {li2020generative}
\bibfield{author}{\bibinfo{person}{Jun Li}, \bibinfo{person}{Ke Niu},
  \bibinfo{person}{Liwei Liao}, \bibinfo{person}{Lijie Wang},
  \bibinfo{person}{Jia Liu}, \bibinfo{person}{Yu Lei}, {and}
  \bibinfo{person}{Minqing Zhang}.} \bibinfo{year}{2020}\natexlab{}.
\newblock \showarticletitle{A Generative Steganography Method Based on
  WGAN-GP}. In \bibinfo{booktitle}{\emph{International Conference on Artificial
  Intelligence and Security}}. Springer, \bibinfo{pages}{386--397}.
\newblock


\bibitem[\protect\citeauthoryear{Li, Liu, Liu, Zhao, and Liu}{Li
  et~al\mbox{.}}{2019}]%
        {li2019neural}
\bibfield{author}{\bibinfo{person}{Naihan Li}, \bibinfo{person}{Shujie Liu},
  \bibinfo{person}{Yanqing Liu}, \bibinfo{person}{Sheng Zhao}, {and}
  \bibinfo{person}{Ming Liu}.} \bibinfo{year}{2019}\natexlab{}.
\newblock \showarticletitle{Neural speech synthesis with transformer network}.
  In \bibinfo{booktitle}{\emph{Proceedings of the AAAI Conference on Artificial
  Intelligence}}, Vol.~\bibinfo{volume}{33}. \bibinfo{pages}{6706--6713}.
\newblock


\bibitem[\protect\citeauthoryear{Li, Chang, and Lyu}{Li et~al\mbox{.}}{2018}]%
        {li2018ictu}
\bibfield{author}{\bibinfo{person}{Yuezun Li}, \bibinfo{person}{Ming-Ching
  Chang}, {and} \bibinfo{person}{Siwei Lyu}.} \bibinfo{year}{2018}\natexlab{}.
\newblock \showarticletitle{In ictu oculi: Exposing ai created fake videos by
  detecting eye blinking}. In \bibinfo{booktitle}{\emph{2018 IEEE International
  Workshop on Information Forensics and Security (WIFS)}}. IEEE,
  \bibinfo{pages}{1--7}.
\newblock


\bibitem[\protect\citeauthoryear{Liang, Lee, Dai, and Xing}{Liang
  et~al\mbox{.}}{2017}]%
        {liang2017dual}
\bibfield{author}{\bibinfo{person}{Xiaodan Liang}, \bibinfo{person}{Lisa Lee},
  \bibinfo{person}{Wei Dai}, {and} \bibinfo{person}{Eric~P Xing}.}
  \bibinfo{year}{2017}\natexlab{}.
\newblock \showarticletitle{Dual motion GAN for future-flow embedded video
  prediction}. In \bibinfo{booktitle}{\emph{Proceedings of the IEEE
  International Conference on Computer Vision}}. \bibinfo{pages}{1744--1752}.
\newblock


\bibitem[\protect\citeauthoryear{Lin and Och}{Lin and Och}{2004}]%
        {lin2004automatic}
\bibfield{author}{\bibinfo{person}{Chin-Yew Lin} {and}
  \bibinfo{person}{Franz~Josef Och}.} \bibinfo{year}{2004}\natexlab{}.
\newblock \showarticletitle{Automatic evaluation of machine translation quality
  using longest common subsequence and skip-bigram statistics}. In
  \bibinfo{booktitle}{\emph{Proceedings of the 42nd Annual Meeting on
  Association for Computational Linguistics}}. Association for Computational
  Linguistics, \bibinfo{pages}{605}.
\newblock


\bibitem[\protect\citeauthoryear{Lin, Shi, and Xue}{Lin et~al\mbox{.}}{2018}]%
        {lin2018idsgan}
\bibfield{author}{\bibinfo{person}{Zilong Lin}, \bibinfo{person}{Yong Shi},
  {and} \bibinfo{person}{Zhi Xue}.} \bibinfo{year}{2018}\natexlab{}.
\newblock \showarticletitle{Idsgan: Generative adversarial networks for attack
  generation against intrusion detection}.
\newblock \bibinfo{journal}{\emph{arXiv preprint arXiv:1809.02077}}
  (\bibinfo{year}{2018}).
\newblock


\bibitem[\protect\citeauthoryear{Liu, Zhang, Liu, Zhang, and Ke}{Liu
  et~al\mbox{.}}{2017b}]%
        {liu2017coverless}
\bibfield{author}{\bibinfo{person}{Ming-ming Liu}, \bibinfo{person}{Min-qing
  Zhang}, \bibinfo{person}{Jia Liu}, \bibinfo{person}{Ying-nan Zhang}, {and}
  \bibinfo{person}{Yan Ke}.} \bibinfo{year}{2017}\natexlab{b}.
\newblock \showarticletitle{Coverless information hiding based on generative
  adversarial networks}.
\newblock \bibinfo{journal}{\emph{arXiv preprint arXiv:1712.06951}}
  (\bibinfo{year}{2017}).
\newblock


\bibitem[\protect\citeauthoryear{Liu, Wang, Sha, Chang, and Sui}{Liu
  et~al\mbox{.}}{2018}]%
        {liu2018table}
\bibfield{author}{\bibinfo{person}{Tianyu Liu}, \bibinfo{person}{Kexiang Wang},
  \bibinfo{person}{Lei Sha}, \bibinfo{person}{Baobao Chang}, {and}
  \bibinfo{person}{Zhifang Sui}.} \bibinfo{year}{2018}\natexlab{}.
\newblock \showarticletitle{Table-to-text generation by structure-aware seq2seq
  learning}. In \bibinfo{booktitle}{\emph{Proceedings of the AAAI Conference on
  Artificial Intelligence}}, Vol.~\bibinfo{volume}{32}.
\newblock


\bibitem[\protect\citeauthoryear{Liu, Ma, Aafer, Lee, Zhai, Wang, and
  Zhang}{Liu et~al\mbox{.}}{2017a}]%
        {liu2017trojaning}
\bibfield{author}{\bibinfo{person}{Yingqi Liu}, \bibinfo{person}{Shiqing Ma},
  \bibinfo{person}{Yousra Aafer}, \bibinfo{person}{Wen-Chuan Lee},
  \bibinfo{person}{Juan Zhai}, \bibinfo{person}{Weihang Wang}, {and}
  \bibinfo{person}{Xiangyu Zhang}.} \bibinfo{year}{2017}\natexlab{a}.
\newblock \showarticletitle{Trojaning attack on neural networks}.
\newblock  (\bibinfo{year}{2017}).
\newblock


\bibitem[\protect\citeauthoryear{Long, Bindschaedler, Wang, Bu, Wang, Tang,
  Gunter, and Chen}{Long et~al\mbox{.}}{2018}]%
        {long2018understanding}
\bibfield{author}{\bibinfo{person}{Yunhui Long}, \bibinfo{person}{Vincent
  Bindschaedler}, \bibinfo{person}{Lei Wang}, \bibinfo{person}{Diyue Bu},
  \bibinfo{person}{Xiaofeng Wang}, \bibinfo{person}{Haixu Tang},
  \bibinfo{person}{Carl~A Gunter}, {and} \bibinfo{person}{Kai Chen}.}
  \bibinfo{year}{2018}\natexlab{}.
\newblock \showarticletitle{Understanding membership inferences on
  well-generalized learning models}.
\newblock \bibinfo{journal}{\emph{arXiv preprint arXiv:1802.04889}}
  (\bibinfo{year}{2018}).
\newblock


\bibitem[\protect\citeauthoryear{Long, Lin, Yang, Gunter, and Li}{Long
  et~al\mbox{.}}{2019}]%
        {long2019scalable}
\bibfield{author}{\bibinfo{person}{Yunhui Long}, \bibinfo{person}{Suxin Lin},
  \bibinfo{person}{Zhuolin Yang}, \bibinfo{person}{Carl~A Gunter}, {and}
  \bibinfo{person}{Bo Li}.} \bibinfo{year}{2019}\natexlab{}.
\newblock \showarticletitle{Scalable differentially private generative student
  model via pate}.
\newblock \bibinfo{journal}{\emph{arXiv preprint arXiv:1906.09338}}
  (\bibinfo{year}{2019}).
\newblock


\bibitem[\protect\citeauthoryear{Long, Wang, Bu, Bindschaedler, Wang, Tang,
  Gunter, and Chen}{Long et~al\mbox{.}}{2020}]%
        {long2020pragmatic}
\bibfield{author}{\bibinfo{person}{Yunhui Long}, \bibinfo{person}{Lei Wang},
  \bibinfo{person}{Diyue Bu}, \bibinfo{person}{Vincent Bindschaedler},
  \bibinfo{person}{Xiaofeng Wang}, \bibinfo{person}{Haixu Tang},
  \bibinfo{person}{Carl~A Gunter}, {and} \bibinfo{person}{Kai Chen}.}
  \bibinfo{year}{2020}\natexlab{}.
\newblock \showarticletitle{A Pragmatic Approach to Membership Inferences on
  Machine Learning Models}. In \bibinfo{booktitle}{\emph{5th IEEE European
  Symposium on Security and Privacy, Euro S and P 2020}}. Institute of
  Electrical and Electronics Engineers Inc., \bibinfo{pages}{521--534}.
\newblock


\bibitem[\protect\citeauthoryear{Lopresti and Raim}{Lopresti and Raim}{2005}]%
        {lopresti2005effectiveness}
\bibfield{author}{\bibinfo{person}{Daniel~P Lopresti} {and}
  \bibinfo{person}{Jarret~D Raim}.} \bibinfo{year}{2005}\natexlab{}.
\newblock \showarticletitle{The effectiveness of generative attacks on an
  online handwriting biometric}. In \bibinfo{booktitle}{\emph{International
  Conference on Audio-and Video-Based Biometric Person Authentication}}.
  Springer, \bibinfo{pages}{1090--1099}.
\newblock


\bibitem[\protect\citeauthoryear{Louizos, Swersky, Li, Welling, and
  Zemel}{Louizos et~al\mbox{.}}{2015}]%
        {louizos2015variational}
\bibfield{author}{\bibinfo{person}{Christos Louizos}, \bibinfo{person}{Kevin
  Swersky}, \bibinfo{person}{Yujia Li}, \bibinfo{person}{Max Welling}, {and}
  \bibinfo{person}{Richard Zemel}.} \bibinfo{year}{2015}\natexlab{}.
\newblock \showarticletitle{The variational fair autoencoder}.
\newblock \bibinfo{journal}{\emph{arXiv preprint arXiv:1511.00830}}
  (\bibinfo{year}{2015}).
\newblock


\bibitem[\protect\citeauthoryear{Lucic, Kurach, Michalski, Gelly, and
  Bousquet}{Lucic et~al\mbox{.}}{2018}]%
        {lucic2018gans}
\bibfield{author}{\bibinfo{person}{Mario Lucic}, \bibinfo{person}{Karol
  Kurach}, \bibinfo{person}{Marcin Michalski}, \bibinfo{person}{Sylvain Gelly},
  {and} \bibinfo{person}{Olivier Bousquet}.} \bibinfo{year}{2018}\natexlab{}.
\newblock \showarticletitle{Are gans created equal? a large-scale study}. In
  \bibinfo{booktitle}{\emph{Advances in neural information processing
  systems}}. \bibinfo{pages}{700--709}.
\newblock


\bibitem[\protect\citeauthoryear{Luo and Huang}{Luo and Huang}{2017}]%
        {luo2017text}
\bibfield{author}{\bibinfo{person}{Yubo Luo} {and} \bibinfo{person}{Yongfeng
  Huang}.} \bibinfo{year}{2017}\natexlab{}.
\newblock \showarticletitle{Text steganography with high embedding rate: Using
  recurrent neural networks to generate chinese classic poetry}. In
  \bibinfo{booktitle}{\emph{Proceedings of the 5th ACM workshop on information
  hiding and multimedia security}}. \bibinfo{pages}{99--104}.
\newblock


\bibitem[\protect\citeauthoryear{Lutz, Sansing~III, Farag, and Ezekiel}{Lutz
  et~al\mbox{.}}{2019}]%
        {lutz2019malware}
\bibfield{author}{\bibinfo{person}{Adam Lutz}, \bibinfo{person}{Victor~F
  Sansing~III}, \bibinfo{person}{Waleed~E Farag}, {and}
  \bibinfo{person}{Soundararajan Ezekiel}.} \bibinfo{year}{2019}\natexlab{}.
\newblock \showarticletitle{Malware classification using fusion of neural
  networks}. In \bibinfo{booktitle}{\emph{Disruptive Technologies in
  Information Sciences II}}, Vol.~\bibinfo{volume}{11013}. International
  Society for Optics and Photonics, \bibinfo{pages}{110130X}.
\newblock


\bibitem[\protect\citeauthoryear{Madry, Makelov, Schmidt, Tsipras, and
  Vladu}{Madry et~al\mbox{.}}{2018}]%
        {madry2018towards}
\bibfield{author}{\bibinfo{person}{Aleksander Madry},
  \bibinfo{person}{Aleksandar Makelov}, \bibinfo{person}{Ludwig Schmidt},
  \bibinfo{person}{Dimitris Tsipras}, {and} \bibinfo{person}{Adrian Vladu}.}
  \bibinfo{year}{2018}\natexlab{}.
\newblock \showarticletitle{Towards Deep Learning Models Resistant to
  Adversarial Attacks}. In \bibinfo{booktitle}{\emph{International Conference
  on Learning Representations}}.
\newblock


\bibitem[\protect\citeauthoryear{Makhzani, Shlens, Jaitly, Goodfellow, and
  Frey}{Makhzani et~al\mbox{.}}{2015}]%
        {makhzani2015adversarial}
\bibfield{author}{\bibinfo{person}{Alireza Makhzani}, \bibinfo{person}{Jonathon
  Shlens}, \bibinfo{person}{Navdeep Jaitly}, \bibinfo{person}{Ian Goodfellow},
  {and} \bibinfo{person}{Brendan Frey}.} \bibinfo{year}{2015}\natexlab{}.
\newblock \showarticletitle{Adversarial autoencoders}.
\newblock \bibinfo{journal}{\emph{arXiv preprint arXiv:1511.05644}}
  (\bibinfo{year}{2015}).
\newblock


\bibitem[\protect\citeauthoryear{Mallenbaum}{Mallenbaum}{2019}]%
        {mallenbaum_2019}
\bibfield{author}{\bibinfo{person}{Carly Mallenbaum}.}
  \bibinfo{year}{2019}\natexlab{}.
\newblock \bibinfo{title}{Bill Hader has NOT seen those deepfake videos he's
  in: 'It's a weird technology, man'}.
\newblock
  \bibinfo{howpublished}{\url{https://www.usatoday.com/story/entertainment/movies/2019/09/02/bill-hader-has-not-seen-those-deepfake-videos/2145558001/}}.
\newblock


\bibitem[\protect\citeauthoryear{Mao, Li, Xie, Lau, Wang, and Paul~Smolley}{Mao
  et~al\mbox{.}}{2017}]%
        {mao2017least}
\bibfield{author}{\bibinfo{person}{Xudong Mao}, \bibinfo{person}{Qing Li},
  \bibinfo{person}{Haoran Xie}, \bibinfo{person}{Raymond~YK Lau},
  \bibinfo{person}{Zhen Wang}, {and} \bibinfo{person}{Stephen Paul~Smolley}.}
  \bibinfo{year}{2017}\natexlab{}.
\newblock \showarticletitle{Least squares generative adversarial networks}. In
  \bibinfo{booktitle}{\emph{Proceedings of the IEEE International Conference on
  Computer Vision}}. \bibinfo{pages}{2794--2802}.
\newblock


\bibitem[\protect\citeauthoryear{Mat~Kiah, Zaidan, Zaidan, Mohammed~Ahmed, and
  Al-Bakri}{Mat~Kiah et~al\mbox{.}}{2011}]%
        {mat2011review}
\bibfield{author}{\bibinfo{person}{ML Mat~Kiah}, \bibinfo{person}{BB Zaidan},
  \bibinfo{person}{AA Zaidan}, \bibinfo{person}{A Mohammed~Ahmed}, {and}
  \bibinfo{person}{Sameer~Hasan Al-Bakri}.} \bibinfo{year}{2011}\natexlab{}.
\newblock \showarticletitle{A review of audio based steganography and digital
  watermarking}.
\newblock \bibinfo{journal}{\emph{International Journal of Physical Sciences}}
  \bibinfo{volume}{6}, \bibinfo{number}{16} (\bibinfo{year}{2011}),
  \bibinfo{pages}{3837--3850}.
\newblock


\bibitem[\protect\citeauthoryear{Mikolov, Karafi{\'a}t, Burget,
  {\v{C}}ernock{\`y}, and Khudanpur}{Mikolov et~al\mbox{.}}{2010}]%
        {mikolov2010recurrent}
\bibfield{author}{\bibinfo{person}{Tom{\'a}{\v{s}} Mikolov},
  \bibinfo{person}{Martin Karafi{\'a}t}, \bibinfo{person}{Luk{\'a}{\v{s}}
  Burget}, \bibinfo{person}{Jan {\v{C}}ernock{\`y}}, {and}
  \bibinfo{person}{Sanjeev Khudanpur}.} \bibinfo{year}{2010}\natexlab{}.
\newblock \showarticletitle{Recurrent neural network based language model}. In
  \bibinfo{booktitle}{\emph{Eleventh annual conference of the international
  speech communication association}}.
\newblock


\bibitem[\protect\citeauthoryear{Minaee and Abdolrashidi}{Minaee and
  Abdolrashidi}{2018}]%
        {minaee2018finger}
\bibfield{author}{\bibinfo{person}{Shervin Minaee} {and}
  \bibinfo{person}{Amirali Abdolrashidi}.} \bibinfo{year}{2018}\natexlab{}.
\newblock \showarticletitle{Finger-gan: Generating realistic fingerprint images
  using connectivity imposed gan}.
\newblock \bibinfo{journal}{\emph{arXiv preprint arXiv:1812.10482}}
  (\bibinfo{year}{2018}).
\newblock


\bibitem[\protect\citeauthoryear{Moffat}{Moffat}{2019}]%
        {moffat2019huffman}
\bibfield{author}{\bibinfo{person}{Alistair Moffat}.}
  \bibinfo{year}{2019}\natexlab{}.
\newblock \showarticletitle{Huffman coding}.
\newblock \bibinfo{journal}{\emph{ACM Computing Surveys (CSUR)}}
  \bibinfo{volume}{52}, \bibinfo{number}{4} (\bibinfo{year}{2019}),
  \bibinfo{pages}{1--35}.
\newblock


\bibitem[\protect\citeauthoryear{Mogren}{Mogren}{2016}]%
        {mogren2016c}
\bibfield{author}{\bibinfo{person}{Olof Mogren}.}
  \bibinfo{year}{2016}\natexlab{}.
\newblock \showarticletitle{C-RNN-GAN: Continuous recurrent neural networks
  with adversarial training}.
\newblock \bibinfo{journal}{\emph{arXiv preprint arXiv:1611.09904}}
  (\bibinfo{year}{2016}).
\newblock


\bibitem[\protect\citeauthoryear{Mohaghegh~Dolatabadi, Erfani, and
  Leckie}{Mohaghegh~Dolatabadi et~al\mbox{.}}{2020}]%
        {dolaadvflow}
\bibfield{author}{\bibinfo{person}{Hadi Mohaghegh~Dolatabadi},
  \bibinfo{person}{Sarah Erfani}, {and} \bibinfo{person}{Christopher Leckie}.}
  \bibinfo{year}{2020}\natexlab{}.
\newblock \showarticletitle{AdvFlow: Inconspicuous Black-box Adversarial
  Attacks using Normalizing Flows}. In \bibinfo{booktitle}{\emph{Advances in
  Neural Information Processing Systems}},
  \bibfield{editor}{\bibinfo{person}{H.~Larochelle},
  \bibinfo{person}{M.~Ranzato}, \bibinfo{person}{R.~Hadsell},
  \bibinfo{person}{M.~F. Balcan}, {and} \bibinfo{person}{H.~Lin}} (Eds.),
  Vol.~\bibinfo{volume}{33}. \bibinfo{publisher}{Curran Associates, Inc.},
  \bibinfo{pages}{15871--15884}.
\newblock
\urldef\tempurl%
\url{https://proceedings.neurips.cc/paper/2020/file/b6cf334c22c8f4ce8eb920bb7b512ed0-Paper.pdf}
\showURL{%
\tempurl}


\bibitem[\protect\citeauthoryear{Monaco, Ali, and Tappert}{Monaco
  et~al\mbox{.}}{2015}]%
        {monaco2015spoofing}
\bibfield{author}{\bibinfo{person}{John~V Monaco}, \bibinfo{person}{Md~Liakat
  Ali}, {and} \bibinfo{person}{Charles~C Tappert}.}
  \bibinfo{year}{2015}\natexlab{}.
\newblock \showarticletitle{Spoofing key-press latencies with a generative
  keystroke dynamics model}. In \bibinfo{booktitle}{\emph{2015 IEEE 7th
  international conference on biometrics theory, applications and systems
  (BTAS)}}. IEEE, \bibinfo{pages}{1--8}.
\newblock


\bibitem[\protect\citeauthoryear{Moraldo}{Moraldo}{2014}]%
        {moraldo2014approach}
\bibfield{author}{\bibinfo{person}{H~Hernan Moraldo}.}
  \bibinfo{year}{2014}\natexlab{}.
\newblock \showarticletitle{An Approach for text steganography based on Markov
  Chains}.
\newblock \bibinfo{journal}{\emph{arXiv preprint arXiv:1409.0915}}
  (\bibinfo{year}{2014}).
\newblock


\bibitem[\protect\citeauthoryear{Nam, Jeon, Kim, and Moon}{Nam
  et~al\mbox{.}}{2020}]%
        {nam2020recurrent}
\bibfield{author}{\bibinfo{person}{Sungyup Nam}, \bibinfo{person}{Seungho
  Jeon}, \bibinfo{person}{Hongkyo Kim}, {and} \bibinfo{person}{Jongsub Moon}.}
  \bibinfo{year}{2020}\natexlab{}.
\newblock \showarticletitle{Recurrent gans password cracker for iot password
  security enhancement}.
\newblock \bibinfo{journal}{\emph{Sensors}} \bibinfo{volume}{20},
  \bibinfo{number}{11} (\bibinfo{year}{2020}), \bibinfo{pages}{3106}.
\newblock


\bibitem[\protect\citeauthoryear{Nasr, Shokri, et~al\mbox{.}}{Nasr
  et~al\mbox{.}}{2020}]%
        {nasr2020improving}
\bibfield{author}{\bibinfo{person}{Milad Nasr}, \bibinfo{person}{Reza Shokri},
  {et~al\mbox{.}}} \bibinfo{year}{2020}\natexlab{}.
\newblock \showarticletitle{Improving Deep Learning with Differential Privacy
  using Gradient Encoding and Denoising}.
\newblock \bibinfo{journal}{\emph{arXiv preprint arXiv:2007.11524}}
  (\bibinfo{year}{2020}).
\newblock


\bibitem[\protect\citeauthoryear{Nataraj, Mohammed, Manjunath, Chandrasekaran,
  Flenner, Bappy, and Roy-Chowdhury}{Nataraj et~al\mbox{.}}{2019}]%
        {nataraj2019detecting}
\bibfield{author}{\bibinfo{person}{Lakshmanan Nataraj},
  \bibinfo{person}{Tajuddin~Manhar Mohammed}, \bibinfo{person}{BS Manjunath},
  \bibinfo{person}{Shivkumar Chandrasekaran}, \bibinfo{person}{Arjuna Flenner},
  \bibinfo{person}{Jawadul~H Bappy}, {and} \bibinfo{person}{Amit~K
  Roy-Chowdhury}.} \bibinfo{year}{2019}\natexlab{}.
\newblock \showarticletitle{Detecting GAN generated fake images using
  co-occurrence matrices}.
\newblock \bibinfo{journal}{\emph{Electronic Imaging}} \bibinfo{volume}{2019},
  \bibinfo{number}{5} (\bibinfo{year}{2019}), \bibinfo{pages}{532--1}.
\newblock


\bibitem[\protect\citeauthoryear{Nguyen, Nguyen, Nguyen, Nguyen, and
  Nahavandi}{Nguyen et~al\mbox{.}}{2019}]%
        {nguyen2019deep}
\bibfield{author}{\bibinfo{person}{Thanh~Thi Nguyen}, \bibinfo{person}{Cuong~M
  Nguyen}, \bibinfo{person}{Dung~Tien Nguyen}, \bibinfo{person}{Duc~Thanh
  Nguyen}, {and} \bibinfo{person}{Saeid Nahavandi}.}
  \bibinfo{year}{2019}\natexlab{}.
\newblock \showarticletitle{Deep Learning for Deepfakes Creation and
  Detection}.
\newblock \bibinfo{journal}{\emph{arXiv preprint arXiv:1909.11573}}
  (\bibinfo{year}{2019}).
\newblock


\bibitem[\protect\citeauthoryear{Nirkin, Keller, and Hassner}{Nirkin
  et~al\mbox{.}}{2019}]%
        {nirkin2019fsgan}
\bibfield{author}{\bibinfo{person}{Yuval Nirkin}, \bibinfo{person}{Yosi
  Keller}, {and} \bibinfo{person}{Tal Hassner}.}
  \bibinfo{year}{2019}\natexlab{}.
\newblock \showarticletitle{FSGAN: Subject agnostic face swapping and
  reenactment}. In \bibinfo{booktitle}{\emph{Proceedings of the IEEE
  international conference on computer vision}}. \bibinfo{pages}{7184--7193}.
\newblock


\bibitem[\protect\citeauthoryear{Odena, Olah, and Shlens}{Odena
  et~al\mbox{.}}{2017}]%
        {odena2017conditional}
\bibfield{author}{\bibinfo{person}{Augustus Odena},
  \bibinfo{person}{Christopher Olah}, {and} \bibinfo{person}{Jonathon Shlens}.}
  \bibinfo{year}{2017}\natexlab{}.
\newblock \showarticletitle{Conditional image synthesis with auxiliary
  classifier gans}. In \bibinfo{booktitle}{\emph{Proceedings of the 34th
  International Conference on Machine Learning-Volume 70}}. JMLR. org,
  \bibinfo{pages}{2642--2651}.
\newblock


\bibitem[\protect\citeauthoryear{Olsson, Bhupatiraju, Brown, Odena, and
  Goodfellow}{Olsson et~al\mbox{.}}{2018}]%
        {olsson2018skill}
\bibfield{author}{\bibinfo{person}{Catherine Olsson}, \bibinfo{person}{Surya
  Bhupatiraju}, \bibinfo{person}{Tom Brown}, \bibinfo{person}{Augustus Odena},
  {and} \bibinfo{person}{Ian Goodfellow}.} \bibinfo{year}{2018}\natexlab{}.
\newblock \showarticletitle{Skill rating for generative models}.
\newblock \bibinfo{journal}{\emph{arXiv preprint arXiv:1808.04888}}
  (\bibinfo{year}{2018}).
\newblock


\bibitem[\protect\citeauthoryear{Oord, Dieleman, Zen, Simonyan, Vinyals,
  Graves, Kalchbrenner, Senior, and Kavukcuoglu}{Oord et~al\mbox{.}}{2016}]%
        {oord2016wavenet}
\bibfield{author}{\bibinfo{person}{Aaron van~den Oord}, \bibinfo{person}{Sander
  Dieleman}, \bibinfo{person}{Heiga Zen}, \bibinfo{person}{Karen Simonyan},
  \bibinfo{person}{Oriol Vinyals}, \bibinfo{person}{Alex Graves},
  \bibinfo{person}{Nal Kalchbrenner}, \bibinfo{person}{Andrew Senior}, {and}
  \bibinfo{person}{Koray Kavukcuoglu}.} \bibinfo{year}{2016}\natexlab{}.
\newblock \showarticletitle{Wavenet: A generative model for raw audio}.
\newblock \bibinfo{journal}{\emph{arXiv preprint arXiv:1609.03499}}
  (\bibinfo{year}{2016}).
\newblock


\bibitem[\protect\citeauthoryear{O'Shea, Pemula, Batra, and Clancy}{O'Shea
  et~al\mbox{.}}{2016}]%
        {o2016radio}
\bibfield{author}{\bibinfo{person}{Timothy~J O'Shea}, \bibinfo{person}{Latha
  Pemula}, \bibinfo{person}{Dhruv Batra}, {and} \bibinfo{person}{T~Charles
  Clancy}.} \bibinfo{year}{2016}\natexlab{}.
\newblock \showarticletitle{Radio transformer networks: Attention models for
  learning to synchronize in wireless systems}. In
  \bibinfo{booktitle}{\emph{2016 50th Asilomar Conference on Signals, Systems
  and Computers}}. IEEE, \bibinfo{pages}{662--666}.
\newblock


\bibitem[\protect\citeauthoryear{Pan, Yu, Yi, Khan, Yuan, and Zheng}{Pan
  et~al\mbox{.}}{2019}]%
        {pan2019recent}
\bibfield{author}{\bibinfo{person}{Zhaoqing Pan}, \bibinfo{person}{Weijie Yu},
  \bibinfo{person}{Xiaokai Yi}, \bibinfo{person}{Asifullah Khan},
  \bibinfo{person}{Feng Yuan}, {and} \bibinfo{person}{Yuhui Zheng}.}
  \bibinfo{year}{2019}\natexlab{}.
\newblock \showarticletitle{Recent progress on generative adversarial networks
  (GANs): A survey}.
\newblock \bibinfo{journal}{\emph{IEEE Access}}  \bibinfo{volume}{7}
  (\bibinfo{year}{2019}), \bibinfo{pages}{36322--36333}.
\newblock


\bibitem[\protect\citeauthoryear{Papernot, Abadi, Erlingsson, Goodfellow, and
  Talwar}{Papernot et~al\mbox{.}}{2016}]%
        {papernot2016semi}
\bibfield{author}{\bibinfo{person}{Nicolas Papernot},
  \bibinfo{person}{Mart{\'\i}n Abadi}, \bibinfo{person}{Ulfar Erlingsson},
  \bibinfo{person}{Ian Goodfellow}, {and} \bibinfo{person}{Kunal Talwar}.}
  \bibinfo{year}{2016}\natexlab{}.
\newblock \showarticletitle{Semi-supervised knowledge transfer for deep
  learning from private training data}.
\newblock \bibinfo{journal}{\emph{arXiv preprint arXiv:1610.05755}}
  (\bibinfo{year}{2016}).
\newblock


\bibitem[\protect\citeauthoryear{Papernot, McDaniel, Goodfellow, Jha, Celik,
  and Swami}{Papernot et~al\mbox{.}}{2017}]%
        {papernot2017practical}
\bibfield{author}{\bibinfo{person}{Nicolas Papernot}, \bibinfo{person}{Patrick
  McDaniel}, \bibinfo{person}{Ian Goodfellow}, \bibinfo{person}{Somesh Jha},
  \bibinfo{person}{Z~Berkay Celik}, {and} \bibinfo{person}{Ananthram Swami}.}
  \bibinfo{year}{2017}\natexlab{}.
\newblock \showarticletitle{Practical black-box attacks against machine
  learning}. In \bibinfo{booktitle}{\emph{Proceedings of the 2017 ACM on Asia
  conference on computer and communications security}}.
  \bibinfo{pages}{506--519}.
\newblock


\bibitem[\protect\citeauthoryear{Papernot, Song, Mironov, Raghunathan, Talwar,
  and Erlingsson}{Papernot et~al\mbox{.}}{2018}]%
        {papernot2018scalable}
\bibfield{author}{\bibinfo{person}{Nicolas Papernot}, \bibinfo{person}{Shuang
  Song}, \bibinfo{person}{Ilya Mironov}, \bibinfo{person}{Ananth Raghunathan},
  \bibinfo{person}{Kunal Talwar}, {and} \bibinfo{person}{{\'U}lfar
  Erlingsson}.} \bibinfo{year}{2018}\natexlab{}.
\newblock \showarticletitle{Scalable private learning with pate}.
\newblock \bibinfo{journal}{\emph{arXiv preprint arXiv:1802.08908}}
  (\bibinfo{year}{2018}).
\newblock


\bibitem[\protect\citeauthoryear{Papineni, Roukos, Ward, and Zhu}{Papineni
  et~al\mbox{.}}{2002}]%
        {papineni2002bleu}
\bibfield{author}{\bibinfo{person}{Kishore Papineni}, \bibinfo{person}{Salim
  Roukos}, \bibinfo{person}{Todd Ward}, {and} \bibinfo{person}{Wei-Jing Zhu}.}
  \bibinfo{year}{2002}\natexlab{}.
\newblock \showarticletitle{BLEU: a method for automatic evaluation of machine
  translation}. In \bibinfo{booktitle}{\emph{Proceedings of the 40th annual
  meeting on association for computational linguistics}}. Association for
  Computational Linguistics, \bibinfo{pages}{311--318}.
\newblock


\bibitem[\protect\citeauthoryear{Park, Gondal, Kamruzzaman, and Oliver}{Park
  et~al\mbox{.}}{2019}]%
        {park2019generative}
\bibfield{author}{\bibinfo{person}{Sean Park}, \bibinfo{person}{Iqbal Gondal},
  \bibinfo{person}{Joarder Kamruzzaman}, {and} \bibinfo{person}{Jon Oliver}.}
  \bibinfo{year}{2019}\natexlab{}.
\newblock \showarticletitle{Generative malware outbreak detection}. In
  \bibinfo{booktitle}{\emph{2019 IEEE International Conference on Industrial
  Technology (ICIT)}}. IEEE, \bibinfo{pages}{1149--1154}.
\newblock


\bibitem[\protect\citeauthoryear{Parmar, Vaswani, Uszkoreit, Kaiser, Shazeer,
  Ku, and Tran}{Parmar et~al\mbox{.}}{2018}]%
        {parmar2018image}
\bibfield{author}{\bibinfo{person}{Niki Parmar}, \bibinfo{person}{Ashish
  Vaswani}, \bibinfo{person}{Jakob Uszkoreit}, \bibinfo{person}{Lukasz Kaiser},
  \bibinfo{person}{Noam Shazeer}, \bibinfo{person}{Alexander Ku}, {and}
  \bibinfo{person}{Dustin Tran}.} \bibinfo{year}{2018}\natexlab{}.
\newblock \showarticletitle{Image Transformer}. In
  \bibinfo{booktitle}{\emph{International Conference on Machine Learning}}.
  \bibinfo{pages}{4055--4064}.
\newblock


\bibitem[\protect\citeauthoryear{Pascanu, Mikolov, and Bengio}{Pascanu
  et~al\mbox{.}}{2013}]%
        {pascanu2013difficulty}
\bibfield{author}{\bibinfo{person}{Razvan Pascanu}, \bibinfo{person}{Tomas
  Mikolov}, {and} \bibinfo{person}{Yoshua Bengio}.}
  \bibinfo{year}{2013}\natexlab{}.
\newblock \showarticletitle{On the difficulty of training recurrent neural
  networks}. In \bibinfo{booktitle}{\emph{International conference on machine
  learning}}. \bibinfo{pages}{1310--1318}.
\newblock


\bibitem[\protect\citeauthoryear{Petrov, Gao, Chervoniy, Liu, Marangonda,
  Um{\'e}, Jiang, RP, Zhang, Wu, et~al\mbox{.}}{Petrov et~al\mbox{.}}{2020}]%
        {petrov2020deepfacelab}
\bibfield{author}{\bibinfo{person}{Ivan Petrov}, \bibinfo{person}{Daiheng Gao},
  \bibinfo{person}{Nikolay Chervoniy}, \bibinfo{person}{Kunlin Liu},
  \bibinfo{person}{Sugasa Marangonda}, \bibinfo{person}{Chris Um{\'e}},
  \bibinfo{person}{Jian Jiang}, \bibinfo{person}{Luis RP},
  \bibinfo{person}{Sheng Zhang}, \bibinfo{person}{Pingyu Wu}, {et~al\mbox{.}}}
  \bibinfo{year}{2020}\natexlab{}.
\newblock \showarticletitle{DeepFaceLab: A simple, flexible and extensible face
  swapping framework}.
\newblock \bibinfo{journal}{\emph{arXiv preprint arXiv:2005.05535}}
  (\bibinfo{year}{2020}).
\newblock


\bibitem[\protect\citeauthoryear{Pevn{\`y}, Filler, and Bas}{Pevn{\`y}
  et~al\mbox{.}}{2010}]%
        {pevny2010using}
\bibfield{author}{\bibinfo{person}{Tom{\'a}{\v{s}} Pevn{\`y}},
  \bibinfo{person}{Tom{\'a}{\v{s}} Filler}, {and} \bibinfo{person}{Patrick
  Bas}.} \bibinfo{year}{2010}\natexlab{}.
\newblock \showarticletitle{Using high-dimensional image models to perform
  highly undetectable steganography}. In
  \bibinfo{booktitle}{\emph{International Workshop on Information Hiding}}.
  Springer, \bibinfo{pages}{161--177}.
\newblock


\bibitem[\protect\citeauthoryear{Pibre, Pasquet, Ienco, and Chaumont}{Pibre
  et~al\mbox{.}}{2016}]%
        {pibre2016deep}
\bibfield{author}{\bibinfo{person}{Lionel Pibre},
  \bibinfo{person}{J{\'e}r{\^o}me Pasquet}, \bibinfo{person}{Dino Ienco}, {and}
  \bibinfo{person}{Marc Chaumont}.} \bibinfo{year}{2016}\natexlab{}.
\newblock \showarticletitle{Deep learning is a good steganalysis tool when
  embedding key is reused for different images, even if there is a cover
  sourcemismatch}.
\newblock \bibinfo{journal}{\emph{Electronic Imaging}} \bibinfo{volume}{2016},
  \bibinfo{number}{8} (\bibinfo{year}{2016}), \bibinfo{pages}{1--11}.
\newblock


\bibitem[\protect\citeauthoryear{Ping, Stoyanovich, and Howe}{Ping
  et~al\mbox{.}}{2017}]%
        {ping2017datasynthesizer}
\bibfield{author}{\bibinfo{person}{Haoyue Ping}, \bibinfo{person}{Julia
  Stoyanovich}, {and} \bibinfo{person}{Bill Howe}.}
  \bibinfo{year}{2017}\natexlab{}.
\newblock \showarticletitle{Datasynthesizer: Privacy-preserving synthetic
  datasets}. In \bibinfo{booktitle}{\emph{Proceedings of the 29th International
  Conference on Scientific and Statistical Database Management}}.
  \bibinfo{pages}{1--5}.
\newblock


\bibitem[\protect\citeauthoryear{Por, Ang, and Delina}{Por
  et~al\mbox{.}}{2008}]%
        {por2008whitesteg}
\bibfield{author}{\bibinfo{person}{Lip~Y Por}, \bibinfo{person}{TF Ang}, {and}
  \bibinfo{person}{B Delina}.} \bibinfo{year}{2008}\natexlab{}.
\newblock \showarticletitle{Whitesteg: a new scheme in information hiding using
  text steganography}.
\newblock \bibinfo{journal}{\emph{WSEAS transactions on computers}}
  \bibinfo{volume}{7}, \bibinfo{number}{6} (\bibinfo{year}{2008}),
  \bibinfo{pages}{735--745}.
\newblock


\bibitem[\protect\citeauthoryear{Porr}{Porr}{2020}]%
        {gpt3blogfool}
\bibfield{author}{\bibinfo{person}{Liam Porr}.}
  \bibinfo{year}{2020}\natexlab{}.
\newblock \bibinfo{title}{My GPT-3 Blog Got 26 Thousand Visitors in 2 Weeks}.
\newblock
  \bibinfo{howpublished}{\url{https://liamp.substack.com/p/my-gpt-3-blog-got-26-thousand-visitors}}.
\newblock


\bibitem[\protect\citeauthoryear{Qian, Dong, Wang, and Tan}{Qian
  et~al\mbox{.}}{2015}]%
        {qian2015deep}
\bibfield{author}{\bibinfo{person}{Yinlong Qian}, \bibinfo{person}{Jing Dong},
  \bibinfo{person}{Wei Wang}, {and} \bibinfo{person}{Tieniu Tan}.}
  \bibinfo{year}{2015}\natexlab{}.
\newblock \showarticletitle{Deep learning for steganalysis via convolutional
  neural networks}. In \bibinfo{booktitle}{\emph{Media Watermarking, Security,
  and Forensics 2015}}, Vol.~\bibinfo{volume}{9409}. International Society for
  Optics and Photonics, \bibinfo{pages}{94090J}.
\newblock


\bibitem[\protect\citeauthoryear{Radford, Metz, and Chintala}{Radford
  et~al\mbox{.}}{2015}]%
        {radford2015unsupervised}
\bibfield{author}{\bibinfo{person}{Alec Radford}, \bibinfo{person}{Luke Metz},
  {and} \bibinfo{person}{Soumith Chintala}.} \bibinfo{year}{2015}\natexlab{}.
\newblock \showarticletitle{Unsupervised representation learning with deep
  convolutional generative adversarial networks}.
\newblock \bibinfo{journal}{\emph{arXiv preprint arXiv:1511.06434}}
  (\bibinfo{year}{2015}).
\newblock


\bibitem[\protect\citeauthoryear{Radford, Wu, Child, Luan, Amodei, and
  Sutskever}{Radford et~al\mbox{.}}{2019}]%
        {radford2019language}
\bibfield{author}{\bibinfo{person}{Alec Radford}, \bibinfo{person}{Jeff Wu},
  \bibinfo{person}{Rewon Child}, \bibinfo{person}{David Luan},
  \bibinfo{person}{Dario Amodei}, {and} \bibinfo{person}{Ilya Sutskever}.}
  \bibinfo{year}{2019}\natexlab{}.
\newblock \showarticletitle{Language Models are Unsupervised Multitask
  Learners}.
\newblock  (\bibinfo{year}{2019}).
\newblock


\bibitem[\protect\citeauthoryear{Raffel, Shazeer, Roberts, Lee, Narang, Matena,
  Zhou, Li, and Liu}{Raffel et~al\mbox{.}}{2019}]%
        {raffel2019exploring}
\bibfield{author}{\bibinfo{person}{Colin Raffel}, \bibinfo{person}{Noam
  Shazeer}, \bibinfo{person}{Adam Roberts}, \bibinfo{person}{Katherine Lee},
  \bibinfo{person}{Sharan Narang}, \bibinfo{person}{Michael Matena},
  \bibinfo{person}{Yanqi Zhou}, \bibinfo{person}{Wei Li}, {and}
  \bibinfo{person}{Peter~J Liu}.} \bibinfo{year}{2019}\natexlab{}.
\newblock \showarticletitle{Exploring the limits of transfer learning with a
  unified text-to-text transformer}.
\newblock \bibinfo{journal}{\emph{arXiv preprint arXiv:1910.10683}}
  (\bibinfo{year}{2019}).
\newblock


\bibitem[\protect\citeauthoryear{Razavi, van~den Oord, and Vinyals}{Razavi
  et~al\mbox{.}}{2019}]%
        {razavi2019generating}
\bibfield{author}{\bibinfo{person}{Ali Razavi}, \bibinfo{person}{Aaron van~den
  Oord}, {and} \bibinfo{person}{Oriol Vinyals}.}
  \bibinfo{year}{2019}\natexlab{}.
\newblock \showarticletitle{Generating diverse high-fidelity images with
  vq-vae-2}. In \bibinfo{booktitle}{\emph{Advances in Neural Information
  Processing Systems}}. \bibinfo{pages}{14837--14847}.
\newblock


\bibitem[\protect\citeauthoryear{Regmi and Borji}{Regmi and Borji}{2018}]%
        {regmi2018cross}
\bibfield{author}{\bibinfo{person}{Krishna Regmi} {and} \bibinfo{person}{Ali
  Borji}.} \bibinfo{year}{2018}\natexlab{}.
\newblock \showarticletitle{Cross-view image synthesis using conditional gans}.
  In \bibinfo{booktitle}{\emph{Proceedings of the IEEE Conference on Computer
  Vision and Pattern Recognition}}. \bibinfo{pages}{3501--3510}.
\newblock


\bibitem[\protect\citeauthoryear{Ren, Ruan, Tan, Qin, Zhao, Zhao, and Liu}{Ren
  et~al\mbox{.}}{2019}]%
        {ren2019fastspeech}
\bibfield{author}{\bibinfo{person}{Yi Ren}, \bibinfo{person}{Yangjun Ruan},
  \bibinfo{person}{Xu Tan}, \bibinfo{person}{Tao Qin}, \bibinfo{person}{Sheng
  Zhao}, \bibinfo{person}{Zhou Zhao}, {and} \bibinfo{person}{Tie-Yan Liu}.}
  \bibinfo{year}{2019}\natexlab{}.
\newblock \showarticletitle{Fastspeech: Fast, robust and controllable text to
  speech}. In \bibinfo{booktitle}{\emph{Advances in Neural Information
  Processing Systems}}. \bibinfo{pages}{3171--3180}.
\newblock


\bibitem[\protect\citeauthoryear{Richardson and Weiss}{Richardson and
  Weiss}{2018}]%
        {richardson2018gans}
\bibfield{author}{\bibinfo{person}{Eitan Richardson} {and}
  \bibinfo{person}{Yair Weiss}.} \bibinfo{year}{2018}\natexlab{}.
\newblock \showarticletitle{On gans and gmms}. In
  \bibinfo{booktitle}{\emph{Advances in Neural Information Processing
  Systems}}. \bibinfo{pages}{5847--5858}.
\newblock


\bibitem[\protect\citeauthoryear{Rifai, Vincent, Muller, Glorot, and
  Bengio}{Rifai et~al\mbox{.}}{2011}]%
        {rifai2011contractive}
\bibfield{author}{\bibinfo{person}{Salah Rifai}, \bibinfo{person}{Pascal
  Vincent}, \bibinfo{person}{Xavier Muller}, \bibinfo{person}{Xavier Glorot},
  {and} \bibinfo{person}{Yoshua Bengio}.} \bibinfo{year}{2011}\natexlab{}.
\newblock \showarticletitle{Contractive auto-encoders: Explicit invariance
  during feature extraction}. In \bibinfo{booktitle}{\emph{Icml}}.
\newblock


\bibitem[\protect\citeauthoryear{Rosenblatt, Liu, Pouyanfar, de~Leon, Desai,
  and Allen}{Rosenblatt et~al\mbox{.}}{2020}]%
        {rosenblatt2020differentially}
\bibfield{author}{\bibinfo{person}{Lucas Rosenblatt}, \bibinfo{person}{Xiaoyan
  Liu}, \bibinfo{person}{Samira Pouyanfar}, \bibinfo{person}{Eduardo de Leon},
  \bibinfo{person}{Anuj Desai}, {and} \bibinfo{person}{Joshua Allen}.}
  \bibinfo{year}{2020}\natexlab{}.
\newblock \showarticletitle{Differentially private synthetic data: Applied
  evaluations and enhancements}.
\newblock \bibinfo{journal}{\emph{arXiv preprint arXiv:2011.05537}}
  (\bibinfo{year}{2020}).
\newblock


\bibitem[\protect\citeauthoryear{Rumelhart, Hinton, and Williams}{Rumelhart
  et~al\mbox{.}}{1986}]%
        {rumelhart1986learning}
\bibfield{author}{\bibinfo{person}{David~E Rumelhart},
  \bibinfo{person}{Geoffrey~E Hinton}, {and} \bibinfo{person}{Ronald~J
  Williams}.} \bibinfo{year}{1986}\natexlab{}.
\newblock \showarticletitle{Learning representations by back-propagating
  errors}.
\newblock \bibinfo{journal}{\emph{nature}} \bibinfo{volume}{323},
  \bibinfo{number}{6088} (\bibinfo{year}{1986}), \bibinfo{pages}{533--536}.
\newblock


\bibitem[\protect\citeauthoryear{Sakurada and Yairi}{Sakurada and
  Yairi}{2014}]%
        {sakurada2014anomaly}
\bibfield{author}{\bibinfo{person}{Mayu Sakurada} {and}
  \bibinfo{person}{Takehisa Yairi}.} \bibinfo{year}{2014}\natexlab{}.
\newblock \showarticletitle{Anomaly detection using autoencoders with nonlinear
  dimensionality reduction}. In \bibinfo{booktitle}{\emph{Proceedings of the
  MLSDA 2014 2nd Workshop on Machine Learning for Sensory Data Analysis}}.
  \bibinfo{pages}{4--11}.
\newblock


\bibitem[\protect\citeauthoryear{Salimans, Goodfellow, Zaremba, Cheung,
  Radford, and Chen}{Salimans et~al\mbox{.}}{2016}]%
        {salimans2016improved}
\bibfield{author}{\bibinfo{person}{Tim Salimans}, \bibinfo{person}{Ian
  Goodfellow}, \bibinfo{person}{Wojciech Zaremba}, \bibinfo{person}{Vicki
  Cheung}, \bibinfo{person}{Alec Radford}, {and} \bibinfo{person}{Xi Chen}.}
  \bibinfo{year}{2016}\natexlab{}.
\newblock \showarticletitle{Improved techniques for training gans}. In
  \bibinfo{booktitle}{\emph{Advances in neural information processing
  systems}}. \bibinfo{pages}{2234--2242}.
\newblock


\bibitem[\protect\citeauthoryear{Salimans, Karpathy, Chen, and Kingma}{Salimans
  et~al\mbox{.}}{2017}]%
        {salimans2017pixelcnn++}
\bibfield{author}{\bibinfo{person}{Tim Salimans}, \bibinfo{person}{Andrej
  Karpathy}, \bibinfo{person}{Xi Chen}, {and} \bibinfo{person}{Diederik~P
  Kingma}.} \bibinfo{year}{2017}\natexlab{}.
\newblock \showarticletitle{Pixelcnn++: Improving the pixelcnn with discretized
  logistic mixture likelihood and other modifications}.
\newblock \bibinfo{journal}{\emph{arXiv preprint arXiv:1701.05517}}
  (\bibinfo{year}{2017}).
\newblock


\bibitem[\protect\citeauthoryear{Samangouei, Kabkab, and Chellappa}{Samangouei
  et~al\mbox{.}}{2018}]%
        {samangouei2018defense}
\bibfield{author}{\bibinfo{person}{Pouya Samangouei}, \bibinfo{person}{Maya
  Kabkab}, {and} \bibinfo{person}{Rama Chellappa}.}
  \bibinfo{year}{2018}\natexlab{}.
\newblock \showarticletitle{Defense-GAN: Protecting Classifiers Against
  Adversarial Attacks Using Generative Models}. In
  \bibinfo{booktitle}{\emph{International Conference on Learning
  Representations}}.
\newblock


\bibitem[\protect\citeauthoryear{Santurkar, Schmidt, and M{\k{a}}dry}{Santurkar
  et~al\mbox{.}}{2017}]%
        {santurkar2017classification}
\bibfield{author}{\bibinfo{person}{Shibani Santurkar}, \bibinfo{person}{Ludwig
  Schmidt}, {and} \bibinfo{person}{Aleksander M{\k{a}}dry}.}
  \bibinfo{year}{2017}\natexlab{}.
\newblock \showarticletitle{A classification-based study of covariate shift in
  gan distributions}.
\newblock \bibinfo{journal}{\emph{arXiv preprint arXiv:1711.00970}}
  (\bibinfo{year}{2017}).
\newblock


\bibitem[\protect\citeauthoryear{Sattigeri, Hoffman, Chenthamarakshan, and
  Varshney}{Sattigeri et~al\mbox{.}}{2019}]%
        {sattigeri2019fairness}
\bibfield{author}{\bibinfo{person}{Prasanna Sattigeri},
  \bibinfo{person}{Samuel~C Hoffman}, \bibinfo{person}{Vijil Chenthamarakshan},
  {and} \bibinfo{person}{Kush~R Varshney}.} \bibinfo{year}{2019}\natexlab{}.
\newblock \showarticletitle{Fairness GAN: Generating datasets with fairness
  properties using a generative adversarial network}.
\newblock \bibinfo{journal}{\emph{IBM Journal of Research and Development}}
  \bibinfo{volume}{63}, \bibinfo{number}{4/5} (\bibinfo{year}{2019}),
  \bibinfo{pages}{3--1}.
\newblock


\bibitem[\protect\citeauthoryear{Scarselli and Tsoi}{Scarselli and
  Tsoi}{1998}]%
        {scarselli1998universal}
\bibfield{author}{\bibinfo{person}{Franco Scarselli} {and}
  \bibinfo{person}{Ah~Chung Tsoi}.} \bibinfo{year}{1998}\natexlab{}.
\newblock \showarticletitle{Universal approximation using feedforward neural
  networks: A survey of some existing methods, and some new results}.
\newblock \bibinfo{journal}{\emph{Neural networks}} \bibinfo{volume}{11},
  \bibinfo{number}{1} (\bibinfo{year}{1998}), \bibinfo{pages}{15--37}.
\newblock


\bibitem[\protect\citeauthoryear{Schum}{Schum}{2001}]%
        {schum2001evidential}
\bibfield{author}{\bibinfo{person}{David~A Schum}.}
  \bibinfo{year}{2001}\natexlab{}.
\newblock \bibinfo{booktitle}{\emph{The evidential foundations of probabilistic
  reasoning}}.
\newblock \bibinfo{publisher}{Northwestern University Press}.
\newblock


\bibitem[\protect\citeauthoryear{Shi, Dong, Wang, Qian, and Zhang}{Shi
  et~al\mbox{.}}{2017}]%
        {shi2017ssgan}
\bibfield{author}{\bibinfo{person}{Haichao Shi}, \bibinfo{person}{Jing Dong},
  \bibinfo{person}{Wei Wang}, \bibinfo{person}{Yinlong Qian}, {and}
  \bibinfo{person}{Xiaoyu Zhang}.} \bibinfo{year}{2017}\natexlab{}.
\newblock \showarticletitle{SSGAN: secure steganography based on generative
  adversarial networks}. In \bibinfo{booktitle}{\emph{Pacific Rim Conference on
  Multimedia}}. Springer, \bibinfo{pages}{534--544}.
\newblock


\bibitem[\protect\citeauthoryear{Shokri, Stronati, Song, and Shmatikov}{Shokri
  et~al\mbox{.}}{2017}]%
        {shokri2017membership}
\bibfield{author}{\bibinfo{person}{Reza Shokri}, \bibinfo{person}{Marco
  Stronati}, \bibinfo{person}{Congzheng Song}, {and} \bibinfo{person}{Vitaly
  Shmatikov}.} \bibinfo{year}{2017}\natexlab{}.
\newblock \showarticletitle{Membership inference attacks against machine
  learning models}. In \bibinfo{booktitle}{\emph{2017 IEEE Symposium on
  Security and Privacy (SP)}}. IEEE, \bibinfo{pages}{3--18}.
\newblock


\bibitem[\protect\citeauthoryear{Solaiman}{Solaiman}{2020}]%
        {solaiman_2020}
\bibfield{author}{\bibinfo{person}{Irene Solaiman}.}
  \bibinfo{year}{2020}\natexlab{}.
\newblock \bibinfo{title}{GPT-2: 1.5B Release}.
\newblock
  \bibinfo{howpublished}{\url{https://openai.com/blog/gpt-2-1-5b-release/}}.
\newblock


\bibitem[\protect\citeauthoryear{Solaiman, Brundage, Clark, Askell,
  Herbert-Voss, Wu, Radford, and Wang}{Solaiman et~al\mbox{.}}{2019}]%
        {solaiman2019release}
\bibfield{author}{\bibinfo{person}{Irene Solaiman}, \bibinfo{person}{Miles
  Brundage}, \bibinfo{person}{Jack Clark}, \bibinfo{person}{Amanda Askell},
  \bibinfo{person}{Ariel Herbert-Voss}, \bibinfo{person}{Jeff Wu},
  \bibinfo{person}{Alec Radford}, {and} \bibinfo{person}{Jasmine Wang}.}
  \bibinfo{year}{2019}\natexlab{}.
\newblock \showarticletitle{Release strategies and the social impacts of
  language models}.
\newblock \bibinfo{journal}{\emph{arXiv preprint arXiv:1908.09203}}
  (\bibinfo{year}{2019}).
\newblock


\bibitem[\protect\citeauthoryear{Solsman}{Solsman}{2020}]%
        {solsman}
\bibfield{author}{\bibinfo{person}{Joan~E. Solsman}.}
  \bibinfo{year}{2020}\natexlab{}.
\newblock \bibinfo{title}{Deepfakes' threat to 2020 US election isn't what
  you'd think}.
\newblock
  \bibinfo{howpublished}{\url{https://www.cnet.com/features/deepfakes-threat-to-the-2020-us-election-isnt-what-youd-think/}}.
\newblock


\bibitem[\protect\citeauthoryear{Song, Eykholt, Evtimov, Fernandes, Li,
  Rahmati, Tramer, Prakash, and Kohno}{Song et~al\mbox{.}}{2018a}]%
        {song2018physical}
\bibfield{author}{\bibinfo{person}{Dawn Song}, \bibinfo{person}{Kevin Eykholt},
  \bibinfo{person}{Ivan Evtimov}, \bibinfo{person}{Earlence Fernandes},
  \bibinfo{person}{Bo Li}, \bibinfo{person}{Amir Rahmati},
  \bibinfo{person}{Florian Tramer}, \bibinfo{person}{Atul Prakash}, {and}
  \bibinfo{person}{Tadayoshi Kohno}.} \bibinfo{year}{2018}\natexlab{a}.
\newblock \showarticletitle{Physical adversarial examples for object
  detectors}. In \bibinfo{booktitle}{\emph{12th $\{$USENIX$\}$ Workshop on
  Offensive Technologies ($\{$WOOT$\}$ 18)}}.
\newblock


\bibitem[\protect\citeauthoryear{Song, Kim, Nowozin, Ermon, and Kushman}{Song
  et~al\mbox{.}}{2017}]%
        {song2017pixeldefend}
\bibfield{author}{\bibinfo{person}{Yang Song}, \bibinfo{person}{Taesup Kim},
  \bibinfo{person}{Sebastian Nowozin}, \bibinfo{person}{Stefano Ermon}, {and}
  \bibinfo{person}{Nate Kushman}.} \bibinfo{year}{2017}\natexlab{}.
\newblock \showarticletitle{Pixeldefend: Leveraging generative models to
  understand and defend against adversarial examples}.
\newblock \bibinfo{journal}{\emph{arXiv preprint arXiv:1710.10766}}
  (\bibinfo{year}{2017}).
\newblock


\bibitem[\protect\citeauthoryear{Song, Shu, Kushman, and Ermon}{Song
  et~al\mbox{.}}{2018b}]%
        {song2018constructing}
\bibfield{author}{\bibinfo{person}{Yang Song}, \bibinfo{person}{Rui Shu},
  \bibinfo{person}{Nate Kushman}, {and} \bibinfo{person}{Stefano Ermon}.}
  \bibinfo{year}{2018}\natexlab{b}.
\newblock \showarticletitle{Constructing unrestricted adversarial examples with
  generative models}. In \bibinfo{booktitle}{\emph{Advances in Neural
  Information Processing Systems}}. \bibinfo{pages}{8312--8323}.
\newblock


\bibitem[\protect\citeauthoryear{Stadler, Oprisanu, and Troncoso}{Stadler
  et~al\mbox{.}}{2021}]%
        {stadler2021synthetic}
\bibfield{author}{\bibinfo{person}{Theresa Stadler}, \bibinfo{person}{Bristena
  Oprisanu}, {and} \bibinfo{person}{Carmela Troncoso}.}
  \bibinfo{year}{2021}\natexlab{}.
\newblock \bibinfo{title}{Synthetic Data -- Anonymisation Groundhog Day}.
\newblock
\newblock
\showeprint[arxiv]{2011.07018}~[cs.LG]


\bibitem[\protect\citeauthoryear{Statt}{Statt}{2019}]%
        {statt_2019}
\bibfield{author}{\bibinfo{person}{Nick Statt}.}
  \bibinfo{year}{2019}\natexlab{}.
\newblock \bibinfo{title}{Thieves are now using AI deepfakes to trick companies
  into sending them money}.
\newblock
  \bibinfo{howpublished}{\url{https://www.theverge.com/2019/9/5/20851248/deepfakes-ai-fake-audio-phone-calls-thieves-trick-companies-stealing-money}}.
\newblock


\bibitem[\protect\citeauthoryear{Suwajanakorn, Seitz, and
  Kemelmacher-Shlizerman}{Suwajanakorn et~al\mbox{.}}{2017}]%
        {suwajanakorn2017synthesizing}
\bibfield{author}{\bibinfo{person}{Supasorn Suwajanakorn},
  \bibinfo{person}{Steven~M Seitz}, {and} \bibinfo{person}{Ira
  Kemelmacher-Shlizerman}.} \bibinfo{year}{2017}\natexlab{}.
\newblock \showarticletitle{Synthesizing obama: learning lip sync from audio}.
\newblock \bibinfo{journal}{\emph{ACM Transactions on Graphics (TOG)}}
  \bibinfo{volume}{36}, \bibinfo{number}{4} (\bibinfo{year}{2017}),
  \bibinfo{pages}{1--13}.
\newblock


\bibitem[\protect\citeauthoryear{Szegedy, Zaremba, Sutskever, Bruna, Erhan,
  Goodfellow, and Fergus}{Szegedy et~al\mbox{.}}{2013}]%
        {szegedy2013intriguing}
\bibfield{author}{\bibinfo{person}{Christian Szegedy},
  \bibinfo{person}{Wojciech Zaremba}, \bibinfo{person}{Ilya Sutskever},
  \bibinfo{person}{Joan Bruna}, \bibinfo{person}{Dumitru Erhan},
  \bibinfo{person}{Ian Goodfellow}, {and} \bibinfo{person}{Rob Fergus}.}
  \bibinfo{year}{2013}\natexlab{}.
\newblock \showarticletitle{Intriguing properties of neural networks}.
\newblock \bibinfo{journal}{\emph{arXiv preprint arXiv:1312.6199}}
  (\bibinfo{year}{2013}).
\newblock


\bibitem[\protect\citeauthoryear{Tantipongpipat, Waites, Boob, Siva, and
  Cummings}{Tantipongpipat et~al\mbox{.}}{2019}]%
        {tantipongpipat2019differentially}
\bibfield{author}{\bibinfo{person}{Uthaipon Tantipongpipat},
  \bibinfo{person}{Chris Waites}, \bibinfo{person}{Digvijay Boob},
  \bibinfo{person}{Amaresh~Ankit Siva}, {and} \bibinfo{person}{Rachel
  Cummings}.} \bibinfo{year}{2019}\natexlab{}.
\newblock \showarticletitle{Differentially Private Mixed-Type Data Generation
  For Unsupervised Learning}.
\newblock \bibinfo{journal}{\emph{arXiv preprint arXiv:1912.03250}}
  (\bibinfo{year}{2019}).
\newblock


\bibitem[\protect\citeauthoryear{Theis, van~den Oord, and Bethge}{Theis
  et~al\mbox{.}}{2016}]%
        {theis2015note}
\bibfield{author}{\bibinfo{person}{L Theis}, \bibinfo{person}{A van~den Oord},
  {and} \bibinfo{person}{M Bethge}.} \bibinfo{year}{2016}\natexlab{}.
\newblock \showarticletitle{A note on the evaluation of generative models}. In
  \bibinfo{booktitle}{\emph{International Conference on Learning
  Representations (ICLR 2016)}}. \bibinfo{pages}{1--10}.
\newblock


\bibitem[\protect\citeauthoryear{Tina~Fang, Jaggi, and Argyraki}{Tina~Fang
  et~al\mbox{.}}{2017}]%
        {fang2017generating}
\bibfield{author}{\bibinfo{person}{Tina Tina~Fang}, \bibinfo{person}{Martin
  Jaggi}, {and} \bibinfo{person}{Katerina Argyraki}.}
  \bibinfo{year}{2017}\natexlab{}.
\newblock \showarticletitle{Generating Steganographic Text with LSTMs}. In
  \bibinfo{booktitle}{\emph{Proceedings of the 55th Annual Meeting of the
  Association for Computational Linguistics-Student Research Workshop}}.
  \bibinfo{pages}{100--106}.
\newblock


\bibitem[\protect\citeauthoryear{Toews}{Toews}{2020}]%
        {toews_2020}
\bibfield{author}{\bibinfo{person}{Rob Toews}.}
  \bibinfo{year}{2020}\natexlab{}.
\newblock \bibinfo{title}{Deepfakes Are Going To Wreak Havoc On Society. We Are
  Not Prepared.}
\newblock
  \bibinfo{howpublished}{\url{https://www.forbes.com/sites/robtoews/2020/05/25/deepfakes-are-going-to-wreak-havoc-on-society-we-are-not-prepared/##5687fc1d7494}}.
\newblock


\bibitem[\protect\citeauthoryear{Torkzadehmahani, Kairouz, and
  Paten}{Torkzadehmahani et~al\mbox{.}}{2019}]%
        {torkzadehmahani2019dp}
\bibfield{author}{\bibinfo{person}{Reihaneh Torkzadehmahani},
  \bibinfo{person}{Peter Kairouz}, {and} \bibinfo{person}{Benedict Paten}.}
  \bibinfo{year}{2019}\natexlab{}.
\newblock \showarticletitle{Dp-cgan: Differentially private synthetic data and
  label generation}. In \bibinfo{booktitle}{\emph{Proceedings of the IEEE/CVF
  Conference on Computer Vision and Pattern Recognition Workshops}}.
  \bibinfo{pages}{0--0}.
\newblock


\bibitem[\protect\citeauthoryear{Touretzky, Mozer, and Hasselmo}{Touretzky
  et~al\mbox{.}}{1996}]%
        {touretzky1996advances}
\bibfield{author}{\bibinfo{person}{David~S Touretzky},
  \bibinfo{person}{Michael~C Mozer}, {and} \bibinfo{person}{Michael~E
  Hasselmo}.} \bibinfo{year}{1996}\natexlab{}.
\newblock \bibinfo{booktitle}{\emph{Advances in Neural Information Processing
  Systems 8: Proceedings of the 1995 Conference}}. Vol.~\bibinfo{volume}{8}.
\newblock \bibinfo{publisher}{Mit Press}.
\newblock


\bibitem[\protect\citeauthoryear{Triastcyn and Faltings}{Triastcyn and
  Faltings}{2018}]%
        {triastcyn2018generating}
\bibfield{author}{\bibinfo{person}{Aleksei Triastcyn} {and}
  \bibinfo{person}{Boi Faltings}.} \bibinfo{year}{2018}\natexlab{}.
\newblock \showarticletitle{Generating artificial data for private deep
  learning}.
\newblock \bibinfo{journal}{\emph{arXiv preprint arXiv:1803.03148}}
  (\bibinfo{year}{2018}).
\newblock


\bibitem[\protect\citeauthoryear{Triastcyn and Faltings}{Triastcyn and
  Faltings}{2020}]%
        {triastcyn2020generating}
\bibfield{author}{\bibinfo{person}{Aleksei Triastcyn} {and}
  \bibinfo{person}{Boi Faltings}.} \bibinfo{year}{2020}\natexlab{}.
\newblock \showarticletitle{Generating higher-fidelity synthetic datasets with
  privacy guarantees}.
\newblock \bibinfo{journal}{\emph{arXiv preprint arXiv:2003.00997}}
  (\bibinfo{year}{2020}).
\newblock


\bibitem[\protect\citeauthoryear{Tulyakov, Liu, Yang, and Kautz}{Tulyakov
  et~al\mbox{.}}{2018}]%
        {tulyakov2018mocogan}
\bibfield{author}{\bibinfo{person}{Sergey Tulyakov}, \bibinfo{person}{Ming-Yu
  Liu}, \bibinfo{person}{Xiaodong Yang}, {and} \bibinfo{person}{Jan Kautz}.}
  \bibinfo{year}{2018}\natexlab{}.
\newblock \showarticletitle{Mocogan: Decomposing motion and content for video
  generation}. In \bibinfo{booktitle}{\emph{Proceedings of the IEEE conference
  on computer vision and pattern recognition}}. \bibinfo{pages}{1526--1535}.
\newblock


\bibitem[\protect\citeauthoryear{Turner, Tsipras, and Madry}{Turner
  et~al\mbox{.}}{2019}]%
        {turner2019label}
\bibfield{author}{\bibinfo{person}{Alexander Turner}, \bibinfo{person}{Dimitris
  Tsipras}, {and} \bibinfo{person}{Aleksander Madry}.}
  \bibinfo{year}{2019}\natexlab{}.
\newblock \showarticletitle{Label-Consistent Backdoor Attacks}.
\newblock \bibinfo{journal}{\emph{arXiv preprint arXiv:1912.02771}}
  (\bibinfo{year}{2019}).
\newblock


\bibitem[\protect\citeauthoryear{van~den Oord and Kalchbrenner}{van~den Oord
  and Kalchbrenner}{2016}]%
        {oord2016pixelrnn}
\bibfield{author}{\bibinfo{person}{A\"aron van~den Oord} {and}
  \bibinfo{person}{Nal Kalchbrenner}.} \bibinfo{year}{2016}\natexlab{}.
\newblock \showarticletitle{Pixel RNN}. In \bibinfo{booktitle}{\emph{ICML}}.
\newblock


\bibitem[\protect\citeauthoryear{Van~den Oord, Kalchbrenner, Espeholt, Vinyals,
  Graves, et~al\mbox{.}}{Van~den Oord et~al\mbox{.}}{2016}]%
        {van2016conditional}
\bibfield{author}{\bibinfo{person}{Aaron Van~den Oord}, \bibinfo{person}{Nal
  Kalchbrenner}, \bibinfo{person}{Lasse Espeholt}, \bibinfo{person}{Oriol
  Vinyals}, \bibinfo{person}{Alex Graves}, {et~al\mbox{.}}}
  \bibinfo{year}{2016}\natexlab{}.
\newblock \showarticletitle{Conditional image generation with pixelcnn
  decoders}. In \bibinfo{booktitle}{\emph{Advances in neural information
  processing systems}}. \bibinfo{pages}{4790--4798}.
\newblock


\bibitem[\protect\citeauthoryear{Vasquez and Lewis}{Vasquez and Lewis}{2019}]%
        {vasquez2019melnet}
\bibfield{author}{\bibinfo{person}{Sean Vasquez} {and} \bibinfo{person}{Mike
  Lewis}.} \bibinfo{year}{2019}\natexlab{}.
\newblock \showarticletitle{Melnet: A generative model for audio in the
  frequency domain}.
\newblock \bibinfo{journal}{\emph{arXiv preprint arXiv:1906.01083}}
  (\bibinfo{year}{2019}).
\newblock


\bibitem[\protect\citeauthoryear{Vaswani, Shazeer, Parmar, Uszkoreit, Jones,
  Gomez, Kaiser, and Polosukhin}{Vaswani et~al\mbox{.}}{2017}]%
        {vaswani2017attention}
\bibfield{author}{\bibinfo{person}{Ashish Vaswani}, \bibinfo{person}{Noam
  Shazeer}, \bibinfo{person}{Niki Parmar}, \bibinfo{person}{Jakob Uszkoreit},
  \bibinfo{person}{Llion Jones}, \bibinfo{person}{Aidan~N Gomez},
  \bibinfo{person}{{\L}ukasz Kaiser}, {and} \bibinfo{person}{Illia
  Polosukhin}.} \bibinfo{year}{2017}\natexlab{}.
\newblock \showarticletitle{Attention is all you need}. In
  \bibinfo{booktitle}{\emph{Advances in neural information processing
  systems}}. \bibinfo{pages}{5998--6008}.
\newblock


\bibitem[\protect\citeauthoryear{Vincent}{Vincent}{2018}]%
        {verge2018deepfake}
\bibfield{author}{\bibinfo{person}{James Vincent}.}
  \bibinfo{year}{2018}\natexlab{}.
\newblock \bibinfo{title}{Watch Jordan Peele use AI to make Barack Obama
  deliver a PSA about fake news}.
\newblock
  \bibinfo{howpublished}{\url{https://www.theverge.com/tldr/2018/4/17/17247334/ai-fake-news-video-barack-obama-jordan-peele-buzzfeed}}.
\newblock


\bibitem[\protect\citeauthoryear{Vincent, Larochelle, Lajoie, Bengio, Manzagol,
  and Bottou}{Vincent et~al\mbox{.}}{2010}]%
        {vincent2010stacked}
\bibfield{author}{\bibinfo{person}{Pascal Vincent}, \bibinfo{person}{Hugo
  Larochelle}, \bibinfo{person}{Isabelle Lajoie}, \bibinfo{person}{Yoshua
  Bengio}, \bibinfo{person}{Pierre-Antoine Manzagol}, {and}
  \bibinfo{person}{L{\'e}on Bottou}.} \bibinfo{year}{2010}\natexlab{}.
\newblock \showarticletitle{Stacked denoising autoencoders: Learning useful
  representations in a deep network with a local denoising criterion.}
\newblock \bibinfo{journal}{\emph{Journal of machine learning research}}
  \bibinfo{volume}{11}, \bibinfo{number}{12} (\bibinfo{year}{2010}).
\newblock


\bibitem[\protect\citeauthoryear{Volkhonskiy, Nazarov, and Burnaev}{Volkhonskiy
  et~al\mbox{.}}{2020}]%
        {volkhonskiy2020steganographic}
\bibfield{author}{\bibinfo{person}{Denis Volkhonskiy}, \bibinfo{person}{Ivan
  Nazarov}, {and} \bibinfo{person}{Evgeny Burnaev}.}
  \bibinfo{year}{2020}\natexlab{}.
\newblock \showarticletitle{Steganographic generative adversarial networks}. In
  \bibinfo{booktitle}{\emph{Twelfth International Conference on Machine Vision
  (ICMV 2019)}}, Vol.~\bibinfo{volume}{11433}. International Society for Optics
  and Photonics, \bibinfo{pages}{114333M}.
\newblock


\bibitem[\protect\citeauthoryear{Walker, Marino, Gupta, and Hebert}{Walker
  et~al\mbox{.}}{2017}]%
        {walker2017pose}
\bibfield{author}{\bibinfo{person}{Jacob Walker}, \bibinfo{person}{Kenneth
  Marino}, \bibinfo{person}{Abhinav Gupta}, {and} \bibinfo{person}{Martial
  Hebert}.} \bibinfo{year}{2017}\natexlab{}.
\newblock \showarticletitle{The pose knows: Video forecasting by generating
  pose futures}. In \bibinfo{booktitle}{\emph{Proceedings of the IEEE
  international conference on computer vision}}. \bibinfo{pages}{3332--3341}.
\newblock


\bibitem[\protect\citeauthoryear{Wallace, Feng, Kandpal, Gardner, and
  Singh}{Wallace et~al\mbox{.}}{2019}]%
        {wallace2019universal}
\bibfield{author}{\bibinfo{person}{Eric Wallace}, \bibinfo{person}{Shi Feng},
  \bibinfo{person}{Nikhil Kandpal}, \bibinfo{person}{Matt Gardner}, {and}
  \bibinfo{person}{Sameer Singh}.} \bibinfo{year}{2019}\natexlab{}.
\newblock \showarticletitle{Universal adversarial triggers for attacking and
  analyzing NLP}.
\newblock \bibinfo{journal}{\emph{arXiv preprint arXiv:1908.07125}}
  (\bibinfo{year}{2019}).
\newblock


\bibitem[\protect\citeauthoryear{Wallace, Zhao, Feng, and Singh}{Wallace
  et~al\mbox{.}}{2020}]%
        {wallace2020customizing}
\bibfield{author}{\bibinfo{person}{Eric Wallace}, \bibinfo{person}{Tony~Z
  Zhao}, \bibinfo{person}{Shi Feng}, {and} \bibinfo{person}{Sameer Singh}.}
  \bibinfo{year}{2020}\natexlab{}.
\newblock \showarticletitle{Customizing Triggers with Concealed Data
  Poisoning}.
\newblock \bibinfo{journal}{\emph{arXiv preprint arXiv:2010.12563}}
  (\bibinfo{year}{2020}).
\newblock


\bibitem[\protect\citeauthoryear{Wang and Gong}{Wang and Gong}{2018}]%
        {wang2018stealing}
\bibfield{author}{\bibinfo{person}{Binghui Wang} {and}
  \bibinfo{person}{Neil~Zhenqiang Gong}.} \bibinfo{year}{2018}\natexlab{}.
\newblock \showarticletitle{Stealing hyperparameters in machine learning}. In
  \bibinfo{booktitle}{\emph{2018 IEEE Symposium on Security and Privacy (SP)}}.
  IEEE, \bibinfo{pages}{36--52}.
\newblock


\bibitem[\protect\citeauthoryear{Wang}{Wang}{2019}]%
        {wang_2019}
\bibfield{author}{\bibinfo{person}{Chenxi Wang}.}
  \bibinfo{year}{2019}\natexlab{}.
\newblock \bibinfo{title}{Deepfakes, Revenge Porn, And The Impact On Women}.
\newblock
  \bibinfo{howpublished}{\url{https://www.forbes.com/sites/chenxiwang/2019/11/01/deepfakes-revenge-porn-and-the-impact-on-women/##1ce404831f53}}.
\newblock


\bibitem[\protect\citeauthoryear{Wang, Wang, Zhang, Owens, and Efros}{Wang
  et~al\mbox{.}}{2020}]%
        {wang2020cnn}
\bibfield{author}{\bibinfo{person}{Sheng-Yu Wang}, \bibinfo{person}{Oliver
  Wang}, \bibinfo{person}{Richard Zhang}, \bibinfo{person}{Andrew Owens}, {and}
  \bibinfo{person}{Alexei~A Efros}.} \bibinfo{year}{2020}\natexlab{}.
\newblock \showarticletitle{CNN-generated images are surprisingly easy to
  spot... for now}. In \bibinfo{booktitle}{\emph{Proceedings of the IEEE/CVF
  Conference on Computer Vision and Pattern Recognition}}.
  \bibinfo{pages}{8695--8704}.
\newblock


\bibitem[\protect\citeauthoryear{Wang, Zhang, and Van De~Weijer}{Wang
  et~al\mbox{.}}{2016}]%
        {wang2016ensembles}
\bibfield{author}{\bibinfo{person}{Yaxing Wang}, \bibinfo{person}{Lichao
  Zhang}, {and} \bibinfo{person}{Joost Van De~Weijer}.}
  \bibinfo{year}{2016}\natexlab{}.
\newblock \showarticletitle{Ensembles of generative adversarial networks}.
\newblock \bibinfo{journal}{\emph{arXiv preprint arXiv:1612.00991}}
  (\bibinfo{year}{2016}).
\newblock


\bibitem[\protect\citeauthoryear{Wang and Bovik}{Wang and Bovik}{2009}]%
        {wang2009mean}
\bibfield{author}{\bibinfo{person}{Zhou Wang} {and} \bibinfo{person}{Alan~C
  Bovik}.} \bibinfo{year}{2009}\natexlab{}.
\newblock \showarticletitle{Mean squared error: Love it or leave it? A new look
  at signal fidelity measures}.
\newblock \bibinfo{journal}{\emph{IEEE signal processing magazine}}
  \bibinfo{volume}{26}, \bibinfo{number}{1} (\bibinfo{year}{2009}),
  \bibinfo{pages}{98--117}.
\newblock


\bibitem[\protect\citeauthoryear{Wang, Bovik, Sheikh, and Simoncelli}{Wang
  et~al\mbox{.}}{2004}]%
        {wang2004image}
\bibfield{author}{\bibinfo{person}{Zhou Wang}, \bibinfo{person}{Alan~C Bovik},
  \bibinfo{person}{Hamid~R Sheikh}, {and} \bibinfo{person}{Eero~P Simoncelli}.}
  \bibinfo{year}{2004}\natexlab{}.
\newblock \showarticletitle{Image quality assessment: from error visibility to
  structural similarity}.
\newblock \bibinfo{journal}{\emph{IEEE transactions on image processing}}
  \bibinfo{volume}{13}, \bibinfo{number}{4} (\bibinfo{year}{2004}),
  \bibinfo{pages}{600--612}.
\newblock


\bibitem[\protect\citeauthoryear{Wen, Zhou, Zhong, and Xue}{Wen
  et~al\mbox{.}}{2019}]%
        {wen2019convolutional}
\bibfield{author}{\bibinfo{person}{Juan Wen}, \bibinfo{person}{Xuejing Zhou},
  \bibinfo{person}{Ping Zhong}, {and} \bibinfo{person}{Yiming Xue}.}
  \bibinfo{year}{2019}\natexlab{}.
\newblock \showarticletitle{Convolutional neural network based text
  steganalysis}.
\newblock \bibinfo{journal}{\emph{IEEE Signal Processing Letters}}
  \bibinfo{volume}{26}, \bibinfo{number}{3} (\bibinfo{year}{2019}),
  \bibinfo{pages}{460--464}.
\newblock


\bibitem[\protect\citeauthoryear{Witten, Neal, and Cleary}{Witten
  et~al\mbox{.}}{1987}]%
        {witten1987arithmetic}
\bibfield{author}{\bibinfo{person}{Ian~H Witten}, \bibinfo{person}{Radford~M
  Neal}, {and} \bibinfo{person}{John~G Cleary}.}
  \bibinfo{year}{1987}\natexlab{}.
\newblock \showarticletitle{Arithmetic coding for data compression}.
\newblock \bibinfo{journal}{\emph{Commun. ACM}} \bibinfo{volume}{30},
  \bibinfo{number}{6} (\bibinfo{year}{1987}), \bibinfo{pages}{520--540}.
\newblock


\bibitem[\protect\citeauthoryear{Wolf, Chaumond, Debut, Sanh, Delangue, Moi,
  Cistac, Funtowicz, Davison, Shleifer, et~al\mbox{.}}{Wolf
  et~al\mbox{.}}{2020}]%
        {wolf2020transformers}
\bibfield{author}{\bibinfo{person}{Thomas Wolf}, \bibinfo{person}{Julien
  Chaumond}, \bibinfo{person}{Lysandre Debut}, \bibinfo{person}{Victor Sanh},
  \bibinfo{person}{Clement Delangue}, \bibinfo{person}{Anthony Moi},
  \bibinfo{person}{Pierric Cistac}, \bibinfo{person}{Morgan Funtowicz},
  \bibinfo{person}{Joe Davison}, \bibinfo{person}{Sam Shleifer},
  {et~al\mbox{.}}} \bibinfo{year}{2020}\natexlab{}.
\newblock \showarticletitle{Transformers: State-of-the-art natural language
  processing}. In \bibinfo{booktitle}{\emph{Proceedings of the 2020 Conference
  on Empirical Methods in Natural Language Processing: System Demonstrations}}.
  \bibinfo{pages}{38--45}.
\newblock


\bibitem[\protect\citeauthoryear{Woodworth, Gunasekar, Ohannessian, and
  Srebro}{Woodworth et~al\mbox{.}}{2017}]%
        {woodworth2017learning}
\bibfield{author}{\bibinfo{person}{Blake Woodworth}, \bibinfo{person}{Suriya
  Gunasekar}, \bibinfo{person}{Mesrob~I Ohannessian}, {and}
  \bibinfo{person}{Nathan Srebro}.} \bibinfo{year}{2017}\natexlab{}.
\newblock \showarticletitle{Learning Non-Discriminatory Predictors}. In
  \bibinfo{booktitle}{\emph{Conference on Learning Theory}}.
  \bibinfo{pages}{1920--1953}.
\newblock


\bibitem[\protect\citeauthoryear{Xiang and Li}{Xiang and Li}{2017}]%
        {xiang2017effects}
\bibfield{author}{\bibinfo{person}{Sitao Xiang} {and} \bibinfo{person}{Hao
  Li}.} \bibinfo{year}{2017}\natexlab{}.
\newblock \showarticletitle{On the effects of batch and weight normalization in
  generative adversarial networks}.
\newblock \bibinfo{journal}{\emph{arXiv preprint arXiv:1704.03971}}
  (\bibinfo{year}{2017}).
\newblock


\bibitem[\protect\citeauthoryear{Xiao, Li, Zhu, He, Liu, and Song}{Xiao
  et~al\mbox{.}}{2018}]%
        {xiao2018generating}
\bibfield{author}{\bibinfo{person}{Chaowei Xiao}, \bibinfo{person}{Bo Li},
  \bibinfo{person}{Jun~Yan Zhu}, \bibinfo{person}{Warren He},
  \bibinfo{person}{Mingyan Liu}, {and} \bibinfo{person}{Dawn Song}.}
  \bibinfo{year}{2018}\natexlab{}.
\newblock \showarticletitle{Generating adversarial examples with adversarial
  networks}. In \bibinfo{booktitle}{\emph{27th International Joint Conference
  on Artificial Intelligence, IJCAI 2018}}. International Joint Conferences on
  Artificial Intelligence, \bibinfo{pages}{3905--3911}.
\newblock


\bibitem[\protect\citeauthoryear{Xie, Lin, Wang, Wang, and Zhou}{Xie
  et~al\mbox{.}}{2018}]%
        {xie2018differentially}
\bibfield{author}{\bibinfo{person}{Liyang Xie}, \bibinfo{person}{Kaixiang Lin},
  \bibinfo{person}{Shu Wang}, \bibinfo{person}{Fei Wang}, {and}
  \bibinfo{person}{Jiayu Zhou}.} \bibinfo{year}{2018}\natexlab{}.
\newblock \showarticletitle{Differentially private generative adversarial
  network}.
\newblock \bibinfo{journal}{\emph{arXiv preprint arXiv:1802.06739}}
  (\bibinfo{year}{2018}).
\newblock


\bibitem[\protect\citeauthoryear{Xu, Ren, Zhang, Zhang, Qin, and Ren}{Xu
  et~al\mbox{.}}{2019a}]%
        {xu2019ganobfuscator}
\bibfield{author}{\bibinfo{person}{Chugui Xu}, \bibinfo{person}{Ju Ren},
  \bibinfo{person}{Deyu Zhang}, \bibinfo{person}{Yaoxue Zhang},
  \bibinfo{person}{Zhan Qin}, {and} \bibinfo{person}{Kui Ren}.}
  \bibinfo{year}{2019}\natexlab{a}.
\newblock \showarticletitle{GANobfuscator: Mitigating information leakage under
  GAN via differential privacy}.
\newblock \bibinfo{journal}{\emph{IEEE Transactions on Information Forensics
  and Security}} \bibinfo{volume}{14}, \bibinfo{number}{9}
  (\bibinfo{year}{2019}), \bibinfo{pages}{2358--2371}.
\newblock


\bibitem[\protect\citeauthoryear{Xu, Yuan, Zhang, and Wu}{Xu
  et~al\mbox{.}}{2018}]%
        {xu2018fairgan}
\bibfield{author}{\bibinfo{person}{Depeng Xu}, \bibinfo{person}{Shuhan Yuan},
  \bibinfo{person}{Lu Zhang}, {and} \bibinfo{person}{Xintao Wu}.}
  \bibinfo{year}{2018}\natexlab{}.
\newblock \showarticletitle{Fairgan: Fairness-aware generative adversarial
  networks}. In \bibinfo{booktitle}{\emph{2018 IEEE International Conference on
  Big Data (Big Data)}}. IEEE, \bibinfo{pages}{570--575}.
\newblock


\bibitem[\protect\citeauthoryear{Xu, Skoularidou, Cuesta-Infante, and
  Veeramachaneni}{Xu et~al\mbox{.}}{2019b}]%
        {xu2019modeling}
\bibfield{author}{\bibinfo{person}{Lei Xu}, \bibinfo{person}{Maria
  Skoularidou}, \bibinfo{person}{Alfredo Cuesta-Infante}, {and}
  \bibinfo{person}{Kalyan Veeramachaneni}.} \bibinfo{year}{2019}\natexlab{b}.
\newblock \showarticletitle{Modeling tabular data using conditional gan}.
\newblock \bibinfo{journal}{\emph{arXiv preprint arXiv:1907.00503}}
  (\bibinfo{year}{2019}).
\newblock


\bibitem[\protect\citeauthoryear{Yan, Yang, Sohn, and Lee}{Yan
  et~al\mbox{.}}{2016}]%
        {yan2016attribute2image}
\bibfield{author}{\bibinfo{person}{Xinchen Yan}, \bibinfo{person}{Jimei Yang},
  \bibinfo{person}{Kihyuk Sohn}, {and} \bibinfo{person}{Honglak Lee}.}
  \bibinfo{year}{2016}\natexlab{}.
\newblock \showarticletitle{Attribute2image: Conditional image generation from
  visual attributes}. In \bibinfo{booktitle}{\emph{European Conference on
  Computer Vision}}. Springer, \bibinfo{pages}{776--791}.
\newblock


\bibitem[\protect\citeauthoryear{Yang, Wu, Li, and Chen}{Yang
  et~al\mbox{.}}{2017}]%
        {yang2017generative}
\bibfield{author}{\bibinfo{person}{Chaofei Yang}, \bibinfo{person}{Qing Wu},
  \bibinfo{person}{Hai Li}, {and} \bibinfo{person}{Yiran Chen}.}
  \bibinfo{year}{2017}\natexlab{}.
\newblock \showarticletitle{Generative poisoning attack method against neural
  networks}.
\newblock \bibinfo{journal}{\emph{arXiv preprint arXiv:1703.01340}}
  (\bibinfo{year}{2017}).
\newblock


\bibitem[\protect\citeauthoryear{Yang, Liu, Kang, Wong, and Shi}{Yang
  et~al\mbox{.}}{2018c}]%
        {yang2018spatial}
\bibfield{author}{\bibinfo{person}{Jianhua Yang}, \bibinfo{person}{Kai Liu},
  \bibinfo{person}{Xiangui Kang}, \bibinfo{person}{Edward~K Wong}, {and}
  \bibinfo{person}{Yun-Qing Shi}.} \bibinfo{year}{2018}\natexlab{c}.
\newblock \showarticletitle{Spatial image steganography based on generative
  adversarial network}.
\newblock \bibinfo{journal}{\emph{arXiv preprint arXiv:1804.07939}}
  (\bibinfo{year}{2018}).
\newblock


\bibitem[\protect\citeauthoryear{Yang, Dai, Yang, Carbonell, Salakhutdinov, and
  Le}{Yang et~al\mbox{.}}{2019a}]%
        {yang2019xlnet}
\bibfield{author}{\bibinfo{person}{Zhilin Yang}, \bibinfo{person}{Zihang Dai},
  \bibinfo{person}{Yiming Yang}, \bibinfo{person}{Jaime Carbonell},
  \bibinfo{person}{Russ~R Salakhutdinov}, {and} \bibinfo{person}{Quoc~V Le}.}
  \bibinfo{year}{2019}\natexlab{a}.
\newblock \showarticletitle{Xlnet: Generalized autoregressive pretraining for
  language understanding}. In \bibinfo{booktitle}{\emph{Advances in neural
  information processing systems}}. \bibinfo{pages}{5753--5763}.
\newblock


\bibitem[\protect\citeauthoryear{Yang, Du, Tan, Huang, and Zhang}{Yang
  et~al\mbox{.}}{2018a}]%
        {yang2018aag}
\bibfield{author}{\bibinfo{person}{Zhongliang Yang}, \bibinfo{person}{Xingjian
  Du}, \bibinfo{person}{Yilin Tan}, \bibinfo{person}{Yongfeng Huang}, {and}
  \bibinfo{person}{Yu-Jin Zhang}.} \bibinfo{year}{2018}\natexlab{a}.
\newblock \showarticletitle{Aag-stega: Automatic audio generation-based
  steganography}.
\newblock \bibinfo{journal}{\emph{arXiv preprint arXiv:1809.03463}}
  (\bibinfo{year}{2018}).
\newblock


\bibitem[\protect\citeauthoryear{Yang, Wang, Li, Huang, and Zhang}{Yang
  et~al\mbox{.}}{2019b}]%
        {yang2019ts}
\bibfield{author}{\bibinfo{person}{Zhongliang Yang}, \bibinfo{person}{Ke Wang},
  \bibinfo{person}{Jian Li}, \bibinfo{person}{Yongfeng Huang}, {and}
  \bibinfo{person}{Yu-Jin Zhang}.} \bibinfo{year}{2019}\natexlab{b}.
\newblock \showarticletitle{TS-RNN: text steganalysis based on recurrent neural
  networks}.
\newblock \bibinfo{journal}{\emph{IEEE Signal Processing Letters}}
  \bibinfo{volume}{26}, \bibinfo{number}{12} (\bibinfo{year}{2019}),
  \bibinfo{pages}{1743--1747}.
\newblock


\bibitem[\protect\citeauthoryear{Yang, Wei, Liu, Huang, and Zhang}{Yang
  et~al\mbox{.}}{2019c}]%
        {yang2019gan}
\bibfield{author}{\bibinfo{person}{Zhongliang Yang}, \bibinfo{person}{Nan Wei},
  \bibinfo{person}{Qinghe Liu}, \bibinfo{person}{Yongfeng Huang}, {and}
  \bibinfo{person}{Yujin Zhang}.} \bibinfo{year}{2019}\natexlab{c}.
\newblock \showarticletitle{GAN-TStega: Text Steganography Based on Generative
  Adversarial Networks.}. In \bibinfo{booktitle}{\emph{IWDW}}.
  \bibinfo{pages}{18--31}.
\newblock


\bibitem[\protect\citeauthoryear{Yang, Xiang, Zhang, Sun, and Huang}{Yang
  et~al\mbox{.}}{2021}]%
        {yang2021linguistic}
\bibfield{author}{\bibinfo{person}{Zhongliang Yang}, \bibinfo{person}{Lingyun
  Xiang}, \bibinfo{person}{Siyu Zhang}, \bibinfo{person}{Xingming Sun}, {and}
  \bibinfo{person}{Yongfeng Huang}.} \bibinfo{year}{2021}\natexlab{}.
\newblock \showarticletitle{Linguistic Generative Steganography With Enhanced
  Cognitive-Imperceptibility}.
\newblock \bibinfo{journal}{\emph{IEEE Signal Processing Letters}}
  \bibinfo{volume}{28} (\bibinfo{year}{2021}), \bibinfo{pages}{409--413}.
\newblock


\bibitem[\protect\citeauthoryear{Yang, Guo, Chen, Huang, and Zhang}{Yang
  et~al\mbox{.}}{2018b}]%
        {yang2018rnn}
\bibfield{author}{\bibinfo{person}{Zhong-Liang Yang},
  \bibinfo{person}{Xiao-Qing Guo}, \bibinfo{person}{Zi-Ming Chen},
  \bibinfo{person}{Yong-Feng Huang}, {and} \bibinfo{person}{Yu-Jin Zhang}.}
  \bibinfo{year}{2018}\natexlab{b}.
\newblock \showarticletitle{RNN-stega: Linguistic steganography based on
  recurrent neural networks}.
\newblock \bibinfo{journal}{\emph{IEEE Transactions on Information Forensics
  and Security}} \bibinfo{volume}{14}, \bibinfo{number}{5}
  (\bibinfo{year}{2018}), \bibinfo{pages}{1280--1295}.
\newblock


\bibitem[\protect\citeauthoryear{Yang, Zhang, Hu, Hu, and Huang}{Yang
  et~al\mbox{.}}{2020}]%
        {yang2020vae}
\bibfield{author}{\bibinfo{person}{Zhong-Liang Yang}, \bibinfo{person}{Si-Yu
  Zhang}, \bibinfo{person}{Yu-Ting Hu}, \bibinfo{person}{Zhi-Wen Hu}, {and}
  \bibinfo{person}{Yong-Feng Huang}.} \bibinfo{year}{2020}\natexlab{}.
\newblock \showarticletitle{VAE-Stega: linguistic steganography based on
  variational auto-encoder}.
\newblock \bibinfo{journal}{\emph{IEEE Transactions on Information Forensics
  and Security}}  \bibinfo{volume}{16} (\bibinfo{year}{2020}),
  \bibinfo{pages}{880--895}.
\newblock


\bibitem[\protect\citeauthoryear{Ye, Jiang, and Huang}{Ye
  et~al\mbox{.}}{2019}]%
        {ye2019heard}
\bibfield{author}{\bibinfo{person}{Dengpan Ye}, \bibinfo{person}{Shunzhi
  Jiang}, {and} \bibinfo{person}{Jiaqin Huang}.}
  \bibinfo{year}{2019}\natexlab{}.
\newblock \showarticletitle{Heard more than heard: An audio steganography
  method based on gan}.
\newblock \bibinfo{journal}{\emph{arXiv preprint arXiv:1907.04986}}
  (\bibinfo{year}{2019}).
\newblock


\bibitem[\protect\citeauthoryear{Yin, Zhu, Liu, Fei, and Zhang}{Yin
  et~al\mbox{.}}{2018}]%
        {yin2018enhancing}
\bibfield{author}{\bibinfo{person}{Chuanlong Yin}, \bibinfo{person}{Yuefei
  Zhu}, \bibinfo{person}{Shengli Liu}, \bibinfo{person}{Jinlong Fei}, {and}
  \bibinfo{person}{Hetong Zhang}.} \bibinfo{year}{2018}\natexlab{}.
\newblock \showarticletitle{An enhancing framework for botnet detection using
  generative adversarial networks}. In \bibinfo{booktitle}{\emph{2018
  International Conference on Artificial Intelligence and Big Data (ICAIBD)}}.
  IEEE, \bibinfo{pages}{228--234}.
\newblock


\bibitem[\protect\citeauthoryear{You, Ying, Ren, Hamilton, and Leskovec}{You
  et~al\mbox{.}}{2018}]%
        {you2018graphrnn}
\bibfield{author}{\bibinfo{person}{Jiaxuan You}, \bibinfo{person}{Rex Ying},
  \bibinfo{person}{Xiang Ren}, \bibinfo{person}{William~L Hamilton}, {and}
  \bibinfo{person}{Jure Leskovec}.} \bibinfo{year}{2018}\natexlab{}.
\newblock \showarticletitle{GraphRNN: Generating Realistic Graphs with Deep
  Auto-regressive Models}. In \bibinfo{booktitle}{\emph{ICML}}.
\newblock


\bibitem[\protect\citeauthoryear{Yousefi-Azar, Varadharajan, Hamey, and
  Tupakula}{Yousefi-Azar et~al\mbox{.}}{2017}]%
        {yousefi2017autoencoder}
\bibfield{author}{\bibinfo{person}{Mahmood Yousefi-Azar},
  \bibinfo{person}{Vijay Varadharajan}, \bibinfo{person}{Len Hamey}, {and}
  \bibinfo{person}{Uday Tupakula}.} \bibinfo{year}{2017}\natexlab{}.
\newblock \showarticletitle{Autoencoder-based feature learning for cyber
  security applications}. In \bibinfo{booktitle}{\emph{2017 International joint
  conference on neural networks (IJCNN)}}. IEEE, \bibinfo{pages}{3854--3861}.
\newblock


\bibitem[\protect\citeauthoryear{Yu, Davis, and Fritz}{Yu
  et~al\mbox{.}}{2019}]%
        {yu2019attributing}
\bibfield{author}{\bibinfo{person}{Ning Yu}, \bibinfo{person}{Larry~S Davis},
  {and} \bibinfo{person}{Mario Fritz}.} \bibinfo{year}{2019}\natexlab{}.
\newblock \showarticletitle{Attributing fake images to gans: Learning and
  analyzing gan fingerprints}. In \bibinfo{booktitle}{\emph{Proceedings of the
  IEEE International Conference on Computer Vision}}.
  \bibinfo{pages}{7556--7566}.
\newblock


\bibitem[\protect\citeauthoryear{Yu and Canales}{Yu and Canales}{2019}]%
        {yu2019conditional}
\bibfield{author}{\bibinfo{person}{Yi Yu} {and} \bibinfo{person}{Simon
  Canales}.} \bibinfo{year}{2019}\natexlab{}.
\newblock \showarticletitle{Conditional lstm-gan for melody generation from
  lyrics}.
\newblock \bibinfo{journal}{\emph{arXiv preprint arXiv:1908.05551}}
  (\bibinfo{year}{2019}).
\newblock


\bibitem[\protect\citeauthoryear{Yu, Gong, Zhong, and Shan}{Yu
  et~al\mbox{.}}{2017}]%
        {yu2017unsupervised}
\bibfield{author}{\bibinfo{person}{Yang Yu}, \bibinfo{person}{Zhiqiang Gong},
  \bibinfo{person}{Ping Zhong}, {and} \bibinfo{person}{Jiaxin Shan}.}
  \bibinfo{year}{2017}\natexlab{}.
\newblock \showarticletitle{Unsupervised representation learning with deep
  convolutional neural network for remote sensing images}. In
  \bibinfo{booktitle}{\emph{International Conference on Image and Graphics}}.
  Springer, \bibinfo{pages}{97--108}.
\newblock


\bibitem[\protect\citeauthoryear{Zemel, Wu, Swersky, Pitassi, and Dwork}{Zemel
  et~al\mbox{.}}{2013}]%
        {zemel2013learning}
\bibfield{author}{\bibinfo{person}{Rich Zemel}, \bibinfo{person}{Yu Wu},
  \bibinfo{person}{Kevin Swersky}, \bibinfo{person}{Toni Pitassi}, {and}
  \bibinfo{person}{Cynthia Dwork}.} \bibinfo{year}{2013}\natexlab{}.
\newblock \showarticletitle{Learning fair representations}. In
  \bibinfo{booktitle}{\emph{International Conference on Machine Learning}}.
  \bibinfo{pages}{325--333}.
\newblock


\bibitem[\protect\citeauthoryear{Zeng, Lu, and Borji}{Zeng
  et~al\mbox{.}}{2017}]%
        {zeng2017statistics}
\bibfield{author}{\bibinfo{person}{Yu Zeng}, \bibinfo{person}{Huchuan Lu},
  {and} \bibinfo{person}{Ali Borji}.} \bibinfo{year}{2017}\natexlab{}.
\newblock \showarticletitle{Statistics of deep generated images}.
\newblock \bibinfo{journal}{\emph{arXiv preprint arXiv:1708.02688}}
  (\bibinfo{year}{2017}).
\newblock


\bibitem[\protect\citeauthoryear{Zhang, Zhou, Shumailov, and Papernot}{Zhang
  et~al\mbox{.}}{2020c}]%
        {zhang2020not}
\bibfield{author}{\bibinfo{person}{Baiwu Zhang}, \bibinfo{person}{Jin~Peng
  Zhou}, \bibinfo{person}{Ilia Shumailov}, {and} \bibinfo{person}{Nicolas
  Papernot}.} \bibinfo{year}{2020}\natexlab{c}.
\newblock \showarticletitle{Not My Deepfake: Towards Plausible Deniability for
  Machine-Generated Media}.
\newblock \bibinfo{journal}{\emph{arXiv preprint arXiv:2008.09194}}
  (\bibinfo{year}{2020}).
\newblock


\bibitem[\protect\citeauthoryear{Zhang, Xu, Li, Zhang, Wang, Huang, and
  Metaxas}{Zhang et~al\mbox{.}}{2017c}]%
        {zhang2017stackgan}
\bibfield{author}{\bibinfo{person}{Han Zhang}, \bibinfo{person}{Tao Xu},
  \bibinfo{person}{Hongsheng Li}, \bibinfo{person}{Shaoting Zhang},
  \bibinfo{person}{Xiaogang Wang}, \bibinfo{person}{Xiaolei Huang}, {and}
  \bibinfo{person}{Dimitris~N Metaxas}.} \bibinfo{year}{2017}\natexlab{c}.
\newblock \showarticletitle{Stackgan: Text to photo-realistic image synthesis
  with stacked generative adversarial networks}. In
  \bibinfo{booktitle}{\emph{Proceedings of the IEEE international conference on
  computer vision}}. \bibinfo{pages}{5907--5915}.
\newblock


\bibitem[\protect\citeauthoryear{Zhang, Chen, Wu, Chen, and Yu}{Zhang
  et~al\mbox{.}}{2019a}]%
        {zhang2019poisoning}
\bibfield{author}{\bibinfo{person}{Jiale Zhang}, \bibinfo{person}{Junjun Chen},
  \bibinfo{person}{Di Wu}, \bibinfo{person}{Bing Chen}, {and}
  \bibinfo{person}{Shui Yu}.} \bibinfo{year}{2019}\natexlab{a}.
\newblock \showarticletitle{Poisoning Attack in Federated Learning using
  Generative Adversarial Nets}. In \bibinfo{booktitle}{\emph{2019 18th IEEE
  International Conference On Trust, Security And Privacy In Computing And
  Communications/13th IEEE International Conference On Big Data Science And
  Engineering (TrustCom/BigDataSE)}}. IEEE, \bibinfo{pages}{374--380}.
\newblock


\bibitem[\protect\citeauthoryear{Zhang, Cormode, Procopiuc, Srivastava, and
  Xiao}{Zhang et~al\mbox{.}}{2017a}]%
        {zhang2017privbayes}
\bibfield{author}{\bibinfo{person}{Jun Zhang}, \bibinfo{person}{Graham
  Cormode}, \bibinfo{person}{Cecilia~M Procopiuc}, \bibinfo{person}{Divesh
  Srivastava}, {and} \bibinfo{person}{Xiaokui Xiao}.}
  \bibinfo{year}{2017}\natexlab{a}.
\newblock \showarticletitle{Privbayes: Private data release via bayesian
  networks}.
\newblock \bibinfo{journal}{\emph{ACM Transactions on Database Systems (TODS)}}
  \bibinfo{volume}{42}, \bibinfo{number}{4} (\bibinfo{year}{2017}),
  \bibinfo{pages}{1--41}.
\newblock


\bibitem[\protect\citeauthoryear{Zhang, Dong, and Liu}{Zhang
  et~al\mbox{.}}{2019b}]%
        {zhang2019invisible}
\bibfield{author}{\bibinfo{person}{Ru Zhang}, \bibinfo{person}{Shiqi Dong},
  {and} \bibinfo{person}{Jianyi Liu}.} \bibinfo{year}{2019}\natexlab{b}.
\newblock \showarticletitle{Invisible steganography via generative adversarial
  networks}.
\newblock \bibinfo{journal}{\emph{Multimedia tools and applications}}
  \bibinfo{volume}{78}, \bibinfo{number}{7} (\bibinfo{year}{2019}),
  \bibinfo{pages}{8559--8575}.
\newblock


\bibitem[\protect\citeauthoryear{Zhang, Yang, Yang, and Huang}{Zhang
  et~al\mbox{.}}{2021}]%
        {zhang2021provably}
\bibfield{author}{\bibinfo{person}{Siyu Zhang}, \bibinfo{person}{Zhongliang
  Yang}, \bibinfo{person}{Jinshuai Yang}, {and} \bibinfo{person}{Yongfeng
  Huang}.} \bibinfo{year}{2021}\natexlab{}.
\newblock \showarticletitle{Provably Secure Generative Linguistic
  Steganography}.
\newblock \bibinfo{journal}{\emph{arXiv preprint arXiv:2106.02011}}
  (\bibinfo{year}{2021}).
\newblock


\bibitem[\protect\citeauthoryear{Zhang, Chen, Gu, and Evans}{Zhang
  et~al\mbox{.}}{2020a}]%
        {zhang2020understanding}
\bibfield{author}{\bibinfo{person}{Xiao Zhang}, \bibinfo{person}{Jinghui Chen},
  \bibinfo{person}{Quanquan Gu}, {and} \bibinfo{person}{David Evans}.}
  \bibinfo{year}{2020}\natexlab{a}.
\newblock \showarticletitle{Understanding the intrinsic robustness of image
  distributions using conditional generative models}. In
  \bibinfo{booktitle}{\emph{International Conference on Artificial Intelligence
  and Statistics}}. PMLR, \bibinfo{pages}{3883--3893}.
\newblock


\bibitem[\protect\citeauthoryear{Zhang, Gan, Fan, Chen, Henao, Shen, and
  Carin}{Zhang et~al\mbox{.}}{2017b}]%
        {zhang2017adversarial}
\bibfield{author}{\bibinfo{person}{Yizhe Zhang}, \bibinfo{person}{Zhe Gan},
  \bibinfo{person}{Kai Fan}, \bibinfo{person}{Zhi Chen},
  \bibinfo{person}{Ricardo Henao}, \bibinfo{person}{Dinghan Shen}, {and}
  \bibinfo{person}{Lawrence Carin}.} \bibinfo{year}{2017}\natexlab{b}.
\newblock \showarticletitle{Adversarial feature matching for text generation}.
  In \bibinfo{booktitle}{\emph{Proceedings of the 34th International Conference
  on Machine Learning-Volume 70}}. \bibinfo{pages}{4006--4015}.
\newblock


\bibitem[\protect\citeauthoryear{Zhang, Liu, Ke, Lei, Li, Zhang, and
  Yang}{Zhang et~al\mbox{.}}{2019c}]%
        {zhang2019generative}
\bibfield{author}{\bibinfo{person}{Zhuo Zhang}, \bibinfo{person}{Jia Liu},
  \bibinfo{person}{Yan Ke}, \bibinfo{person}{Yu Lei}, \bibinfo{person}{Jun Li},
  \bibinfo{person}{Minqing Zhang}, {and} \bibinfo{person}{Xiaoyuan Yang}.}
  \bibinfo{year}{2019}\natexlab{c}.
\newblock \showarticletitle{Generative steganography by sampling}.
\newblock \bibinfo{journal}{\emph{IEEE Access}}  \bibinfo{volume}{7}
  (\bibinfo{year}{2019}), \bibinfo{pages}{118586--118597}.
\newblock


\bibitem[\protect\citeauthoryear{Zhang, Song, and Qi}{Zhang
  et~al\mbox{.}}{2018}]%
        {zhang2018decoupled}
\bibfield{author}{\bibinfo{person}{Zhifei Zhang}, \bibinfo{person}{Yang Song},
  {and} \bibinfo{person}{Hairong Qi}.} \bibinfo{year}{2018}\natexlab{}.
\newblock \showarticletitle{Decoupled learning for conditional adversarial
  networks}. In \bibinfo{booktitle}{\emph{2018 IEEE Winter Conference on
  Applications of Computer Vision (WACV)}}. IEEE, \bibinfo{pages}{700--708}.
\newblock


\bibitem[\protect\citeauthoryear{Zhang, Wang, Li, Honorio, Backes, He, Chen,
  and Zhang}{Zhang et~al\mbox{.}}{2020b}]%
        {zhang2020privsyn}
\bibfield{author}{\bibinfo{person}{Zhikun Zhang}, \bibinfo{person}{Tianhao
  Wang}, \bibinfo{person}{Ninghui Li}, \bibinfo{person}{Jean Honorio},
  \bibinfo{person}{Michael Backes}, \bibinfo{person}{Shibo He},
  \bibinfo{person}{Jiming Chen}, {and} \bibinfo{person}{Yang Zhang}.}
  \bibinfo{year}{2020}\natexlab{b}.
\newblock \showarticletitle{PrivSyn: Differentially Private Data Synthesis}.
\newblock \bibinfo{journal}{\emph{arXiv preprint arXiv:2012.15128}}
  (\bibinfo{year}{2020}).
\newblock


\bibitem[\protect\citeauthoryear{Zhao, Song, and Ermon}{Zhao
  et~al\mbox{.}}{2017}]%
        {zhao2017towards}
\bibfield{author}{\bibinfo{person}{Shengjia Zhao}, \bibinfo{person}{Jiaming
  Song}, {and} \bibinfo{person}{Stefano Ermon}.}
  \bibinfo{year}{2017}\natexlab{}.
\newblock \showarticletitle{Towards deeper understanding of variational
  autoencoding models}.
\newblock \bibinfo{journal}{\emph{arXiv preprint arXiv:1702.08658}}
  (\bibinfo{year}{2017}).
\newblock


\bibitem[\protect\citeauthoryear{Zhao, Dua, and Singh}{Zhao
  et~al\mbox{.}}{2018}]%
        {zhao2017generating}
\bibfield{author}{\bibinfo{person}{Zhengli Zhao}, \bibinfo{person}{Dheeru Dua},
  {and} \bibinfo{person}{Sameer Singh}.} \bibinfo{year}{2018}\natexlab{}.
\newblock \showarticletitle{Generating Natural Adversarial Examples}. In
  \bibinfo{booktitle}{\emph{International Conference on Learning
  Representations}}.
\newblock


\bibitem[\protect\citeauthoryear{Zheng, Zhou, Sheng, Xue, and Chen}{Zheng
  et~al\mbox{.}}{2018}]%
        {zheng2018generative}
\bibfield{author}{\bibinfo{person}{Yu-Jun Zheng}, \bibinfo{person}{Xiao-Han
  Zhou}, \bibinfo{person}{Wei-Guo Sheng}, \bibinfo{person}{Yu Xue}, {and}
  \bibinfo{person}{Sheng-Yong Chen}.} \bibinfo{year}{2018}\natexlab{}.
\newblock \showarticletitle{Generative adversarial network based telecom fraud
  detection at the receiving bank}.
\newblock \bibinfo{journal}{\emph{Neural Networks}}  \bibinfo{volume}{102}
  (\bibinfo{year}{2018}), \bibinfo{pages}{78--86}.
\newblock


\bibitem[\protect\citeauthoryear{Zhou, Gordon, Krishna, Narcomey, Fei-Fei, and
  Bernstein}{Zhou et~al\mbox{.}}{2019}]%
        {zhou2019hype}
\bibfield{author}{\bibinfo{person}{Sharon Zhou}, \bibinfo{person}{Mitchell
  Gordon}, \bibinfo{person}{Ranjay Krishna}, \bibinfo{person}{Austin Narcomey},
  \bibinfo{person}{Li~F Fei-Fei}, {and} \bibinfo{person}{Michael Bernstein}.}
  \bibinfo{year}{2019}\natexlab{}.
\newblock \showarticletitle{Hype: A benchmark for human eye perceptual
  evaluation of generative models}. In \bibinfo{booktitle}{\emph{Advances in
  Neural Information Processing Systems}}. \bibinfo{pages}{3449--3461}.
\newblock


\bibitem[\protect\citeauthoryear{Zhou, Peng, Yang, Wen, Xue, and Zhong}{Zhou
  et~al\mbox{.}}{2021}]%
        {zhou2021linguistic}
\bibfield{author}{\bibinfo{person}{Xuejing Zhou}, \bibinfo{person}{Wanli Peng},
  \bibinfo{person}{Boya Yang}, \bibinfo{person}{Juan Wen},
  \bibinfo{person}{Yiming Xue}, {and} \bibinfo{person}{Ping Zhong}.}
  \bibinfo{year}{2021}\natexlab{}.
\newblock \showarticletitle{Linguistic steganography based on adaptive
  probability distribution}.
\newblock \bibinfo{journal}{\emph{IEEE Transactions on Dependable and Secure
  Computing}} (\bibinfo{year}{2021}).
\newblock


\bibitem[\protect\citeauthoryear{Zhu, Kr{\"a}henb{\"u}hl, Shechtman, and
  Efros}{Zhu et~al\mbox{.}}{2016}]%
        {zhu2016generative}
\bibfield{author}{\bibinfo{person}{Jun-Yan Zhu}, \bibinfo{person}{Philipp
  Kr{\"a}henb{\"u}hl}, \bibinfo{person}{Eli Shechtman}, {and}
  \bibinfo{person}{Alexei~A Efros}.} \bibinfo{year}{2016}\natexlab{}.
\newblock \showarticletitle{Generative visual manipulation on the natural image
  manifold}. In \bibinfo{booktitle}{\emph{European Conference on Computer
  Vision}}. Springer, \bibinfo{pages}{597--613}.
\newblock


\bibitem[\protect\citeauthoryear{Zhu, Park, Isola, and Efros}{Zhu
  et~al\mbox{.}}{2017}]%
        {zhu2017unpaired}
\bibfield{author}{\bibinfo{person}{Jun-Yan Zhu}, \bibinfo{person}{Taesung
  Park}, \bibinfo{person}{Phillip Isola}, {and} \bibinfo{person}{Alexei~A
  Efros}.} \bibinfo{year}{2017}\natexlab{}.
\newblock \showarticletitle{Unpaired image-to-image translation using
  cycle-consistent adversarial networks}. In
  \bibinfo{booktitle}{\emph{Proceedings of the IEEE international conference on
  computer vision}}. \bibinfo{pages}{2223--2232}.
\newblock


\bibitem[\protect\citeauthoryear{Ziegler, Deng, and Rush}{Ziegler
  et~al\mbox{.}}{2019}]%
        {ziegler2019neural}
\bibfield{author}{\bibinfo{person}{Zachary~M Ziegler}, \bibinfo{person}{Yuntian
  Deng}, {and} \bibinfo{person}{Alexander~M Rush}.}
  \bibinfo{year}{2019}\natexlab{}.
\newblock \showarticletitle{Neural linguistic steganography}.
\newblock \bibinfo{journal}{\emph{arXiv preprint arXiv:1909.01496}}
  (\bibinfo{year}{2019}).
\newblock


\end{thebibliography}
}

\end{document}